\newcommand{\titel}{Noisy Networks, Nosy Neighbors: Inferring Privacy Invasive Information from Encrypted Wireless Traffic}
\newcommand{\art}{Bachelorarbeit} 
\newcommand{\ort}{Leipzig}

\newcommand{\fachgebiet}{Professur für Data Privacy and Security}
\newcommand{\fakultaet}{Fakultät für Mathematik und Informatik}
\newcommand{\institut}{Institut für Informatik}
\newcommand{\autor}{Bartosz Wojciech Burgiel}
\newcommand{\matrikelnr}{3777136}
\newcommand{\erstbetreuer}{Prof. Dr. Erik Buchmann}
\newcommand{\zweitbetreuer}{Victor Jüttner}
\newcommand{\jahr}{2025}

\newcommand{\eingereicht}{xx.xx.xxxx}

\documentclass[
	11pt,					
	DIV=10,
	ngerman,				
	a4paper,				
	oneside,				
	titlepage,				
	parskip=half,			
	headings=normal, 
	numbers=withendperiod, 
	listof=totoc,				
	bibliography=totoc,				
	index=totoc,				
	captions=tableheading,		
	final					
]{scrreprt}

\usepackage[
	automark,			
	headsepline,	
	ilines				
]{scrlayer-scrpage}
\usepackage{scrhack} 

\usepackage{pseudocode}
\usepackage{nicefrac}

\usepackage{subfigure}

\usepackage{dsfont}
\usepackage[english]{babel}

\usepackage[utf8]{inputenc}
\usepackage[T1]{fontenc}
\usepackage{textcomp} 
\usepackage{lmodern} 

\usepackage{multicol}

\usepackage[dvips,final]{graphicx}
\usepackage{wrapfig}
\graphicspath{{Bilder/}} 

\usepackage{amsmath,amsfonts,amsthm}

\usepackage{makeidx}

\usepackage{setspace}
\usepackage{geometry}

\usepackage{rotating} 

\usepackage[intoc]{nomencl}

\usepackage{floatflt}

\usepackage{listings}
\usepackage{xcolor} 
\definecolor{hellgelb}{rgb}{1,1,0.9}
\definecolor{colKeys}{rgb}{0,0,1}
\definecolor{colIdentifier}{rgb}{0,0,0}
\definecolor{colComments}{rgb}{1,0,0}
\definecolor{colString}{rgb}{0,0.5,0}
\usepackage{url}
\usepackage{multirow}

\usepackage[autocite=inline, sorting=none]{biblatex}
\bibliography{quellen.bib} 

\usepackage{csquotes} 

\usepackage{caption}

\usepackage[
bookmarks,
bookmarksopen=true,
pdftitle={\titel},
pdfauthor={\autor},
pdfcreator={\autor},
pdfsubject={\titel},
pdfkeywords={\titel},
colorlinks=true,
linkcolor=black, 
anchorcolor=black,
citecolor=black, 
filecolor=black, 
menucolor=black, 
urlcolor=black, 
plainpages=false,
pdfpagelabels,
hypertexnames=false,
linktocpage 
]{hyperref}

\usepackage{chngcntr}

\usepackage{longtable}
\usepackage{array}
\usepackage{ragged2e}
\usepackage{lscape}

\usepackage{supertabular}

\newcolumntype{w}[1]{>{\raggedleft\hspace{0pt}}p{#1}}

\usepackage{paralist}

\usepackage{tablefootnote}
\usepackage{lipsum}

\makeindex
\makenomenclature

\ihead{\textit{\headmark}}
\ohead{}
\setheadwidth[0pt]{textwithmarginpar} 

\ifoot{\autor \\ \matrikelnr}
\ofoot{\pagemark}
\setfootwidth[0pt]{text}

\definecolor{gray}{rgb}{0.9,0.9,0.9}

\counterwithout{footnote}{chapter}

\begin{document}
\setcounter{secnumdepth}{3}
\setcounter{tocdepth}{3}

\ofoot{}
\thispagestyle{plain}
\begin{titlepage}

\begin{center}
\includegraphics[height=7cm]{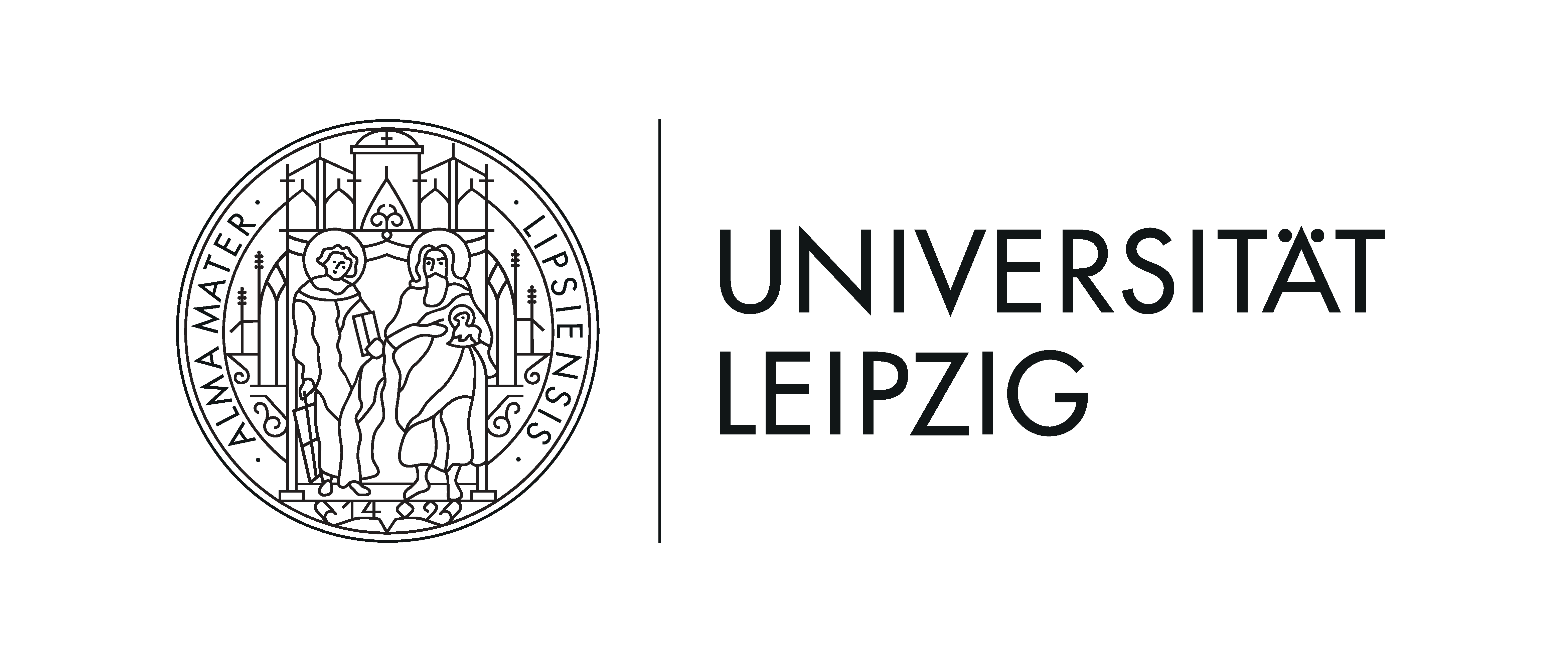}\\[2.5ex]

\institut\\
\fakultaet\\
\fachgebiet\\[6ex]

\textbf{\large\titel}\\[1.5ex]
\art\\[6ex]

\normalsize
vorgelegt von:\\
\autor\\[1.5ex]
Matrikelnummer:\\
\matrikelnr\\[1.5ex]
Betreuer:\\
\erstbetreuer\\
\zweitbetreuer\\[1.0ex]
\end{center}


\begin{center}
\copyright\ \jahr\\[1.0ex]
\end{center}

\singlespacing
\small
\noindent Dieses Werk einschließlich seiner Teile ist \textbf{urheberrechtlich geschützt}. Jede Verwertung außerhalb der engen Grenzen des Urheberrechtgesetzes ist ohne Zustimmung des Autors unzulässig und strafbar. Das gilt insbesondere für Vervielfältigungen, Übersetzungen, Mikroverfilmungen sowie die Einspeicherung und Verarbeitung in elektronischen Systemen.

\end{titlepage}

\section*{Abstract}
\label{sec:Abstract}

This thesis explores the extent to which passive observation of wireless traffic in a smart home environment can be used to infer privacy-invasive information about its inhabitants. Using a setup that mimics the capabilities of a nosy neighbor in an adjacent flat, we analyze raw 802.11 packets and Bluetooth Low Energy advertisemets. From this data, we  identify devices, infer their activity states and approximate their location using RSSI-based trilateration. Despite the encrypted nature of the data, we demonstrate that it is possible to detect active periods of multimedia devices, infer common activities such as sleeping, working and consuming media, and even approximate the layout of the neighbor's apartment. Our results show that privacy risks in smart homes extend beyond traditional data breaches: a nosy neighbor behind the wall can gain privacy-invasive insights into the lives of their neighbors purely from encrypted network traffic.  
\section*{Acknowledgments}
\label{sec:Danksagung}

I would like to express my gratitude to my supervisor Victor Jüttner and Prof. Buchmann for the patience and extremely detailed feedback over the course of the thesis supervision. 

I would not finish my journey without the support of my friends and collegues. Most importantly I would like to thank you Beata for your patience and support during the most challenging moments of the thesis. I am also grateful to Annika and Maya for their participation the experiment. A special thanks goes to, my coworkers Adrian, Marcel, Louica, Kai and Lukas for their help and encouragement along the way. 

\newpage
\ofoot{\pagemark}

\pagenumbering{Roman}

\tableofcontents			

\printnomenclature

\listoffigures					
\listoftables					


\clearpage
\pagenumbering{arabic}

\chapter{Introduction}
\label{cha:Introduction}
\section{Motivation}
\label{sec:motivation}

Smart home devices are ubiquitous by design. To ensure this property, most of them communicate wirelessly. Home automation significantly increases the quality of life and convenience, but also imposes a severe risk for privacy. Consider the following scenario: an inhabitant of a smart home wakes up, puts on their smart watch and turns on a smart light bulb to simulate the sunrise. They play some music through their Bluetooth speaker and begin their day. After getting ready, they turn off the lamp and music, activate security cameras, turn on a vacuum cleaner for the daily routine and before they leave for work, they apply a new wireless insulin patch. At the end of the day, they come back from work and prepare a meal using their smart cooking assistant. In the evening, they watch some television on their smart TV before playing some online games on their gaming console with their friends. 

Meanwhile, a noisy neighbor passively monitors the wireless traffic without actively interacting with the payloads or devices themselves in an adjacent apartment. They have three cheap WiFi antennas which are spatially distributed along the shared wall. Since the emitted WiFi traffic is encrypted, the attacker is unable to see what their neighbor is doing. Instead, they observe when network activity occurs and can measure its intensity by observing how many network packets are being sent and received by each device. Then, since their antennas are spatially distributed, they can estimate the position of victim's devices by performing trilateration calculations using the varying signal strengths. Due to the physical obstructions and a black-box nature of this scenario, the attacker would not be able to approximate the exact position of their neighbor's devices. However, they can still see the general area or direction in which the devices are located. This is sufficient to split the apartment into semantic regions such as kitchen, bathroom or an office. If the neighbor wears smart wearable devices around the house, their approximated location can be tracked nearly in real time. Lastly, since many typical smart home devices have Bluetooth capabilities, the nosy neighbor can listen to device's advertisements to enumerate them. 

Having all of this information, the passive observer can infiltrate the secrecy of a home and peek inside to see what their neighbor is doing. By correlating activity of different devices, they can deduce specific behaviors of their victim. For example, increased network activity in their cooking assistant tells that they are most likely in the kitchen preparing a meal. Then, consistent traffic from their computer at usual working times implies that the neighbor is working from home. Detecting new devices in their network which produce similar network traffic fingerprints to a smartphone is an indicator for them having visitors. 

This scenario presents an opportunity for a nosy neighbors to acquire information about their victims that they likely do not intend to disclose. This knowledge can be gained with absolutely no interaction or contact with the victim, as long as they are withing the range of their WiFi antennas. It is as if the walls between flats were made of frosted glass - neighbor could observe and approximate what they are doing and where they are. 

Aside from privacy risks, security concerns arise in this scenario. An attacker who acquired a set of devices, which are installed in a smart home, can use their knowledge to execute dedicated attacks. For one, they would know the specific companies which produced the devices. This opens up potential for spear-phishing attacks, if the attacker already has some contact information about the victim. This prerequisite is not far fetched, if the attacker and the victim are neighbors. Even if that's not the case, victims' email or full name could be found out from OSINT, since the attacker could know their surname from the intercom label. For instance, the adversary would deduce that the smart home environment includes an LG Smart TV. Then, by incorporating the specific version and hostname of the TV, they could construct a very convincing phishing email.  

Knowing the exact device model and version of models can aid the adversaries in narrowing down a list of vulnerabilities of the smart home. For example if a smart lock reveals its manufacturer or model in the broadcasted device advertisement data, the attacker could research for weak points of this specific door lock. Similarly, security systems such as cameras or motion sensors can also be compromised. The received signal strength from these sensors can be used by the attackers to localize these devices and find blind spots in surveillance coverage. 

Admittedly, the presented scenarios rely on many prerequisites such as effective inventory attack which reveals all of the devices, their approximate location and that they disclose their device information. Moreover, the adversary would have to remain undetected during the recon phase, especially when probing for location information, which could be difficult. However, the described scenarios are nonetheless realistic and can aid criminals in their attacks. Security breaches in IT rarely rely on one weakness or vulnerability of a system component. It is always a chain of poor decisions in the system design, where the attackers systematically exploit them and progress step by step into the system. So a burglar might not deliver a no-click exploit to remotely open a smart lock, but knowing all of the other security-sensitive information, it might get them the first foothold into that smart home environment. Every seemingly minor exposure contributes to a larger attack surface.

\section{Problem Statement}

Spying on inhabitants of a smart home using encrypted wireless traffic has been done before. Privacy analysis of smart homes and traffic based human activity recognition are established and active research fields. In other words, an attacker with the goal of learning the most about a smart home inhabitant using side-channel information, has many methods and techniques to choose from. Currently, no study has considered a passive eavesdropper in an adjacent flat. This work investigates how well these methods can be applied in the context of a neighbor behind the wall. Taking obstructions and real-life limitations into account, we will try to answer the following question: "What types of information can a nosy neighbor behind a wall infer from wireless traffic through multiple side-channel indicators?". To better approach this problem, we subdivide this question into the following research questions: "What can your neighbor learn about your smart home?", "Can your neighbor know where you are?" and finally: "What can your neighbor learn about you?". 

To evaluate this threat, a practical experiment will be conducted in a simulated smart home environment. One room of a five-room apartment will be utilized as a monitoring room where three WiFi and BLE sniffers will be installed. Then, remaining rooms will be equipped with smart home devices such as light bulbs and plugs. Each device will be used according to the typical usage patterns. This setup will mimic living in a smart home environment where a neighbor behind the wall is monitoring the traffic. 

To determine what a nosy neighbor could learn about its victims from their encrypted wireless traffic, first we determine which devices are installed in the smart home. Knowing the devices, we will analyze their traffic to derive their state. Then, we will analyze the received signal strength from these devices and estimate how accurately this information can be used to infer the location of individual devices and to derive a floor plan of the apartment. Lastly, having the information about devices and their approximate location, human activity recognition will be performed. 

For completeness, general-purpose devices such as smartphones, tablets and laptops will also be considered in this work. While these devices would typically not be classified as smart devices, they still leak valuable information. An attacker with the motivation to learn the most about their neighbor could infer facts about the neighbors' life beyond their smart home.

The experiment will adhere to the black-box nature of this scenario - despite the fact that devices are known beforehand, analysis will only consider information that was acquired using OSINT research. For completeness, wherever applicable, a white-box analysis will also be completed to show what the most successful attacker could determine. 
 
\section{Structure}

First, necessary technical background information as well as all related work is presented. This section goes into detail about the specifics of the wireless technologies which will be used to conduct the analysis. After that the current state of research and related work will be presented. 

Having the necessary foundations and context for the research, we will present the methodology, i.e. what is the process to set up the monitoring environment and the specific smart home setup. Then, this chapter will cover all the used network traffic capturing technique and the data processing pipeline. 

After that, the results of the analyses will be presented. The research question will be divided into three subquestions, each focusing on one aspect of the research question. First we will present what the neighbor can learn about the smart home itself, then we will present all about localization and lastly we'll perform human activity recognition. The chapter concludes with a case study presenting reconstruction of events during a visit. 

Discussion of the results follows after. In that chapter, the concrete findings are interpreted in the context of the original research question. We summarize the key findings of this work and discuss their impact and realism of the nosy neighbor attack. Then, limitations and future research possibilities are presented.  

\chapter{Background and Related Work}
\label{cha:Background}

To properly frame the research conducted in this work, it is necessary to present a general introduction of the relevant concepts and prior work. This section will cover the necessary technical foundations of wireless technologies. Later, we will present the current state of research to introduce existing attacks on privacy in smart home contexts. This will highlight the research gap, into which this work contributes. 

\section{Background and Related Work}

This section lays out the conceptual foundations of this work. First, the relevant wireless technologies, will be introduced. Specifically, we will cover the technical background for WiFi and Bluetooth Low Energy (BLE). Since these technologies are very complex, this section will focus solely on the relevant aspects for this work.

To present this technology, we will begin by elaborating the motivation for using WiFi in smart home environments as well as its general concepts. Then we will present the necessary specifics of this standard beginning with the Layer 1 aspects of the OSI Model, i.e. radio wave transmission and the Radiotap Header. After that, a general structure of a WiFi frame will be presented together with some selected frame types which are especially relevant for future analysis. Having the necessary foundations about WiFi, we will provide a brief introduction into Bluetooth Low Energy (BLE) technology with its relevant features.

\subsection{WiFi}
\label{sec:wireless}

Wireless Fidelity (WiFi) technologies are essential for the ubiquitous nature of smart home devices. Wired connections are infeasible for certain device types such as smart light bulbs. Additionally, some sensors need to be placed flexibly throughout the home, making a wired connection to a gateway or router impractical. Hence, most smart home devices use wireless protocols for communication. This section provides the necessary background information about WiFi.

The following sections use the OSI Model \cite{1094702} to describe different elements of WiFi. It is important to reiterate what this network layer classification states and how it relates to this research. Table \ref{tab:osi_layer} presents what each layer of the OSI Model represents.

\begin{table}[h]
	\centering
	\renewcommand{\arraystretch}{1.3} 
	\captionsetup{skip=1ex}
	\setlength{\tabcolsep}{8pt}
	\begin{tabular}{|c|c|p{0.55\linewidth}|}
		\hline
		\textbf{Layer} & \textbf{Description} & \textbf{Interpretation in this work} \\ \hline
		1 & Physical Layer & Radio transmission, radiotap header \\ \hline
		2 & Data Link Layer & WiFi traffic, addressing and plaintext information \\ \hline
		3 - 7 & Payload & Encrypted \\ \hline
	\end{tabular}
	
	\vspace{1em}
	\caption{Interpretation of the layers in the OSI Model in the context of this work.}
	\label{tab:osi_layer}
\end{table}

\subsubsection{General Information}
\label{cha:wifi}

WiFi is a suitable communication protocol for smart home devices for a number of reasons. For one, it operates with low power consumption and offers a high bandwidth communication. The typical range of WiFi is generally sufficient for smart home environments, with no need for deploying range extenders. Then, most households already have WiFi in place, which provides easy integration of devices into the network. Lastly, WiFi enables connection to the internet, which allows remote access. This is especially useful for devices such as cameras or other smart security enhancements. 

WiFi technology is based on the well established 802.11 standard \cite{9363693} which has been in use since the early 2000s. This technology has evolved into multiple generations (WiFi-\{0..8\}) as is subdivided into different standards (802.11\{b,a,g,n,ac,ax,be,bn\}). They all differ in their supported bandwidth, utilized radio frequency ranges and capabilities. This research will focus on the 802.11n standard, also known as WiFi 4. As of 2019, it is the most popular WiFi standard in IoT systems \cite{8847082}. This WiFi variant operates in 2.4 GHz and 5 GHz frequency bands. Those differ not only in the maximum possible bandwidth, where 5 GHz is generally faster, but also in the interference robustness, legacy device compatibility and utilized frequency range.

\begin{figure}[h]
	\centering
	\includegraphics[width=\linewidth]{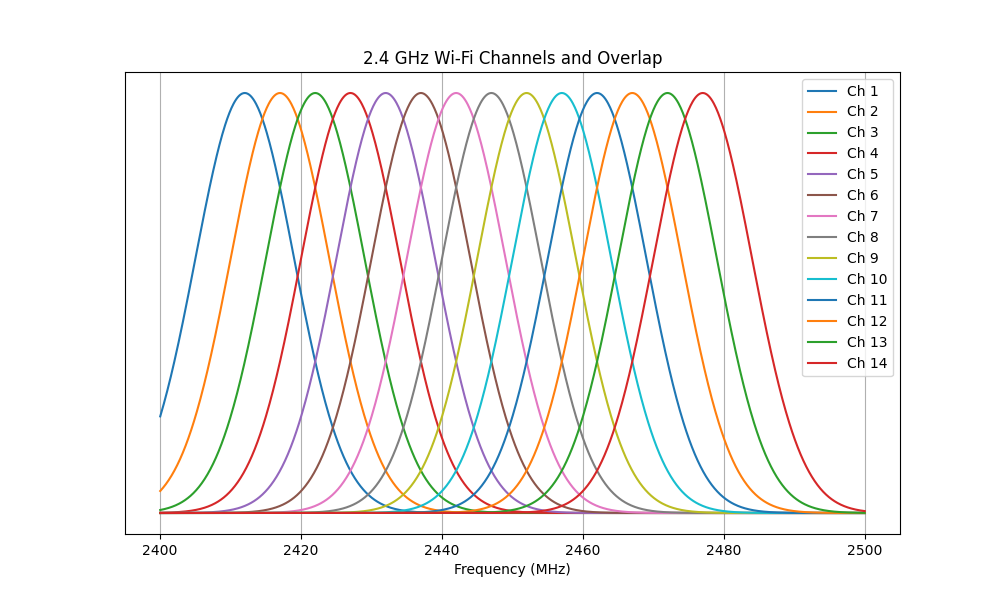}
	\caption{Visualization of all channels in a 2.4 GHz spectrum with 20 MHz channel width.}
	\label{fig:channel_centers_24ghz}
\end{figure}

Wi-Fi uses channelization \cite{7786995}, a common technique in telecommunication technology, which separates the communication channels in order to avoid collisions, and reduce network congestion. In the 2.4 GHz band, a station (STA), i.e. 802.11 capable device, can officially operate across 14 channels, each one being 20 MHz or 40 MHz wide \cite{7786995}. The exact frequency range differs slightly in some countries like Japan and the US, due to regulatory requirements \cite{7786995}. Europe utilizes channels 1-13. Each channel center is separated by 5 MHz, having the center of channel 1 at 2412 MHz, channel 2 at 2417 MHz and so on \cite{7786995}. Figure \ref{fig:channel_centers_24ghz} shows the channels in this frequency range. It is visible how the channels overlap.

\begin{figure}[h]
	\centering
	\includegraphics[width=\linewidth]{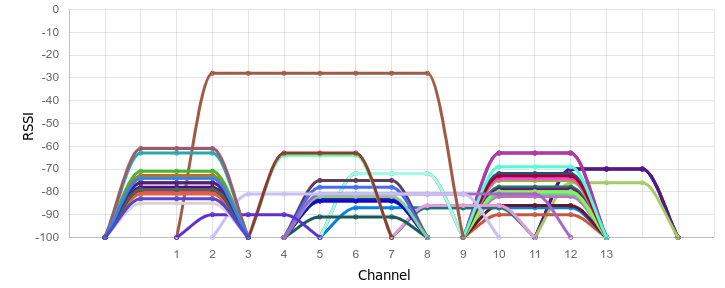}
	\caption{Distribution of used channels in a routers vicinity. Each unique color represents a different WiFi router. The width of each shape describes the channel width, the narrower being 20 MHz and the wider being 40 MHz.}
	\label{fig:local_channel_distribution}
\end{figure}

This overlap phenomenon is what causes the quality of WiFi to degrade in densely populated areas, with numerous WiFi networks operating at the same frequency. Modern WiFi routers are capable of dynamically detecting busy channels and steering the communication to other, less populated channels. This mechanism is sometimes referred to as channel-hopping. Figure \ref{fig:local_channel_distribution} displays a scan performed by a Vodafone Station WiFi router, which reveals a relatively uneven distribution of used channels in the router's vicinity.  

\begin{figure}[h]
	\centering
	\includegraphics[width=\linewidth]{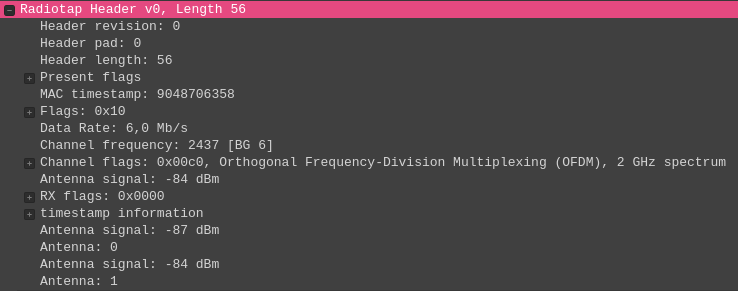}
	\caption{\textit{Radiotap Header} of a 802.11n frame viewed in \textit{Wireshark}.}
	\label{fig:wireshark_radio}
\end{figure}

The specifics about the Layer 1 transport of the WiFi packets, such as the channel, data rate and signal strength, can be decoded using the \textit{Radiotap} \cite{radiotap, radiotap-manpage} standard. This data can then be decoded and visualized using the network analysis tool \textit{Wireshark} \cite{wireshark}. Information, which this header provides, is not included in any 802.11 frame and is derived from the raw radio signal using the system's WiFi interface. Figure \ref{fig:wireshark_radio} shows the typical contents of a \textit{Radiotap Header} as shown in \textit{Wireshark}. 

The most valuable information for this research, present in the Radiotap Header, is the antenna signal, also known as Received Signal Strength Indicator (RSSI). This value represents the power of the signal received by the radio antenna, which can be used to locate a device using multi-laterization algorithms. Moreover, the header also contains the channelization information, i.e. at which frequency (=channel) the communication occurs. Last important information for this research is the data rate, i.e. the transmission rate of the sent data.

\subsubsection{802.11n}
\label{cha:80211n}
The 802.11n media access control frames, i.e. individual packets, are categorized into three main types: management, control and data. Management frames (i.e. beacon, probe request/response, authentication) handle network discovery and association. Control frames (i.e., ACK, RTS/CTS, PS-Poll) are used for medium access and transmission coordination. Data frames carry the actual payload between STAs. The structure of individual types or subtypes is not relevant for this research, as we will solely focus on selected frame fields, which are generally present in every frame type. These fields include the addressing fields as well as the raw payload.

Unlike higher-level communication protocols such as IP or TCP, which always contain only one value for the source and destination of the packet (either IP address or port), 802.11n addressing of individual frames is dynamic and depends on frame control header. For different configurations and frame types, the source and destination addresses can be composed of two different mac addresses. Generally, we distinguish the following address types in the 802.11n addressing:

\begin{multicols}{2}
	\begin{itemize}
		\item Source Address (SA)
		\item Destination Address (DA)
		\item Transmitter Address (TA)
		\item Receiver Address (RA)
	\end{itemize}
\end{multicols}

\begin{figure}[h]
	\centering
	\includegraphics[width=0.7\linewidth]{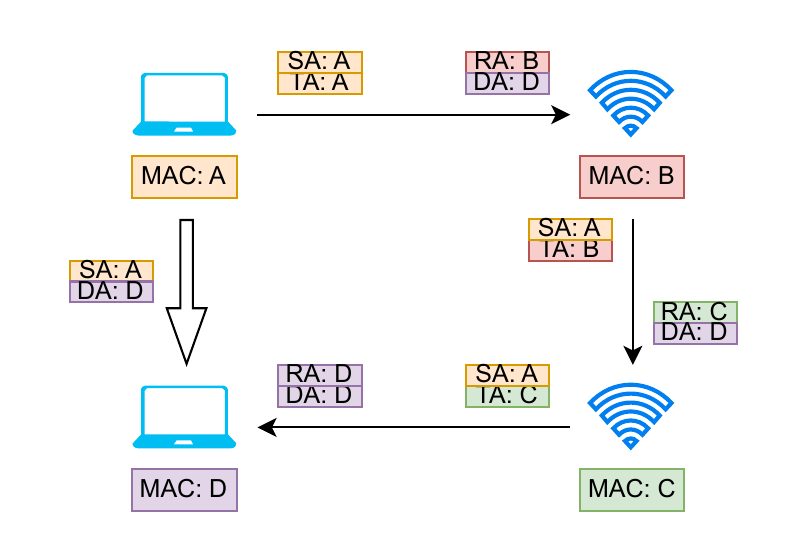}
	\caption{Communication workflow between two devices, routed through two relays. Big arrow from device A to D represents the higher level IP traffic. The values for TA/RA changes at every node, whereas SA/DA remains constant.  }
	\label{fig:relay_communication}
\end{figure}

In this research we will consider the source (SA) and transmitter (TA) addresses as well as destination (DA) and receiver (RA) addresses to be synonymous. The distinction for these address types is relevant only for complex WiFi topologies such as \textit{Mesh WiFi} or networks with repeaters. There, the higher-level SA and DA fields correspond with the mac addresses of the STAs which communicate with each other, while the lower-level TA and RA fields hold the mac addresses of the relays which connect the two devices. This utilization of the relay addresses is analogous to the IP address/mac address of LAN networks. Figure \ref{fig:relay_communication} shows the values for each address types in a WiFi network topology with two relays. Moreover, if a STA has multiple WiFi antennas, the TA/RA addresses will differ from the SA/DA and will correspond to the specific WiFi interface which is being used. This work will focus on the star topology, where no relays are installed. 

\begin{figure}[h]
	\centering
	\includegraphics[width=\linewidth]{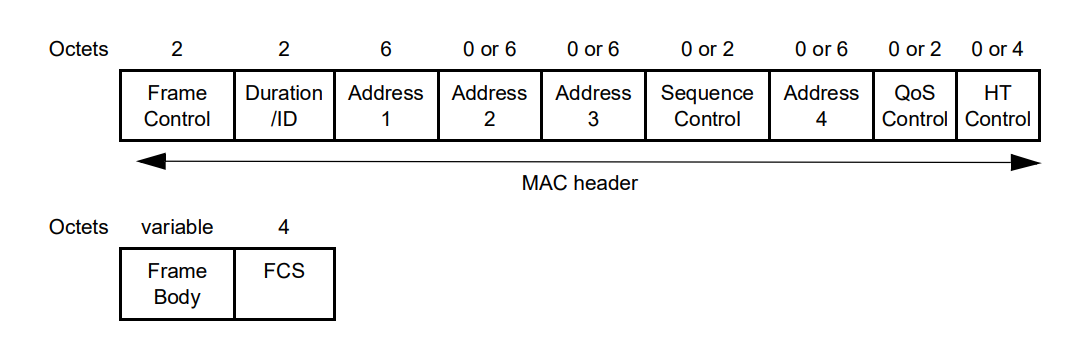}
	\caption{General 802.11n frame format. Adopted from \cite{9363693}.}
	\label{fig:80211_general_frame}
\end{figure}

A general frame structure is illustrated in Figure \ref{fig:80211_general_frame}. Some fields, such as \textit{Address 2}, are optional, because they are not required in every frame type. For example, a beacon management frame, which is used for AP advertisement, does not have a destination/receiver address (commonly found in the \textit{Address 3} field), because it is a broadcast frame meant to reach every device in its range. While the detailed specifics of each 802.11n frame type extend beyond the scope of this research, it is still important to introduce selected frame types which will be analyzed in this work. These frame types include the data frame type and the probe request frames.

\subsubsection{Data Frame}
\label{cha:data_frame}

\begin{figure}[h]
	\centering
	\includegraphics[width=\linewidth]{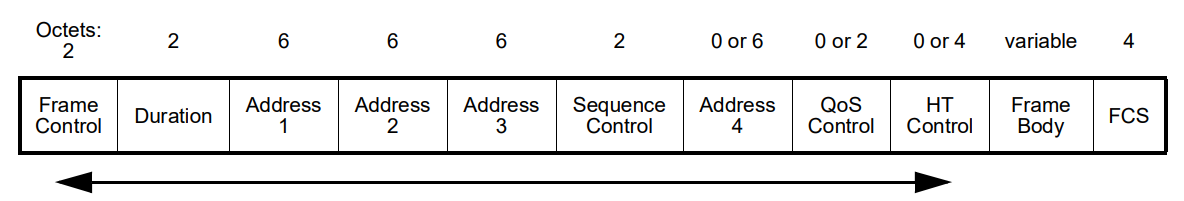}
	\caption{General 802.11n data frame format. Adopted from \cite{9363693}.}
	\label{fig:80211_data_frame}
\end{figure}

Data frames \cite{9363693} are the packets which are most useful in this work's context. These frames convey the actual payload, i.e. encrypted raw data which the device sends to its receiver. Typically the content of these data frames consists of raw IP packets. However, due to our assumed scenario of a passive eavesdropper, we ignore the specific contents of the payload field and consider only observable features such as length or throughput over time.

A general 802.11n data frame format is defined as shown in Figure \ref{fig:80211_data_frame}. The structure does not deviate from the general format, except for the first three \textit{Address} fields not being optional. The frame control field's last two bits (bitmask=0x3) describe the Distribution System (DS) \cite{9363693} status flag which determines the direction of a data frame. This flag holds values for the To DS and From DS indicators. The combinations of these two values can be interpreted as presented in Table \ref{tab:ds_status_flag}. 

\vspace{3ex}

\begin{table}[h]
	\centering
	\renewcommand{\arraystretch}{1.3} 
	\captionsetup{skip=1ex}
	\setlength{\tabcolsep}{8pt}
	\begin{tabular}{|c|c|p{0.55\linewidth}|}
		\hline

		\textbf{From DS / To DS} & \textbf{Hex Value} & \textbf{Interpretation in this work} \\ \hline
		0 / 0 & 0x0 & Exchange between two devices in a peer-to-peer fashion, or broadcast. \\ \hline
		0 / 1 & 0x1 & Data frame is sent from STA to AP, i.e., uplink traffic. \\ \hline
		1 / 0 & 0x2 & Data is sent from the AP to STA, i.e., downlink traffic.  \\ \hline
		1 / 1 & 0x3 & Communication between two APs. \\ \hline
	\end{tabular}

	\vspace{1ex}
	\caption{Interpretation of the DS status flag in 802.11n.}
	\label{tab:ds_status_flag}
\end{table}

The correlation of individual \textit{Address} fields to the different address types depends on the values of the DS status flag as well as other frame control fields. This mapping is complicated and not directly relevant for this research. Therefore an overview of all possible mappings is be omitted. However, network packet analysis tools such as \textit{Wireshark} and \textit{scapy} \cite{scapy} automatically interpret and assign the appropriate address types, simplifying the analysis process.

\begin{figure}[h]
	\centering
	\includegraphics[width=0.8\linewidth]{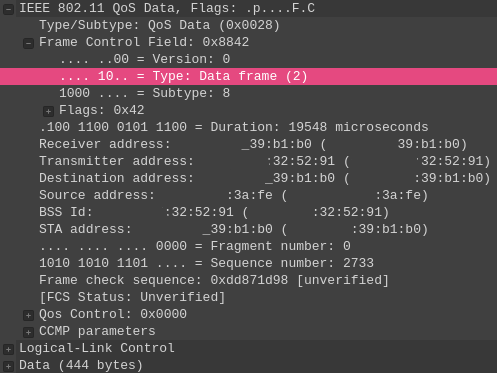}
	\caption{Observed 802.11n data frame viewed using \textit{Wireshark}. Parts of mac addresses are obfuscated for privacy.  }
	\label{fig:wireshark_data_frame}
\end{figure}

A raw data frame, viewed using \textit{Wireshark}, presents itself as shown in Figure \ref{fig:wireshark_data_frame}. The order of the address fields in the visual representation  in \textit{Wireshark} deviates from the general 802.11n definition presented in Figure \ref{fig:80211_data_frame}. The respective \textit{Wireshark} values correspond to the following 802.11n fields:

\begin{itemize}
	\item Receiver/Destination/STA address corresponds to \textit{Address 1}
	\item Transmitter/BSS Id address corresponds to \textit{Address 2}
	\item Source address corresponds to \textit{Address 3}.
\end{itemize}

\begin{figure}[h]
\centering
\includegraphics[width=0.6\linewidth]{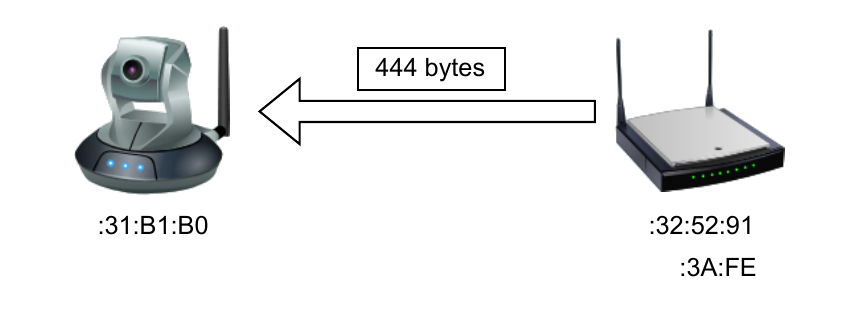}
\caption{Visualization of the observed data frame in \textit{Wireshark}. }
\label{fig:wireshark_communication}
\end{figure}

Based on the data frame represented in Figure \ref{fig:wireshark_data_frame} and its DS status, which can be derived from the \textit{Flags} field in the frame control using bitwise arithmetic (0x42 \& 0x3 = 0x2), we see that the frame represents downlink traffic. The STA with mac address \textbf{:31:B1:B0} (DA) receives 444 bytes from its AP with mac addresses \textbf{:3A:FE} (SA) and \textbf{:32:52:91}. The data transfer is visualized in Figure \ref{fig:wireshark_communication}. 

Every data frame contains the Basic Service Set Identifier (BSSID) field, which holds the mac address of its AP.

\subsubsection{Probe request}
\label{cha:probe_request}
\begin{figure}[h]
	\centering
	\includegraphics[width=0.6\linewidth]{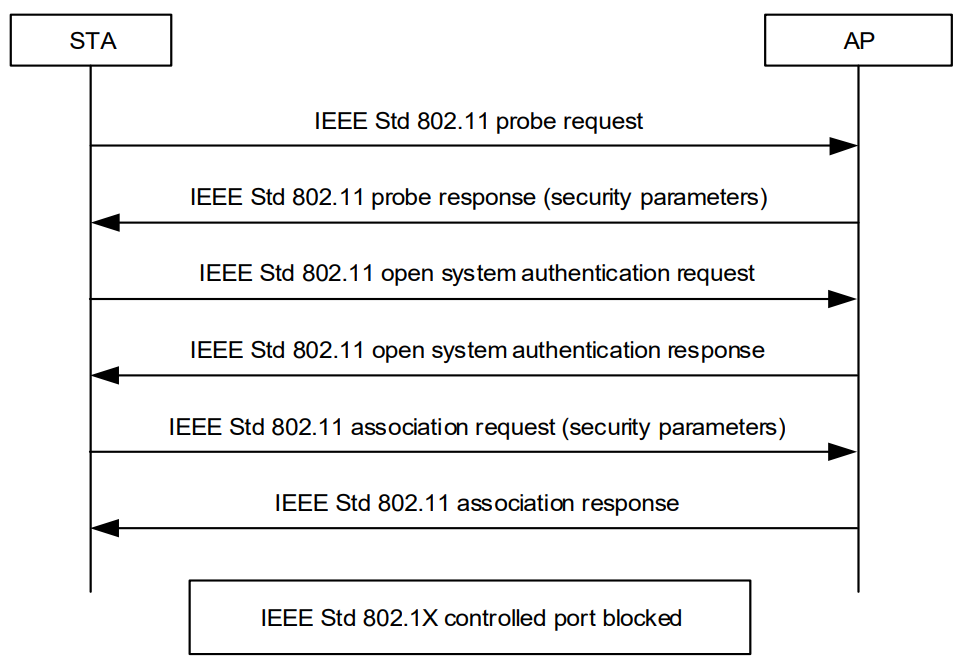}
	\caption{Association handshake in 802.11n. The last field of this sequence diagram represents data exchange. Adopted from \cite{9363693}. }
	\label{fig:80211_assoc}
\end{figure}

Another frame type which will be directly analyzed in this research is a subtype of the management frames - probe request frame \cite{9363693}. Its general purpose is AP discovery. A STA sends probe requests to search for trusted APs in its proximity. It usually contains information about the probed AP, such as its network name, i.e. the Service Set Id (SSID) and capabilities. In this research we will focus on the SSID element of this frame. If the probed AP receives the probe request, it responds with a probe response, which again contains the SSID and all capabilities of the AP. After this exchange, the STA decides to initiate the association (connection) handshake to the AP based on, for example, the offered security or its RSSI \cite{android-wifi}. Figure \ref{fig:80211_assoc} visualizes the handshake. 

This information has privacy implications, since it leaks information about the APs which a STA trusts, i.e. has connected to before. Depending on the WiFi driver or the operating system of the AP, a device will periodically send probe requests depending on the state of the device. For example, Android devices decide when to conduct a connectivity scan, i.e. send probe requests, based on the screen being on or off and the device having an active WiFi connection \cite{android-wifi}. Typically, an Android STA scans for saved networks \cite{android-wifi}. Since this packet's contents are transmitted as plain-text, a passive eavesdropper is able to sniff the SSIDs which a STA probes for. 

In reality, it is a difficult task to obtain a full list of saved APs of a device from the probed SSIDs. Android provides a suggestion API for WiFi discovery \cite{android-suggestion}, so smartphone manufacturers can define their own strategies for network discovery. Then, devices typically do not probe for all of the saved networks. Developers can optimize the algorithm for scanning the most often used networks or correlate the SSIDs to their approximate location. The 802.11 specification recognizes probing for saved network as a privacy risk \cite{9363693} and recommends periodically randomizing the mac address of the STA. This would make profiling and subsequent tracking of probing devices more challenging.

\begin{figure}[h]
	\centering
	\includegraphics[width=\linewidth]{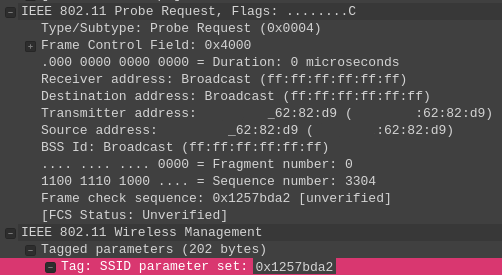}
	\caption{Probe request frame in \textit{Wireshark}. The mac addresses as well as the SSID are obfuscated for privacy. }
	\label{fig:probe_request_wireshark}
\end{figure}

A captured probe request, displayed in \textit{Wireshark}, presents itself as shown in Figure \ref{fig:probe_request_wireshark}. Here, the destination address and SSID is set to \textbf{FF:FF:FF:FF:FF:FF}. This address is reserved for broadcast, such that it is received by all nearby WiFi-capable devices. Then, the source address \textbf{:62:82:D9} represents the actual mac address of the device probing for the given network. Lastly, we can see the SSID of the network, here covered with the hex representation of the frame check sequence. The network name is represented in clear text with no encoding or encryption.

\subsubsection{Side-channel information}

Nowadays, WiFi traffic is usually encrypted using secure protocols such as WPA\{2,3\}. WiGLE.net \cite{WiGLE}, a wardriving database provides insight into the popularity of encryption protocols over time. Wardriving describes the act of recording broadcasted AP advertisements and their GPS position while driving a car. As of 2025, at least 75\% of observed WiFi Access Points (APs) offer WPA2 or WPA3 encryption protocols \cite{WiGLE-Encr}. They are considered secure, because WPA2 utilizes the AES primitive for encryption with a unique 128bit key. WPA3 improves it's predecessor's security by offering longer encryption keys (196-bit) for its AES cipher and offering a more secure key exchange handshake. However, despite WiFi traffic being relatively secret, the communication over-the-ether inherently leaks metadata and side-channel information about the transmitted payloads. An adversary who is passively sniffing on the encrypted WiFi traffic, can not feasibly decipher anything past Layer 2 of the OSI Model from the recorded frames. Specifically, encryption obscures everything except the WiFi protocol headers and the radiotap information. 

As discussed in the previous paragraphs, 802.11 discloses protocol information in its encrypted traffic. This information can be aggregated to infer patterns and deduce traits about the transmitter payloads and the device itself. For one, timing and throughput of the traffic can be used to describe the activity of a device. Despite the data frame packets being encrypted, they still reveal the content's length. While it does not correlate exactly to the size of transmitted payloads due to the cryptographic padding, but it is enough to classify the device's state. Then, RSSI correlates to the real-life distance between the sender and receiver. Section \ref{cha:rssi_localization} covers the applicability of RSSI to localization in detail.

\subsection{Bluetooth Low Energy}
\label{cha:ble}
Bluetooth~\cite{Bluetooth4.0} and its variants is a very widespread communication protocol for IoT devices. Its most commonly used for peer-to-peer connections, where the two devices communicate together without third-party like a router or gateway. A typical example is Bluetooth speakers and headphones, where only the encoded audio signal is transmitted from the sender (smartphone) to the receiver (wireless headphones). No connection to other devices or the internet is required for this configuration.

\subsubsection{General Information}

In this research we focus on Bluetooth Low Energy (BLE), a variant tailored for low power IoT devices. The key difference in the context of this research between BLE and Bluetooth Classic is the use advertisement packets, which are present in the BLE variant. Data transmission and other protocol-specific concepts will not be discussed. Interestingly, there exists a great overlap in Bluetooth and WiFi at Layer 1. As discussed in Section \ref{cha:wifi}, Bluetooth Classic and BLE also utilize the 2.4 GHz frequency, they also use channelization and have share similar communication concepts such as encryption and association. In fact, WiFi and Bluetooth is so similar on the lower level that many embedded devices, such as ESP32, use the same hardware for Bluetooth Classic, BLE and WiFi \cite{ESP-Specs}. A BLE driver is also able to provide the RSSI of any BLE packet.

Monitoring Bluetooth traffic is a notoriously difficult task. A paper from 2018 \cite{8556096} proposes a novel approach in sniffing Bluetooth traffic which yields 90\% package capture rate, which demonstrates how difficult this challenge is. Bluetooth, similarly to WiFi, utilizes channel-hopping, so the full channel sequence is necessary in order to reconstruct the packet flow. The algorithm responsible of channel coordination, communicates over an encrypted channel, so a passive eavesdropper must guess the subsequent channels or monitor all channels. Contrary to WiFi, Bluetooth operates over 78 different channels, which makes monitoring the whole frequency band much more challenging. Lastly, sniffing Bluetooth Classic packets requires specialized hardware, because regular Bluetooth chips do not have monitoring capabilities.

\begin{table}[h]
	\centering
	\renewcommand{\arraystretch}{1.3} 
	\captionsetup{skip=1ex}
	\setlength{\tabcolsep}{8pt}
	\begin{tabular}{|c|c|p{0.45\linewidth}|}
		\hline
		
		\textbf{Value} & \textbf{Name} & \textbf{Interpretation in this work} \\ \hline
		0x2-0x7& Service UUID & Device type and capabilities \\ \hline
		0x8/0x9& Short/Complete Local Name & Device name, for example: Alice's Headphones \\ \hline
		0x14-0x15 & Service Solicitation& Query for devices supporting the provided services. \\ \hline
		0xFF & Manufacturer Data & First two octets often contain the company identifier, reminder is proprietary manufacturer data. \\ \hline
	\end{tabular}
	
	\vspace{1ex}
	\caption{Most relevant BLE advertisement packet fields \cite{Bluetooth4.0}.}
	\label{tab:ble_advertisement}
\end{table}

\subsubsection{BLE advertisement}

The key element of BLE is its advertisement~\cite{Bluetooth4.0} functionality. This is analogous to the beacon frames in 802.11, i.e. packets that are used to announce the BLE device's presence and capabilities~\cite{Bluetooth4.0}. These packets are broadcasted and recording them is trivial, since they can be monitored with any Bluetooth capable device. In fact, this is possible by design as this technology is used for device discovery. The structure of an advertisement packet is dynamic and there are no mandatory fields or data which a packet must contain. However, the most relevant fields which are commonly present in the packet, are presented in Table \ref{tab:ble_advertisement}. Every piece of information in an advertisement packet is represented as a tuple with an index representing the type of data, for example the Complete Local Name and the field's value. There are other fields which can be used for fingerprinting or to localize the device, like the TX power field. This information can be used to enhance localization algorithms, since the transmission power can improve the RSSI-based distance estimation algorithm, however this research will only consider the device's name, services and manufacturer information. 

\begin{figure}[h]
	\centering
	\includegraphics[width=\linewidth]{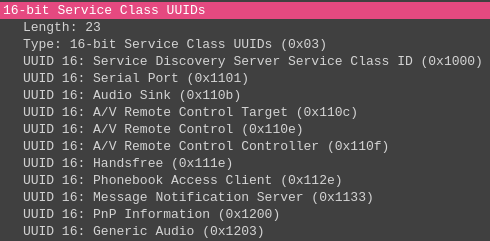}
	\caption{BLE Service UUID of wireless headphones, decoded using Wireshark.}
	\label{fig:ble_service_wireshark}
\end{figure}

The BLE advertisement service Universal Unique Identifier (UUID) stores the information about the device type and its capabilities. This data is represented in three different lengths: 16bit, 32bit and 128bit, with the 16bit being the most common. Bluetooth lists all numbers, i.e. capabilities, in the Service representation in \cite{BluetoothAssignedNumbers}. Figure \ref{fig:ble_service_wireshark} represents a decoded Service UUID of an advertisement packet of wireless headphones. In transit, the service UUID field is represented as follows: \textbf{0x10001101110b110c110e112e113312001203}.

The final relevant piece of information in a BLE advertisement packet, is the manufacturer-specific data. It is a field which contains proprietary data defined by the manufacturer, however the first two bytes of this field often represent the company identifier~\cite{Bluetooth4.0}. It is infeasible to decode the data after this information, so it will be ignored.

The time interval or conditions for when an advertisement packet should be sent by a BLE device is not specified in the official specification. It depends on the device itself and the firmware implementation. Typically, advertisements packets are sent when the device is idle, or at least not connected to any device. When a device is having an active connection, advertisement for its intended purpose is not necessary. However, as mention before, this behavior is implemented by the firmware manufacturer and no general conclusion can be drawn about the advertisement behavior for all devices.

\subsection{Related Work}

This section presents a general overview of the related concepts as well as the current state of research. Specifically, we introduce the concept of fingerprinting. We show how this can generally be done and elaborate which of the presented techniques are most relevant for this work. Then, RSSI-based localization is introduced in a similar fashion. Lastly, relevant works on human activity recognition in similar contexts will be presented to show the possible outcomes of this work and the current state of research in this field.

Some of the presented techniques are only applicable in this work's context under specific circumstances. The goal of this section is to present the possibilities which most successful attackers would have to spy on their neighbors. This review highlights how seemingly inconspicuous data can yield privacy invasive insights.

\subsubsection{Fingerprinting techniques}
\label{cha:fingerprinting}

Fingerprinting, sometimes profiling or inference attack, refers to the method of using a set of non-identifiable, observable attributes to uniquely identify a target. It can be anything from a natural person or devices to abstract concepts like protocols or behaviors. The set of these attributes is called a fingerprint. For example, a uniquely identifiable attribute for a person is their government ID number, or the insurance number. Non-identifiable attributes such as hair color, height, weight and whether they wear glasses or not, can still potentially identify an individual. This phenomenon is based on the fact, that although the individual attributes may not identify a person on their own, their union presents a unique profile that maps to a single person. Profiling in the context of digital privacy can be performed on different levels - transport, communication and application, as discussed in \cite{10.1007/978-3-642-20769-3_16}. Table \ref{tab:profiling_levels} provides an overview of protocols and technologies associated with different profiling levels. 

\begin{table}[h]
	\centering
	\renewcommand{\arraystretch}{1.3} 
	\captionsetup{skip=1ex}
	\setlength{\tabcolsep}{8pt}
	\begin{tabular}{|c|c|c|}
		\hline
		
		\textbf{Layer} & \textbf{Technologies} & \textbf{OSI layer} \\ \hline
		Transport & RFID, radio waves, channel state information & Layer 1 \\ \hline
		Communication & VoIP, TCP, IP & Layer 2-4 \\ \hline
		Application & DNS, HTTP(s), OpenVPN & Layer 4-7\\ \hline

	\end{tabular}
	
	\vspace{1em}
	\caption{Overview of technologies at different layers of profiling attacks.}
	\label{tab:profiling_levels}
\end{table}

The concept of fingerprinting devices based on the attributes of their wireless communication is referred to as Radio-Frequency Fingerprinting (RFF). WiFi-capable devices can be profiled based on the subtle differences in the radio wave transmission, which can be captured using a Software Defined Radio (SDR). In a study \cite{10.1145/2939918.2939936}, researchers were able to distinguish between network card models with a 95\% certainty based on, for example, the frequency offset of the sent packets. A more recent paper from 2024~\cite{10793404} presents a framework which uses Channel State Information (CSI) of WiFi signals to uniquely identify different network cards with a 99.53\% success rate. There, researchers developed a convolutional neural network which classified the signals based on its propagation variables (phase, amplitude and frequency). Another work \cite{10.1145/1409944.1409959} examines WiFi network cards by analyzing the fluctuations in the signal transmission and achieved a near-perfect accuracy of over 99\%.  

\begin{figure}[h]
	\centering
	\includegraphics[width=\linewidth]{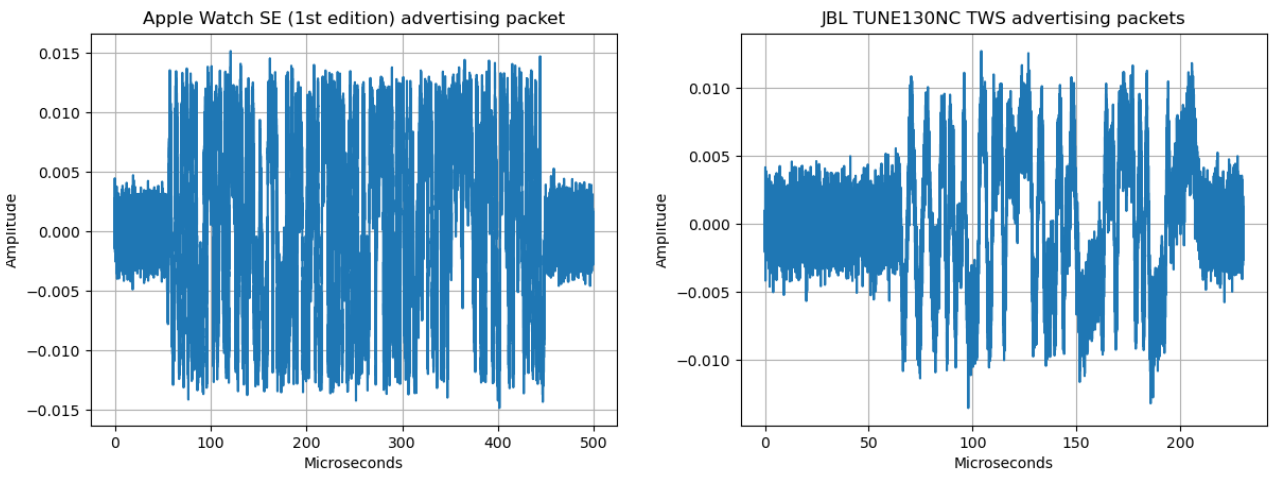}
	\caption{Radio wave fingerprints of different Bluetooth capable devices. Adapted from \cite{rušiņš2024experimentalstudyrffingerprinting}.}
	\label{fig:ble_wireless_fingerprint}
\end{figure}

A study from 2019 \cite{8691737} applied RFF techniques to Bluetooth-enabled devices and was able to distinguish between Bluetooth devices with a certainty of up to 90\%. The researchers applied an established RFF method, the \textit{Hilbert-Huang transform}, a technique from the signal processing domain~\cite{SOUZA2022103292}, to create fingerprints of Bluetooth transmission signal of the individual devices. Another study from 2024 \cite{rušiņš2024experimentalstudyrffingerprinting} used similar techniques to the ones presented in \cite{10.1145/2939918.2939936}, but on Bluetooth devices. There, researchers examined raw Bluetooth signals using an SDR and were able to achieve profiling accuracy of 84\%. Figure \ref{fig:ble_wireless_fingerprint} represents fingerprints of BLE advertisement packets sent by different devices. 

As elaborated in \ref{cha:wifi}, WiFi and Bluetooth technologies are not confidential on all layers and leak protocol information. This data, as well as properties of the traffic itself, can be used to profile wireless devices. It is an established field of research, which began in the early 2010s. A paper from 2013 \cite{6258210} reviewed attributes in the 802.11 protocol, such as data rate switching or probe request transmission pattern, to profile WiFi-enabled devices. Moreover, they performed traffic analysis by examining the inter-arrival time of the packets. They found that the transmission time, i.e. time from sending and reception of the packet, can be used to differentiate between devices with a likelihood of at least 80\%. A more established work from 2007 \cite{wright2007language} reviewed how observable patterns in encrypted Voice over IP (VoIP) traffic can be used to determine the spoken language. There, the researchers discovered that the bitrate of the encoded signal varies among the languages. Overall, they were able to construct a general classifier for 21 languages which differentiates any pair of languages with a success rate of 86\%. For most languages, they could classify them with an accuracy greater than 90\%. 

\begin{figure}[h]
	\centering
	\includegraphics[width=\linewidth]{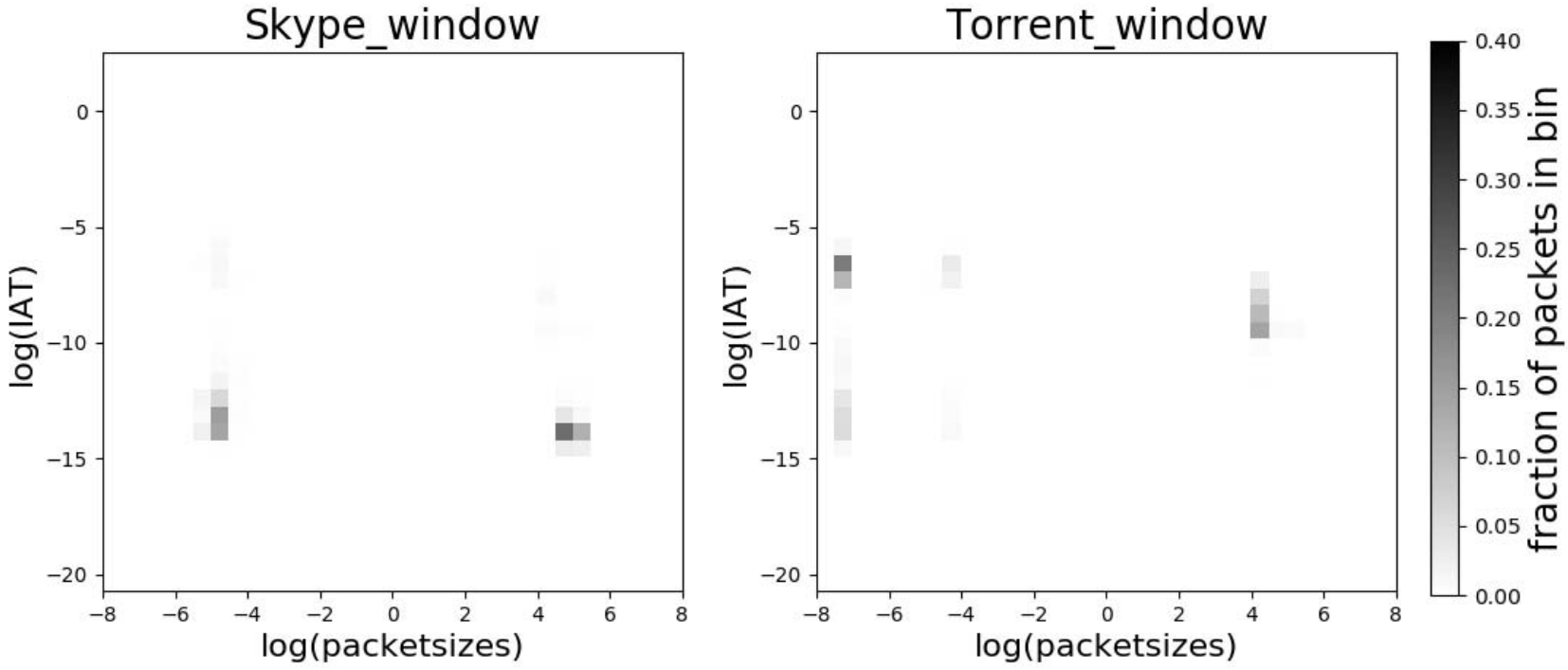}
	\caption{Fingerprints of traffic generated by Skype and torrenting a file. Adapted from \cite{8406218}.}
	\label{fig:traffic_fingerprint}
\end{figure}

A more recent work from 2018 \cite{8406218} researched the fingerprinting possibilities of classifying network traffic types from encrypted traffic. They examined the throughput features such as inter-arrival time and packet bursts with the goal of differentiating between web browsing, VoIP, video streaming, and P2P traffic. Even if the traffic was sent using additional encryption layers such as IPSec, Tor or VPN, they were still able to achieve over 80\% accuracy. Figure \ref{fig:traffic_fingerprint} shows how different the fingerprints of network activities are. Analyzing traffic on Layer 4, i.e. the TCP and UPD transport layer, allows for protocol fingerprinting. Researchers in \cite{280012} examined unencrypted parts of the TCP headers as well as the patterns in the ACK-packets to detect OpenVPN traffic with over 85\% success rate, even if typical fingerprinting countermeasures such as XOR-ing the payloads are deployed. 

When considering the observable protocol-specific unencrypted data, it usually results in the most privacy-invasive and effective fingerprinting. In the context of this work, an eavesdropper would be able to apply any of the previously mentioned fingerprinting analyses and would have the access to some plaintext information. When considering internet traffic, they would see information such as DNS queries, IP addresses and Server Name Indication extension of the TLS1.2 ClientHello handshake. While  TLS1.3~\cite{cryptoeprint:2019/749,239072} and Encrypted DNS (DNS-over-HTTPS or DNS-over-Tor)~\cite{10.1145/3320269.3384728} exist and are supported by all modern browsers, many websites and services do not use it. Hence, DNS and Server Name Indication (SNI) in the TLS ClientHello handshake is often observed in plaintext. This leads to privacy invasive user profiling attack possibilities, like presented in \cite{HERRMANN201317}. This work analyzed patterns in DNS traffic of users and achieved a 76\% accuracy in user profiling. These techniques can not only be applied to active user activity, but possibly also to passive background tasks of their browsers \cite{5958207}. Additionally, even if the DNS traffic is encrypted, it is still possible to use this information for inference attacks~\cite{siby2019encrypteddnsprivacy}, for example to profile Android Apps \cite{10.1145/3465481.3465764}.

Plaintext data of 802.11 can allow for inference attacks as well. A paper from 2016 \cite{REDONDI20181} built knowledge graphs from data included in the probe request packets. They were able to differentiate between different social groups (employees from their university faculty, passerby's, visitors) based on the temporal patterns and contents of the probe requests. Another study from 2014 \cite{CUNCHE201456} studied the overlap between the probed SSIDs by different devices and was able to detect relationships between individual persons. Deploying probe request analysis on a large scale, opens possibilities for deanonymization. In this paper \cite{7524459}, researchers were able to approximate the political orientation of smartphone users from their probe requests.

Fingerprinting of wireless devices and their users can be performed on any communication layer. First, at the raw signal transmission level, where fluctuations of the signals and CSI can be used to distinguish between devices. Then, by analyzing the individual 802.11 frames, an adversary could determine the traffic type of a wireless device using by analyzing the throughput and the size of the packets. Lastly, having the access to some unencrypted elements of their traffic such as DNS or SNI, or in the case of WiFi, probe request packets, it is possible to create profiles of individual persons based on the patterns within that plaintext data. Due to the fact that WiFi, as elaborated in \ref{cha:wifi}, is broadcasted over-the-aether and can be trivially monitored, these findings present a severe threat to WiFi user's privacy.

\subsubsection{RSSI - Based localization}
\label{cha:rssi_localization}

\begin{figure}[h]
	\centering
	\includegraphics[width=0.7\linewidth]{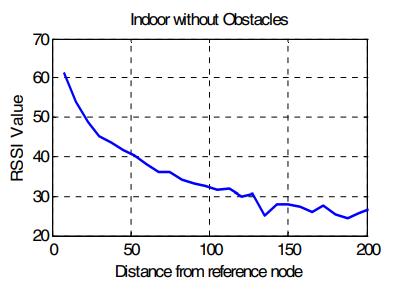}
	\caption{Signal strength decay over distance indoors with no obstructions. Adopted from \cite{5972367}.}
	\label{fig:rssi_over_distance}
\end{figure}

As per the path-loss equation, radio waves decay over distance approximating the long-tail curve \cite{5972367}. Solving this simplified equation for distance results in the following relation: $distance \approx (p_s/p_r)^{1/n}$ , where $n$ represents a path loss coefficient (normally between 2 and 4) and $p_s$ and $p_r$ represent the sent and received signal power. This approximation is visualized in Figure \ref{fig:rssi_over_distance}. However, due to numerous unknown variables, such as the transmission power of the sender device ($p_s$), obstacles in the path between the sender and receiver, interference with other signals in the same frequency, it is infeasible to reliably calculate the distance between two devices based on received signal strength alone. Moreover, because this relationship is non-linear, it becomes even more challenging to determine the beyond after a certain threshold, even when all variables are known.

There exists research~\cite{https://doi.org/10.1002/dac.70008}, which utilizes machine learning algorithms to improve the correlation between distance and the RSSI values. There the researchers were able to correlate the RSSI to real-life euclidean distance with the median average error being as low as 1m. In other words, by utilizing most optimal algorithms it is feasible to locate a device within one meter accuracy. However, they only considered devices in the same room without physical obstructions.

Given that the distance-based localization techniques are not reliable with RSSI, different approaches for RSSI-based localization have been proposed. This challenge can be seen as a fingerprinting problem - instead of measuring the euclidean distance between senders and receivers, we compare the RSSI fingerprints to pre-recorded location markers. 

Such analysis typically consists of two phases, as elaborated in \cite{s21082769}. First phase is analogous to the offline training phase for machine learning models, where the fingerprints of the RSSI reference points are computed. Depending on the context, this phase may require human interaction to supply the necessary domain knowledge into this dataset, for example assigning labels to some pre-defined location grids or setting up receivers to map the space. Then, having a fingerprint database, the online, i.e. the inference phase takes place where the tracked devices' RSSI fingerprints are matched with the fingerprints in the database. Here, clustering algorithms are used to find a cluster of fingerprints, which represents the device's fingerprint. Depending on the context, such cluster can be reduced to a single vector, for example its centroid, which could represent the real-world position of the device. Alternatively, the labeled of the fingerprint from the database with the smallest distance to the device's fingerprint can be used as its location.

\begin{figure}[h]
	\centering
	\includegraphics[width=\linewidth]{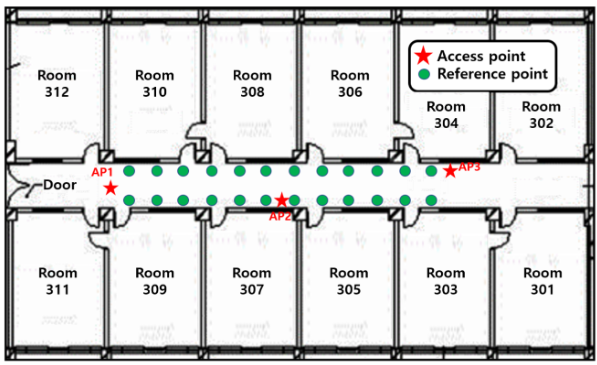}
	\caption{Spatial setup of transmitters (access point) and receivers (reference point) in \cite{s21082769} during the offline phase. Adopted from \cite{s21082769}.}
	\label{fig:rssi_obstructions}
\end{figure}

A work on RSSI-based localization from 2021~\cite{s21030971} considered obstructions in their research. They placed up to 30 RSSI antennas (BLE receivers) and recorded how each of the sensors receives signals from all of the other receivers. Having their relative signal strengths, they were able to construct an obstruction map of the space where the localization would be performed. Then, having only three receivers and the obstruction map, they were able to achieve an accuracy of 2.28m. Figure \ref{fig:rssi_obstructions} shows the setup of their sniffers during the offline phase.

A paper from 2021 \cite{9540865} treats this kind of localization as a clustering problem. In a 720m$^2$ indoor area, over 140 RSSI reference points were calculated and then used in clustering algorithms were to localize the senders. A reference point, i.e. a fingerprint, was represented as a vector of the RSSI values of every RSSI sender at any point in space. Then, localization was performed by clustering the sender's location fingerprint with the pre-calculated reference points. This approach yielded an accuracy of almost 2.5m. 

Another paper from 2024 \cite{10744048} applied similar techniques, but on a larger scale. They used RSSI-based fingerprinting to locate devices in smart city environments. Their dataset consisted of a set of 5m x 5m grids and their classifier matched the fingerprint of each grid to the sender's location fingerprint. Best classifier resulted in a 96\% accuracy. 

Localization based on the received signal strength is possible and there are many ways to do so. For one, because RSSI roughly correlates to the real-world distance, it is possible to develop a model which takes the obstruction into account, yet these obstructions must be known beforehand. The most suitable method for this research is the fingerprint-based approach, due to the poor quality of the RSSI readings, which are caused by the obstructions. Nevertheless it is a feasible approach. Also, for the context of this research, it is sufficient to know the rough location of a smart home device - its enough to know that a smart sensor is mounted in the kitchen, rather than that it is 2.3 meters from an other sensor.

\subsubsection{Human Activity Recognition}
\label{sec:har}

Human activity recognition (HAR) is an established interdisciplinary field of  research. As the name suggests, its is concerned with classifying human actions from information media such as videos, images, sounds or side-channel information. HAR typically requires advanced machine learning techniques to extract the relevant features. This paper from 2015 \cite{article_har_review} provides a detailed overview of techniques used for various HAR analyses in general. In this research, we will focus on human activity recognition in IoT and smart home environments. In the HAR categorization proposed in \cite{article_har_review} such analyses fall in the unimodal, i.e rule-based and space-time-based, as well as multimodal, i.e. behavioral, categories. 

\begin{figure}[h]
	\centering
	\includegraphics[width=\linewidth]{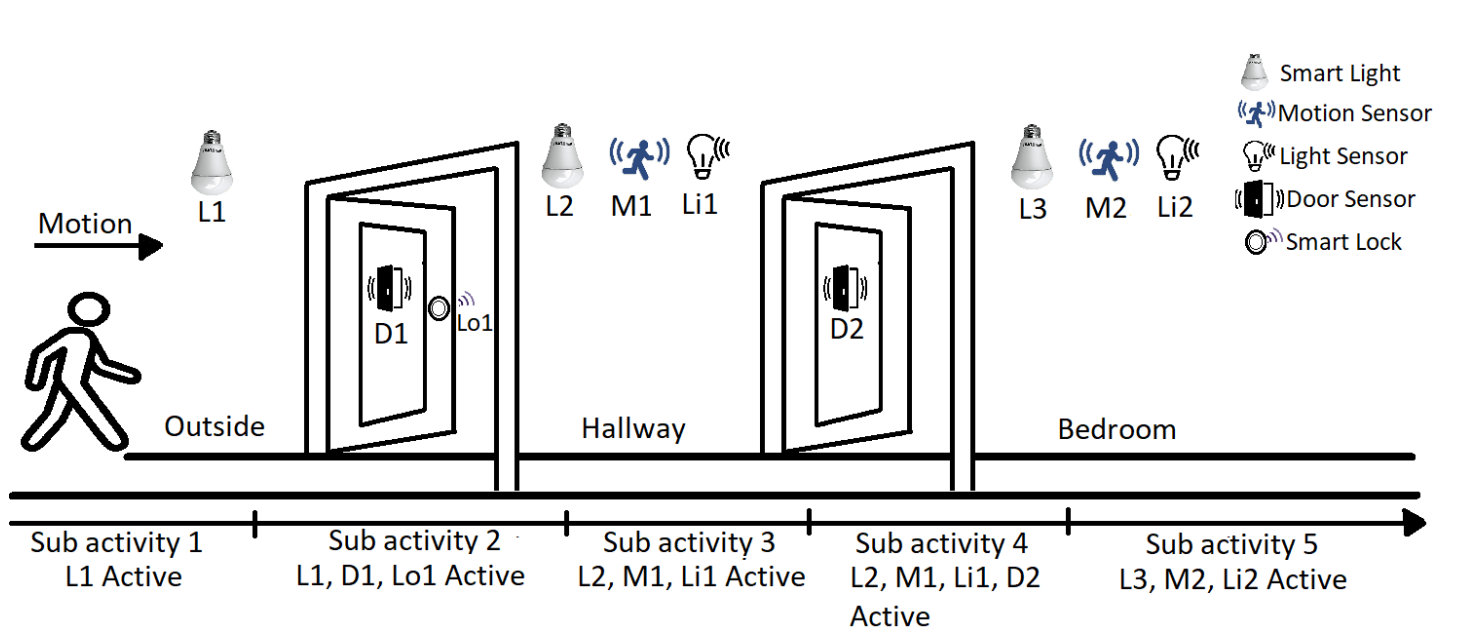}
	\caption{Sequence of device activities when entering a smart home. Adopted from \cite{10.1145/3395351.3399421}.}
	\label{fig:activity_sequence}
\end{figure}

A paper from 2020 \cite{10.1145/3395351.3399421} considers a very similar scenario to the one of this work. There, researchers perform device fingerprinting and human activity recognition based on wireless encrypted WiFi traffic as well as BLE and Zigbee. Their multistage step-by-step approach infiltrates the smart home's privacy by learning about the devices and their usage patterns. Finally, they combine their findings to perform human activity recognition by detecting sequences in activities which correlate to certain actions. For example, as showed in Figure \ref{fig:activity_sequence}, they find that a specific sequence of door sensors activating correlate with a smart home inhabitant entering the smart home environment using hidden markov models. 

A more recent work from 2023 \cite{10.1145/3580890} considered a scenario in which an adversary placed various WiFi sniffers inside of the victim's apartment. Then, researchers developed a rule-based framework for activity recognition. It used contextual clues such as localization, which they inferred through the signal strength of the senders, device types and device activity. One of the findings of their work is the possibility of differentiating between different inhabitants based on device activity, which they use. By creating a behavioral fingerprint consisting of the sequence device activities, they could distinguish between three different smart home inhabitants. 

Another paper \cite{apthorpe2017smarthomecastleprivacy} examined privacy risks in IoT by monitoring traffic on the IP level. The internet service providers as well as malicious actors who compromised the security of a WiFi network have access to this data. It is worth noting that despite the fact that WPA2, the most commonly WiFi security standard, is considered secure, it is trivial to break its encryption if a weak password is used. The researchers in that paper were able to identify the devices with great accuracy using their DNS traffic. Similarly to previously presented works, they analysed the patterns in the activity of said devices and were able to infer how they interoperate.

A more established paper from 2016 \cite{10.1145/2873587.2873594} examined the privacy risks of BLE advertisement. Their findings include insight into human activity recognition, where they concluded that the signal strength of the wearable devices can be used to fingerprint user's activity. Most notably, they discovered that a person's gait can be fingerprinted which allows for tracking, since the gait is unique feature which allows for differentiating between humans. 

\begin{figure}[h]
	\centering
	\includegraphics[width=\linewidth]{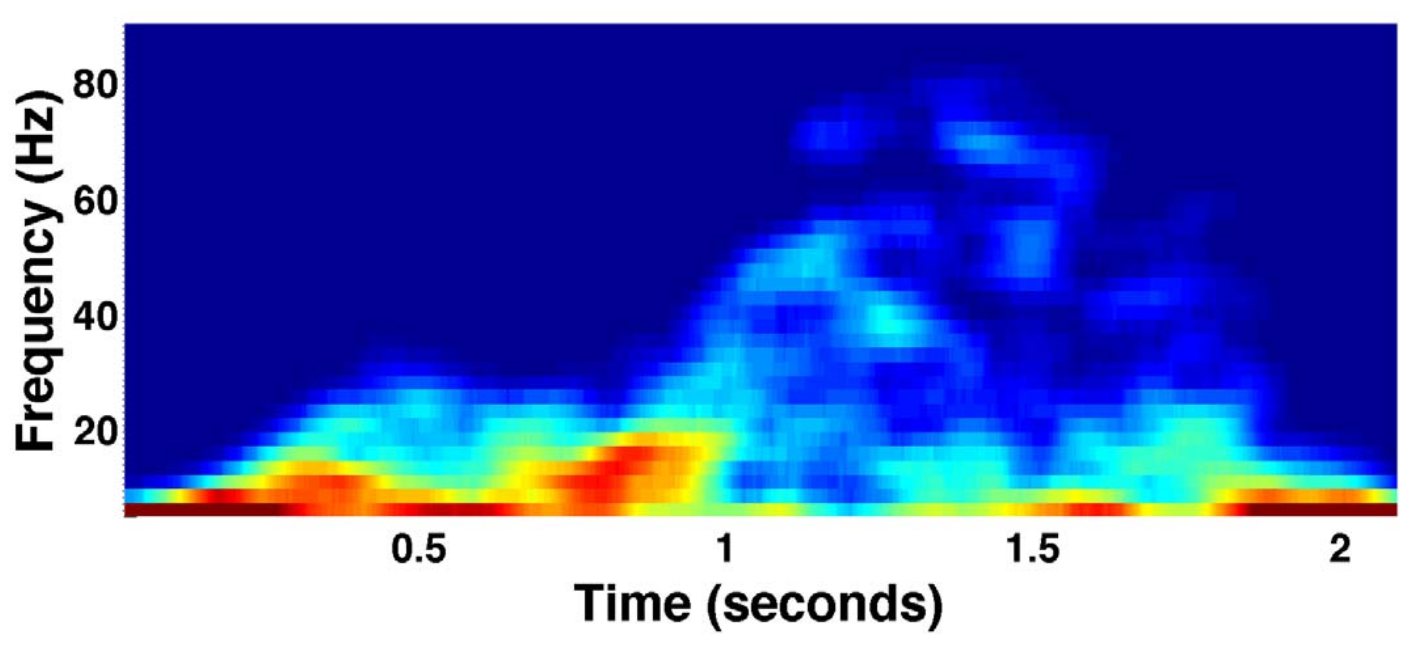}
	\caption{Activity of a human falling reflected on a CSI spectrograph. Adopted from \cite{7875148}.}
	\label{fig:csi_fall}
\end{figure}

Channel state information of a WiFi signal is the most commonly used data medium to conduct WiFi-based human activity recognition. It gets attention due to high resolution of the data and its ubiquity - the CSI of any WiFi device can be used for human activity recognition \cite{8011040}. Since radio waves are evenly sent in all directions, they reach the CSI sensor with varying phase shifts, because they take different paths to finally reach the receiver. This phenomenon is also referred to as multipath propagation and has been known and utilized already in the 1970s \cite{j.1538-7305.1972.tb01923.x,1146527, }. Its application has overlaps with computer vision-based HAR methods. Human bodies consist mostly of water, which interacts with the radio waves in a significant way and is visible in the CSI, as shown in Figure \ref{fig:csi_fall}. Following papers present possibilities for CSI-based HAR \cite{9205901, 8011040, 9747420, 8422895, 8391737, 7875148}. Most notable work is \cite{10.1145/3191755}, in which the researchers developed a model which could read sign language using CSI. 

Human activity recognition can be conducted on many different levels having a varying degree of contextual information. However most HAR methods in the context of this work are rule and behavior based. This means that the human actions are inferred by observing patterns in the activities of the smart home devices. For example, if a device identified as a cooking utensil, such as a Thermomix and a humidity sensor in the same room is active, then an inhabitant of that smart home is most certainly cooking. Then, by observing sequences and patterns in device activities, more complex actions can be derived. For example a specific sequence of activation of door sensors and motion sensors can indicate that an inhabitant entered a smart home.

\subsubsection{Research Gap}

The presented literature survey presents how many possibilities an adversary has, if they want to spy on their neighbors using wireless traffic. Fluctuations and patterns in data streams aid the eavesdropper in identifying devices and their activity states. Then, using variations in the received signal strength between devices, methods have been discovered which allow for localization. Lastly, patterns in different device states and their location can be correlated to various human activities. An attacker with the goal of spying on their neighbors by passively monitoring the communication of their smart devices years of prior research to their disposal. 

Most of the presented works focuses on one aspect of the privacy-invasive analysis. They present ways of determining device activity using patterns in data streams \cite{wright2007language, 8406218,apthorpe2017smarthomecastleprivacy} or different approaches for device localization using the signal strength \cite{s21030971, 10744048}. Few studies combined these research fields and applied them to smart home settings. Most notably, this study \cite{10.1145/3395351.3399421} examined how encrypted WiFi traffic can be used for human activity recognition by examining the device states. Only one work \cite{10.1145/3580890} utilized location fingerprints in their traffic-based human activity recognition in smart home contexts. Still, their experiments were conducted in scenarios which only few attackers would be capable of inducing. For example, in \cite{10.1145/3580890} the researchers assumed a situation in which an adversary placed the WiFi sniffers inside of the victim's apartment. While this results in the greatest device positioning accuracy and least packet loss during sniffing, due to minimized interference and obstructions, it is significantly more challenging to setup than this work's presumption. This work assumes a more realistic scenario, that of a nosy neighbor which monitors behind the wall. Then, the selection of smart devices is more minimal and resembles an environment of a more average inhabitant, whereas their research considers 27 distinct smart devices. Finally, for an even more realistic setup, this work also examines multimedia devices such as smartphones and laptops, together with their unique 802.11 artifacts such as probe requests. 

This work's contribution to the existing research is its practical red-teaming like approach to inferring information about the neighbors. As presented before, human activity recognition, RSSI-based localization as well as device recognition and fingerprinting are active research areas with novel techniques being published each year. However, no prior work considered a practical approach to conduct this multi-stage human activity recognition on data with inferred labels.   
 
\chapter{Methodology}
\label{cha:Methodology}

This chapter presents details of situation which this work assumes - neighbor behind a wall passively monitors their victim's traffic. First the experimental setup will be presented. Here, smart home devices and their location will be introduced to show the target smart home environment. Then, the sniffing setup will be presented. Lastly we present the specific techniques which were used to derive conclusions about data generated by the sniffing setup. 

\section{Experiment Setup}
\label{cha:experiment}

The goal of this work is to research how well established techniques can be applied to a real-life scenario and what the neighbor can learn about their victim from encrypted WiFi and broadcasted BLE traffic. To answer this question, this exact situation will be simulated. A household will be equipped with smart home devices and their wireless traffic will be monitored. The monitoring setup will be deployed in a separate room, to simulate the neighbor behind a wall. The constant monitoring will span across three weeks. During this time, the devices will be installed, setup and used according to the usual usage patterns. The inhabitants of this household will live according to their regular schedule. The data from preliminary experiments will also be used. It was recorded in similar fashion and reveals more information about the daily schedules of the inhabitants, since it spans across a larger time frame and captures more realistic daily schedules of the inhabitants. 

\subsection{Data collection}

In this work, we explore the possibilities which a nosy neighbor has at three levels: we perform device reconnaissance, localization and human activity recognition. Our goal is to breach the secrecy of a smart home by inferring information about the smart home itself, its spatial features and ultimately how inhabitants' behavior can be observed through these fingerprints. Each of these aspects is investigated in an exploratory manner, beginning with the most fundamental properties and progressively incorporating previously inferred contextual information.

First, we conduct reconnaissance in which we enumerate the devices associated with the neighbor's smart home. Specifically, we use OSINT techniques and devices' network traffic behavior to identify them. We monitor the installation and configuration process of new smart devices (see Table \ref{tab:smart_home_devices}) in the reconnaissance.  Then, by thoroughly examining their network traffic fingerprints, we try to determine the devices' current activity state. This process answers the question of: "What can your neighbor learn about your smart home?"

Afterwards, we perform localization analysis by exploring the resolution achievable through RSSI-based techniques. First, we will analyze if mobile and stationary can be differentiated based on their RSSI readings. Then, we aim for more detailed localization analysis in which we try to differentiate between spatially separated devices. This insight allows us to construct a map of semantic areas in the smart home environment, so that we can, for example, distinguish the kitchen from the bedroom area. Lastly, having these distinct areas of the apartment, we fill in the gaps with previously acquired knowledge to classify these areas. This process answers the question of: "Can your neighbor know where you are?"

Lastly, having the context information about the smart home itself and its spacial allocation, we apply this information to gain knowledge about the victim. First, their probe requests will be analyzed to obtain an overview over their trusted networks. Then, we perform human activity recognition based on all of the acquired information. Specifically, insight into the inhabitant's behavior can be derived from the combination of their location and device activity. For example, if an inhabitant is localized in the living room and their smart TV generates high downlink traffic, it is likely that they are watching TV. In order to explore to which extent human activity recognition is possible, events of one day will be reconstructed while taking all of the previously learned side-channel information. This process answers the question of: "What can your neighbor learn about you?"

\subsection{Smart home}
\label{sec:smart_home_setup}
\begin{figure}[h]
	\centering
	\includegraphics[width=\linewidth]{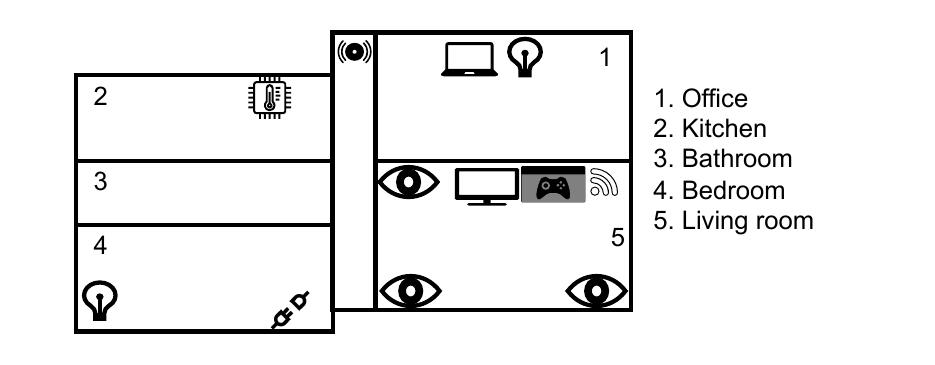}
	\caption{WiFi sniffers, represented by eyes, and smart home device setup. Each of the rooms is numbered. The empty space in the middle represents the corridor.}
	\label{fig:sniffer_setup}
\end{figure}

To simulate living in a household with smart home devices, three out of five rooms in an apartment will be equipped with at least one smart home device as well as wireless devices. One room will be used for recording the WiFi traffic. The remaining room 3, i.e. the bathroom, has no smart home devices due to logistical reasons. Figure \ref{fig:sniffer_setup} visualizes the spacial allocation of the WiFi sniffers and the smart home devices.

The setup has been designed such that each room has at least one mounted device. The RSSI of these devices will act as a hotspot or anchor for where each room is. In other words, the fingerprint of their localization signals will represent the location fingerprint for each room in the smart home. Since it is infeasible to calculate the euclidean distance between the devices based on obstructed RSSI signal, precise measurements of the apartment are not necessary. 

The devices located in the room with the sniffers will not be considered in location analysis. Table \ref{tab:smart_home_devices} shows the devices which are used in the experiment. All smart devices are provided by the Leipzig University. 

\begin{table}[h]
	\centering
	\renewcommand{\arraystretch}{1.3} 
	\captionsetup{skip=1ex}
	\setlength{\tabcolsep}{8pt}
	\begin{tabular}{|c|>{\raggedright}p{8.5cm}|c|c|}
		\hline
		\textbf{Mac} & \textbf{Device} & \textbf{Capabilities} & \textbf{Room}\\ \hline
		d8:f1 & Tuya light bulb & WiFi & 1 \\ \hline
		08:b6 & Shelly temperature \& humidity sensor  & WiFi \& Bluetooth & 2 \\ \hline
		6c:5a & Tapo light bulb & WiFi & 4 \\ \hline
		54:af & Tapo smart plug & WiFi \& Bluetooth & 4 \\ \hline 
		8c:f6 & Shelly motion sensor & WiFi & corridor \\ \hline
		\hline		
		9c:fc & Laptop (Intel) & WiFi \& Bluetooth & 1 \\ \hline
		24:2f & TP-Link router & WiFi  & 5 \\ \hline
		20:28 & LG Smart TV & WiFi \& Bluetooth & 5 \\ \hline
		60:1a & Nintendo Switch console & WiFi & 5 \\ \hline
		a4:45 & Xiaomi smartphone & WiFi \& Bluetooth & mobile \\ \hline

	\end{tabular}
	\vspace{1em}
	\caption{Devices used in the smart home setting. Mac addresses are shortened and randomized for privacy. Top half of the table represents new devices, and the bottom half devices which are already present and integrated into the network. }
	\label{tab:smart_home_devices}
\end{table}

The devices have been chosen for their popularity and wireless capabilities. All devices are at least WiFi-capable, since this work focuses on WiFi analysis. Same methods can also be applied to Bluetooth-capable devices. However, as explained in Section \ref{cha:ble} monitoring Bluetooth traffic is more challenging and does not generate as much traffic as WiFi, due to different applications. While Bluetooth is used almost exclusively for P2P communication, WiFi can be used for P2P and internet connection. The greater data throughput improves the quality of traffic-based fingerprinting. 

\subsection{Traffic sniffing}

While monitoring WiFi and BLE traffic is relatively simple, it is not a trivial task. Technical know-how is necessary to put the Bluetooth and WiFi interfaces into monitoring states, and dedicated antennas may be necessary to obtain more reliable results. 

Then, the attacker must determine which AP belongs to their victim. This task can be challenging in densely populated apartment buildings. Almost every home has at least one WiFi AP. For instance, the flat in which the experiment is conducted, picks up signals from more than 15 APs. Most of them have generic names such as Vodafone-12AB or Fritz!Box-1234-AB, which makes mapping a specific AP to a certain neighbor difficult. Hiding a WiFi network, i.e. disabling the beacon packets does not make the reconnaissance any more challenging - the SSID field is still present in the data packets, but the network name is unknown.  

There are two approaches which can aid the neighbor in correlating the AP to their victim. For one, the neighbor might know a specific device which the neighbor owns, either Bluetooth or WiFi. Then, seeing that device connect to or leave a network, the attacker could deduce that it is the neighbor's AP. For example, the attacker could learn that the neighbor has an iPhone. Then, candidates for this device could be identified within the observable devices by looking up the manufacturer from the mac address. Apple devices which leave a network at the same time as the neighbor, would reveal which AP the belongs to them. Second approach is to use the signal strength of the AP. By viewing the RSSI of the APs in real time, the location of AP can be approximated. 

It is more challenging to determine which Bluetooth-capable devices are deployed in a smart home, since it is challenging to monitor their traffic to see with which devices they communicate. One way to enumerate them is to consider context clues such as correlation of activity with other devices or localization. For example a Bluetooth lamp would almost never be active if the neighbor is not home. Analogously, the RSSI-based method is also applicable to Bluetooth-capable devices.

\subsubsection{WiFi}
\label{sec:wifi-setup}
\begin{figure}[h]
	\centering
	\includegraphics[width=\linewidth]{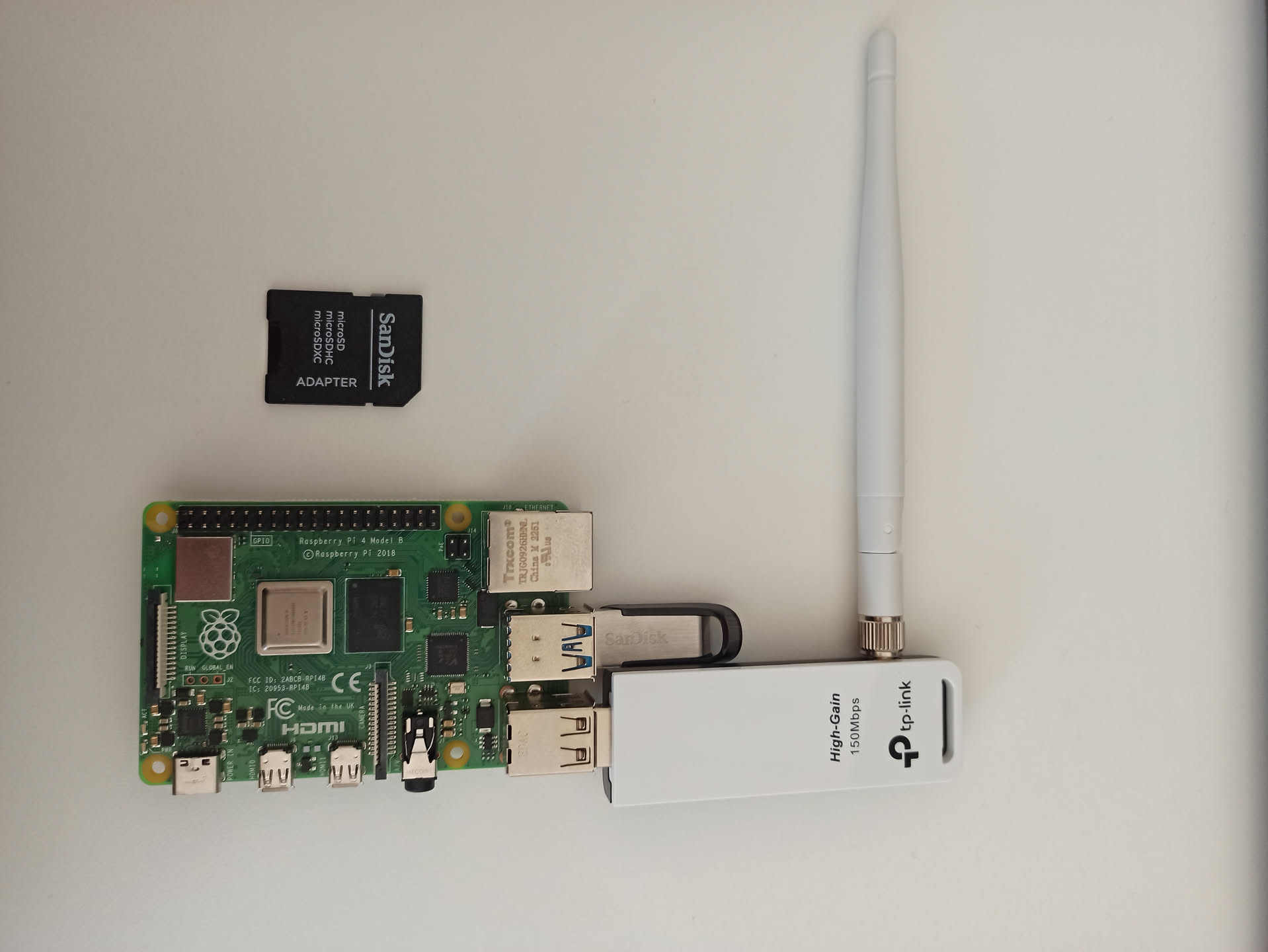}
	\caption{WiFi sniffer based on a Raspberry Pi with the TP-Link antenna. SD Card adapter for scale.}
	\label{fig:wifi_sniffer}
\end{figure}

To monitor the WiFi traffic with the focus on their RSSI, a three-sensor sniffing setup has been deployed. Each sensor consists of a Raspberry Pi single board computer with an USB TP-Link TL-WN722N WiFi interface, provided by the Leipzig University. Figure \ref{fig:wifi_sniffer} shows one of the three sniffers. 

Each device had 4 GB RAM and Raspberry Pi OS Lite installed. To ensure fault-tolerant data capture, each of the sniffers was equipped with an additional 128 GB USB drive for data storage. During capture, the software was configured to write directly onto the USB drive into rotating pcap files, changing after writing 100 MB. Each WiFi sniffer needs to be prepared for receiving packets which are not intended for them. In this case, specific drivers \cite{wifi_drivers} for the TP-Link antenna must be installed and then this interface must be put in monitor mode. Then, the packets can be recorded for analysis using tcpdump \cite{tcpdump}. It's a very optimal WiFi monitoring setup due to its low power requirements, small size and ease of setup.

Since the channel used by an AP is in the broadcasted beacon frame, an eavesdropper can tune to this specific channel in order to monitor the traffic in this WiFi network. The WiFi sniffers will only monitor the channel 8 with the band width of 20 Hz. This configuration has been applied to the router, to force all packets to go through this channel. It is the least busy channel in the apartment's proximity. Forcing communication through this channel results in the least packet loss due to interference with other WiFi APs. In this case, the sniffers should receive the vast majority of the packets, with some packet lost due to interference or other technical issues. During monitoring, no packet filters will be applied to ensure that no data is lost by filter misconfiguration. This results in packets from other devices being included in the dataset, however they will be discarded during the data pre-processing. This task is trivial if the target network is known beforehand, otherwise it can be challenging to pin-point a specific WiFi network to a certain neighbor, since

In preparation for this research, a simplified version of the sniffing setup was performed. That experiment monitored fewer devices with one sniffer, over a span of three weeks. This data is valuable, as it highlights regularities in device activity and daily routines of the inhabitants over time. The setup was not optimized for maximum data capture rate, so the data set is incomplete at certain times. 

\subsubsection{BLE}

Sniffing BLE signals is straightforward and there is no specialized hardware or device drivers required to monitor BLE advertisement. A passive eavesdropper has many options when it comes to hardware as well as software. An important BLE monitor property to consider is its version - using BLE 5.0 over 4.x is recommended due to greater compatibility with devices. Higher versions can receive traffic of devices communicating using lower versions, but not other way round. Unlike WiFi sniffing, there is no complexity drawback in monitoring newer version. 

\begin{figure}[h]
	\centering
	\includegraphics[width=\linewidth]{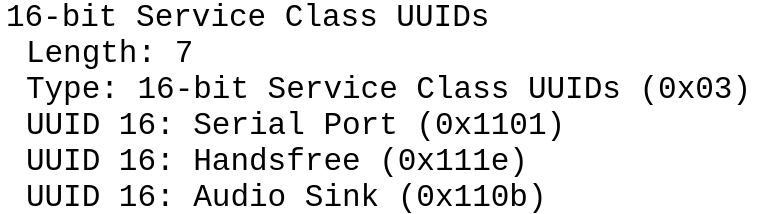}
	\caption{BLE advertisement packet's service field decoded using \textit{Wireshark}.}
	\label{fig:ble_wireshark}
\end{figure}

Most modern laptops and smartphones \cite{ble_scanner_app} support monitoring BLE traffic, however using a laptop is recommended due to wider selection of sniffing software. \textit{Wireshark} supports monitoring BLE traffic out-of-the-box, which is the recommended method, since it has capabilities to automatically decode the advertised services. Alternatively, it is possible to use the host controller interface to record the advertisements directly. During the sniffing phase of the research, both \textit{Wireshark} and \textit{bluepy} \cite{ble_drivers} were used to record the advertisement packets. Figure \ref{fig:ble_wireshark} shows the 16-bit service field of a Bluetooth audio device, where \textit{Wireshark} automatically resolved the capability numbers. 

Using \textit{bluepy} over \textit{Wireshark} has the benefit of working within a \textit{Python} environment, which makes tasks of automation, pre-processing and filtering easier. It does not support automatic service resolution, however for small datasets, it is feasible to look them up manually. Bluepy provides the most minimal example for a BLE advertisement sniffer in their documentation \cite{bluepy_minimal}.

\subsubsection{Probe Requests}

An attacker with the motivation of learning the most about their neighbor from broadcasted network traffic, will use probe requests sent by their victim's mobile devices. As discussed in Section \ref{cha:fingerprinting}, privacy invasive information can be obtained from this dataset such as political orientation, frequented places and social relationships. While it is arguably the most privacy invasive data source, as it possibly discloses traits about neighbor's life beyond the smart home, it is also the most difficult one to exploit in our assumed scenario. 

\begin{figure}[h]
	\centering
	\includegraphics[width=\linewidth]{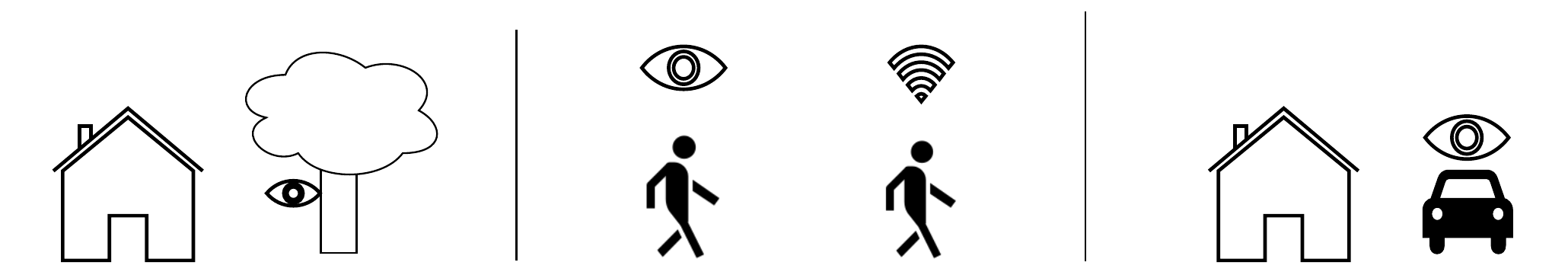}
	\caption{Enhancements to the sniffing setup to capture more probe requests.}
	\label{fig:probe_sniffing}
\end{figure}

When devices are already connected to a WiFi network, probing for other networks is not necessary. Monitoring probe requests through the wall is not effective and yields practically meaningless results. However, the sniffing setup can be enhanced without violating the passive listening premise to record some probe request. Figure \ref{fig:probe_sniffing} visualizes possible techniques to capture these requests. The easiest enhancement is installing an additional WiFi sniffer outside of the neighbor's AP range, for example near the entrance to the apartment building. A motivated attacker can mount their sniffers in a mailbox, trash cans or in the garage. Then, probe requests sent after leaving the apartment would be captured. Alternatively, tailgating outside of the house with a mobile WiFi sniffer can be performed. There, an attacker could leave the apartment building at the same time as the neighbor and pretend to head in the same direction as their victim. Lastly, a nosy neighbor can just wait outside, for example in their car, to get in range of the victim's device.

It is important to note that even with enhanced sniffing strategies, it is not guaranteed that they will yield meaningful results. Then, privacy-enhancing techniques such as mac address randomization is used by many smartphone manufacturers. While there are ways to circumvent such counter-measures \cite{bravenec2022explorationuserprivacy80211, 8747391, 10.1145/2766498.2766517}, they are complex and there is no guarantee that the de-anonymized results are valuable. Then, phone manufacturers utilize different probing strategies, as discussed in \ref{cha:probe_request}, so the observed network names may not necessarily be the most relevant ones. 

Nevertheless, a successful attacker can potentially find information in descriptive probes for SSIDs. Specifically they can learn about neighbor's interests (CleverFit-WiFi, Ceramic-Studio-Mila-WiFi, Guitar-Store), workspace (Check24-WiFi, Rossmann-Intern) or places they frequent (Cafe-Luke, Hotel Aquarius, Asian Restaurant Free WiFi). Generic and unique network names (Fritz!Box - AB123, Vodafone-AB321) can potentially reveal households which an attacker visits. Then, the WiGLE database can be used to lookup the GPS location of discovered networks to learn where the neighbor's social relations live.

\subsection{Data Pre-processing}
This monitoring setup generates a dataset of the raw 802.11 packets, which the three sniffers have picked up. Since no packet filters were applied during the recording, this dataset contains many irrelevant data points. For one, the sniffing setup picks up packets from nearby APs and devices not included in this experiment. Furthermore, not all 802.11 frame types are relevant for the analysis. Lastly, as discussed in Section \ref{sec:wireless}, the payloads of a data packet are irrelevant for this research. and not all fields of the data frames are relevant. 

Hence, the first step of the data pre-processing is to filter out the communication outside of the examined WiFi network. It is trivial, since each relevant 802.11 frame contains the Basic SSID (BSSID) address field which reveals the mac address of AP over which the communication occurs. Having the packets of the relevant devices, the dataset can be split into three further datasets: traffic data, signal strength vectors and probe request packets. The 802.11 packets contain all relevant information for splitting the frames by type, since the frame type is a part of the 802.11 header. This pre-processing was performed using tshark, the command line interface counterpart of \textit{Wireshark}. This powerful tool is compatible with \textit{Wireshark}'s packet capture filters and offers selection of 802.11 packet fields. Using this tool, the raw pcap files were filtered and exported into CSV files for compatibility with other technologies. 

The traffic dataset can be minimized by aggregating the transmitted payload to its length, which significantly reduces the size of the dataset. Then, both traffic and the RSSI dataset can be grouped into one second time intervals. Preliminary experiments show that this not only reduces the size of the dataset, but also removes some noise. Since at this point, the data is in the CSV format, Pandas \cite{reback2020pandas} can be used to process this data. This framework offers all necessary aggregation functionalities out of the box and with its intuitive language designed paired with the integration into the \textit{Python} environment, it is the perfect tool for this task. Then, during the evaluation phase, matplotlib \cite{Hunter:2007} was used for creating the visualizations.

\chapter{Results}
\label{cha:results}

This chapter shows the results of analyses conducted in this research. We will begin with the lower-level findings which establish the context-clues. Specifically, we begin by analysing the smart home setup itself where we show how accurately the devices and their state can be identified and what conclusion can be drawn about the smart home inhabitants. Then, we analyze the RSSI readings from the sensors to establish localization information. We examine how accurately we can distinguish stationary devices from mobile devices, then how this information can be used to derive a floor plan of the smart home. Finally we apply the floor plan to track mobile device's movements to see where the neighbor is currently in the house. Lastly, we use the gathered contextual information to perform high-level human activity recognition. 

\section{What can your neighbor learn about your smart home?}

\subsection{Device identification}
\label{sec:dev-id}
It is trivial to determine which WiFi-capable devices are connected to the neighbor's network once we know their SSID. All 802.11 data frames carry the information about which AP the device is connected to, so a list of all devices associated with the smart home can be generated using a single packet filter. Every WiFi-capable device which was used in this experiment could be observed in the network traffic dump. 

However, this information on its own is not of great value for the attacker, without their type or any other nearer classification. In this section, we analyze the side-channel information of the smart devices to identify them. Specifically, we present how their activity patterns can provide insight into their type. Then, we apply OSINT to the disclosed information in the 802.11 protocol and BLE to classify them. 

\begin{figure}[h]
	\centering
	\includegraphics[width=\linewidth]{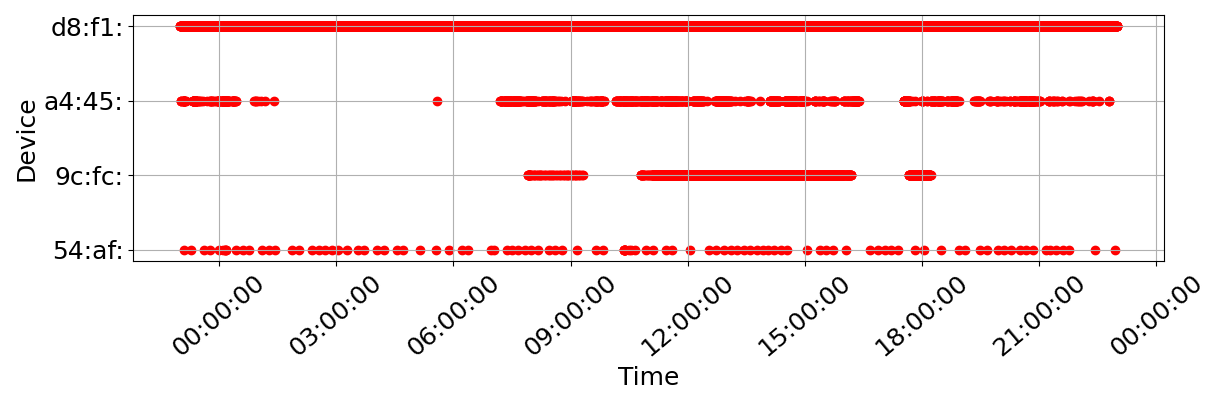}
	\caption{Visualization of presence of selected devices in the network on a specific day.}
	\label{fig:inventory}
\end{figure}

Plotting the presence in the network over time reveals patterns in the daily schedules of the smart home inhabitants as well as first insights into the device types. Figure \ref{fig:inventory} shows the usage of four selected devices in one day. In the usage of a smartphone, laptop and two smart devices. We observe, the first and last device are active at all times, which implies that these devices operate on their own, without human interaction. This is a typical trait of smart devices, especially devices such as sensors which periodically send their readings around the clock. Figure \ref{tab:smart_vs_human} presents the results of this identification. Every device could bee correctly classified as smart or manually-controlled.

\begin{table}[h]
	\centering
	\renewcommand{\arraystretch}{1.3} 
	\captionsetup{skip=1ex}
	\setlength{\tabcolsep}{8pt}
	\begin{tabular}{|c|c|}
		\hline
		\textbf{Mac address} & \textbf{Type} \\ \hline
		9c:fc & manually-controlled \\ \hline
		20:28 & manually-controlled \\ \hline
		60:1a & manually-controlled \\ \hline
		a4:45 & manually-controlled \\ \hline
		d8:f1 & smart \\ \hline
		08:b6 & smart \\ \hline
		6c:5a & smart \\ \hline
		54:af & smart \\ \hline		
		24:2f & smart \\ \hline
		8c:f6 & smart \\ \hline
	\end{tabular}
	\vspace{1em}
	\caption{Device classification as smart device or manually-controlled device based on network activity. }
	\label{tab:smart_vs_human}
\end{table}

%

The BLE advertisement packets are observable for all Bluetooth-capable devices. They disclose at least the Complete Local Name field for all devices. Table \ref{tab:shelly_ble_data} shows the BLE advertisement information of the air quality sensor. The device name of every smart device is descriptive enough to be used as a search query to get specific information about each device. Using Google search to lookup the broadcasted names yields the correct product page as the first result. Lack of service information or the manufacturer field is compensated by the very descriptive name which leads to the product page.

\begin{table}[h]
	\centering
	\renewcommand{\arraystretch}{1.3} 
	\captionsetup{skip=1ex}
	\setlength{\tabcolsep}{8pt}
	\begin{tabular}{|c|c|}
		\hline
		\textbf{Field} & \textbf{Value} \\ \hline
		Complete Local Name & ShellyPlusHT-08B6 \\ \hline
		Manufacturer & a90b010 \\ \hline
	\end{tabular}
	\vspace{1em}
	\caption{Fields of a BLE advertisement packet sent by the air sensor. Values are obfuscated for privacy. }
	\label{tab:shelly_ble_data}
\end{table}

In the case of this experiment, it did not occur that a device broadcasted its services, but not its name. In such case, it still would be possible determine the device type by decoding the service UUID using Wireshark, as presented in Figure \ref{fig:ble_wireshark}. 

Overall, using the Bluetooth Low Energy advertisements, three devices could be precisely identified. Out of five Bluetooth-capable devices, three of them disclosed their full device name in the advertisements, leading to identification.

If the device does not broadcast BLE advertisement packets or if its name is not present, the mac address and usage patterns can be used to narrow down possible device types. All devices which are considered in this experiment revealed their manufacturer through the OUI prefix in their mac address. Considering the activity shown in Figure \ref{fig:inventory} we can see that the device with  mac address beginning with \textbf{d8:f1} operates at times when the inhabitants are possibly sleeping - it might be a smart device. The OUI record for this mac address prefix is presented in Figure \ref{fig:esp_oui}. We see that it was produced by Espressif Inc, a popular IoT device manufacturer. Combination of these two facts implies that it is likely a smart device. However it is not possible to determine the specific type of the device without further contextual clues.

\begin{figure}[h]
	\centering
	\begin{tabular}{@{}ll@{}}
		\texttt{D8-F1-5B (hex)}   & Espressif Inc. \\[5pt]
		\texttt{D8F15B (base 16)}  & Espressif Inc. \\[5pt]
		& Room 204, Building 2, 690 Bibo Rd, Pudong New Area \\
		& Shanghai 201203, CN \\
	\end{tabular}
	\vspace{1em}
	\caption{Information about the OUI of the d8:f1 mac address. Adapted from \cite{mac-lookup}.}
	\label{fig:esp_oui}
\end{figure}

Every single device, except the movement sensor, could be at least partly identified using the OUI information in their mac address. This information, paired with their traffic patterns, such as those in Figure \ref{fig:inventory}, narrows down the possible device type. For example, a device with OUI field pointing to a manufacturer of IoT devices and exhibiting a constant traffic, is most likely a smart device.

The source of information that undoubtedly revealed the most insight into a device, is the installation process of a WiFi-capable devices. When a new smart device with no Bluetooth capabilities is coupled with its respective companion app, it creates an unencrypted WiFi network. In order to complete the pairing process, the smartphone must connect to this network. There, the device and the smartphone exchange configuration information and the open network closes. The unencrypted connection is open for not longer than three minutes, yet it is enough for a passive eavesdropper to gain insights into the specifics of the device. This process occurred during the installation of two devices.

This information is only available if we assume that the neighbor would detect and record this communication. In reality, it is a rare event - only devices which are only WiFi-capable create their own network for configuration. Then, this usually occurs only once per device. 

\begin{figure}[h]
	\centering
	\includegraphics[width=\linewidth]{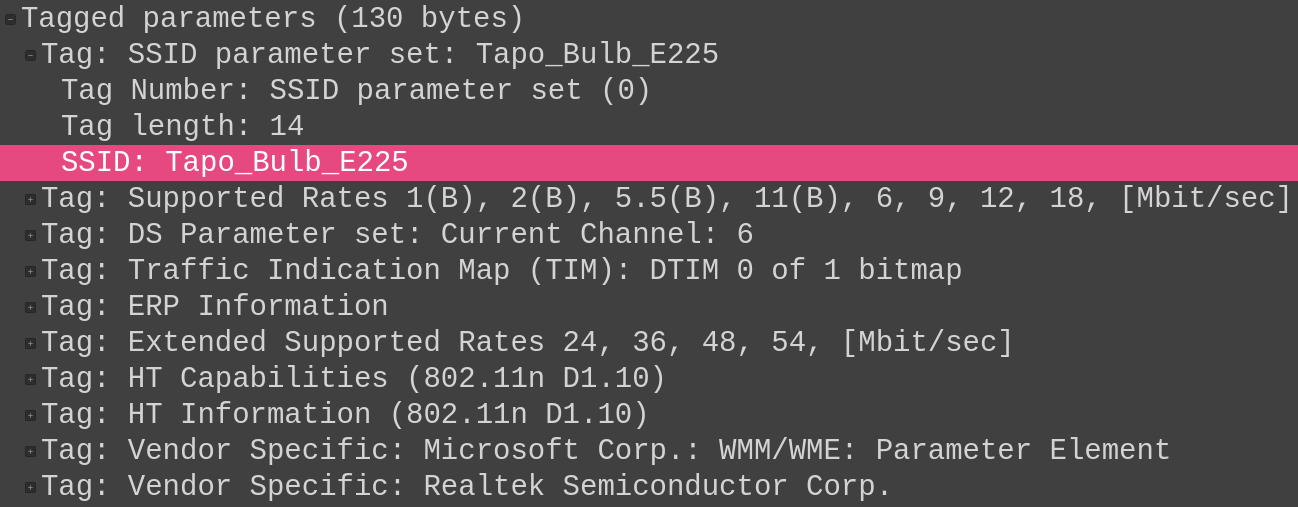}
	\caption{Contents of the beacon frame of the Tapo Light Bulb E225.}
	\label{fig:tapo_installation}
\end{figure}

First device which opened its own insecure network was the Tapo Light Bulb E225 smart light bulb. This was immediately visible in the network traffic by the presence of its beacon frames which dislosed the device name, as shown in Figure \ref{fig:tapo_installation}. Beacon frames reveal all of the necessary information about the device in order to determine its type, manufacturer and model. Its SSID, being the device name, can be used to find its product page. Sniffing the unencrypted traffic reveals information about the device's internal technologies such as the HTTP endpoints and open ports, but no relevant technical or personal information is disclosed. Despite the traffic being transmitted over HTTP, the POST request bodies seem to be encrypted, given their high Shannon entropy value of over 7 for two selected POST bodies.

Second device which opened its own network for configuration was the Shelly Motion sensor. This process was analogous to the previous device, however it disclosed more information about the device. Similarly to the first installation process, device name was visible in the beacon frame. Then, during the coupling process, the device communicated to the smartphone over HTTP, which revealed the device's configuration settings as well as its used technologies. 

\begin{table}[h]
	\centering
	\renewcommand{\arraystretch}{1.3} 
	\captionsetup{skip=1ex}
	\setlength{\tabcolsep}{8pt}
	\begin{tabular}{|c|c|}
		\hline
		\textbf{Information} & \textbf{Value} \\ \hline
		Firmware & 20220801-153139/v2.1.8@bbbe9821+ \\ \hline
		Web Server & lwIP/2.1.2 \\ \hline
		SNTP Server & time.google.com \\ \hline
	\end{tabular}
	\vspace{1em}
	\caption{Information about the Shelly Motion sensor disclosed during the coupling process.}
	\label{tab:shelly_information}
\end{table}

The motion sensor hosts a configuration web-interface at 192.168.33.1:80. Because the web server operates without SSL, some of the requests and responses could be observed. Table \ref{tab:shelly_information} shows which device information were disclosed in the HTTP traffic. Web server transmitted critical device configuration such as blind time, i.e. how much time must pass before the motion sensor detects another reading. A burglar could use this information to circumvent this sensor's security features. 

\begin{figure}[h]
	\centering
	\includegraphics[width=\linewidth]{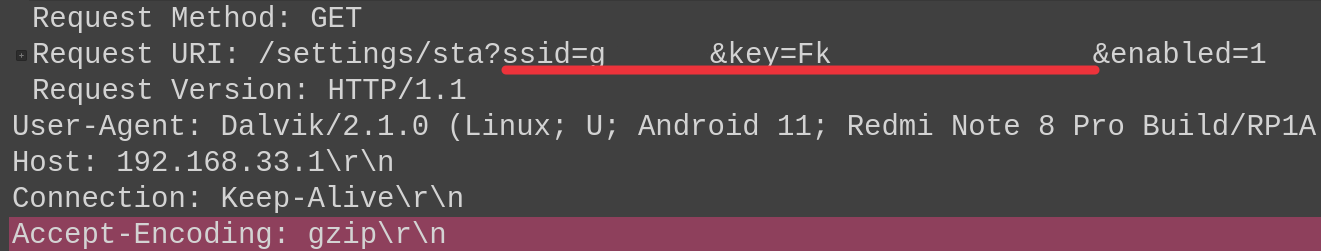}
	\caption{WiFi credentials disclosed in plain text by the companion application. SSID and password are obfuscated. }
	\label{fig:wifi_creds}
\end{figure}

The last HTTP request sent by the smartphone to  configuration interface discloses the most critical information of the entire coupling process, after which the open WiFi network closed. There, smartphone submits the WiFi credentials to the device as parameters in a GET request over HTTP. Figure \ref{fig:wifi_creds} shows the contents of this request. Recording this information would grant a nosy neighbor unrestricted access to the victim's network. After intercepting the four-way EAPOL encryption handshake of any device, the adversary could then decrypt the network traffic for that device. This allows them to perform profiling attacks on the application layer as discussed in Section \ref{cha:fingerprinting}. Those attacks disclose important information which aid the device identification. For example, DNS traffic of these devices may disclose the manufacturer, as discovered in \cite{apthorpe2017smarthomecastleprivacy}.

\begin{table}[h]
	\centering
	\renewcommand{\arraystretch}{1.3} 
	\captionsetup{skip=1ex}
	\setlength{\tabcolsep}{8pt}
	\begin{tabular}{|c|c|}
		\hline
		\textbf{Information} & \textbf{Value} \\ \hline
		User Agent & Dalvik/2.1.0 (Linux; U; Android 11; Redmi Note 8 Pro Build/RP1A.200720.011) \\ \hline
		Device Name & Readme Note 8 Pro \\ \hline
		Services & mi\_connect, spotify-connect, spotify-social-listening, googlecast \\ \hline
	\end{tabular}
	\vspace{1em}
	\caption{Information about the smartphone disclosed during the coupling process.}
	\label{tab:phone_information}
\end{table}

\begin{figure}[h]
	\centering
	\includegraphics[width=\linewidth]{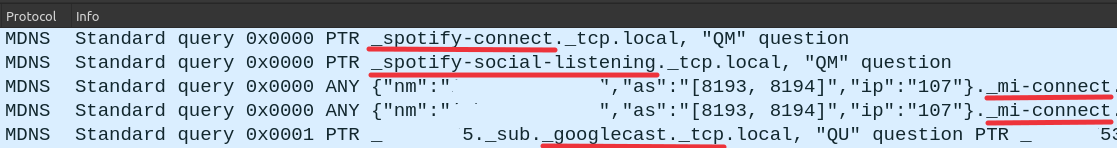}
	\caption{MDNS traffic generated by the smartphone in an unencrypted network. Service names are highlighted with the red line. Device name and Googlecast ID are obfuscated. }
	\label{fig:mdns_traffic}
\end{figure}

While the smartphone was connected to a unencrypted network, its mDNS traffic is present in the network dumps. This protocol tends to reveal privacy sensitive information such as the name of the owners, apps they have installed on their phones and most users are not aware of this \cite{6569062}. Table \ref{tab:phone_information} shows which information about the smartphone could be retrieved during the coupling process of both devices from the mDNS queries and the HTTP traffic. Figure \ref{fig:mdns_traffic} shows a fragment of the observed mDNS packets sent by the companion smartphone. 

Table \ref{tab:device_identification} shows what information about each installed smart home device could be determined and what was the source of this information. We could identify the general type of half of the devices. Mac address all of the devices reveals its manufacturer, however it can be sometimes misleading as discussed in Section \ref{cha:ble}. In this case, the times at which the device is present in the network can aid the neighbor in determining a possible device type. The correlation between devices being Bluetooth-capable and the amount of information they disclosed is great, because the BLE advertisements discloses much information about the device, by design. Lastly, if we exclude the observations from the coupling process, the models of the smartphone, light bulb and the motion sensor would be unknown. However using patterns in the activity, it would still be possible to approximate their type as multimedia and smart devices respectively. 

\begin{table}[h]
	\centering
	\renewcommand{\arraystretch}{1.3} 
	\captionsetup{skip=1ex}
	\setlength{\tabcolsep}{8pt}
	\begin{tabular}{|c|c|c|c|c|c|}
		\hline
		\textbf{Mac} & \textbf{Type} & \textbf{Manufacturer} & \textbf{Model} & \textbf{Source} \\ \hline
		9c:fc & Multimedia & Intel & Unknown & mac address \\ \hline
		d8:f1 & Smart device & Intel & Espressif & mac address \\ \hline
		08:b6 &  Air sensor & Shelly & Shelly Plus HT & BLE \\ \hline
		6c:5a &  Light Bulb & Tapo  & Tapo Bulb E225 & installation \\ \hline
		54:af & Smart device & TP-Link & Unknown & mac address \\ \hline		
		24:2f &  Router & TP-Link & Unknown & mac address \& addressing \\ \hline
		20:28 &  Smart TV & LG & TV UQ75009LF & BLE \\ \hline
		60:1a &  Gaming console & Nintendo & Nintendo Switch & mac address \\ \hline
		a4:45 &  Smartphone & Xiaomi & Readme Note 8 Pro & installation \\ \hline
		8c:f6 & Motion sensor & Shelly & Shelly Motion 2 & installation \\ \hline
	\end{tabular}
	\vspace{1em}
	\caption{Results of the OSINT research for device identification based on their broadcasted traffic.}
	\label{tab:device_identification}
\end{table}

\subsection{Device state recognition}
\label{sec:dev-state}
\begin{figure}[h]
	\centering
	\includegraphics[width=\linewidth]{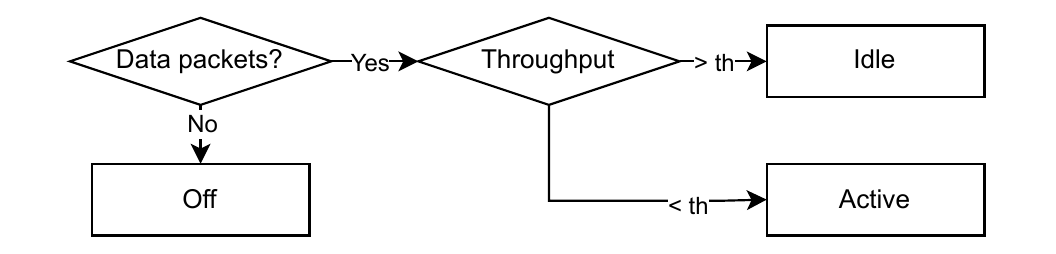}
	\caption{Flowchart of the device state classification process. The distinction whether a device is idle or active is based on a individual threshold (th) for each device. }
	\label{fig:state_class}
\end{figure}

Knowing which devices and their approximate types, their activity can be analyzed to gain more insight into the smart home. We can categorize the traffic of each device into three states: off, idle and active. A device is considered off if no 802.11 traffic from this device has been observed for a some amount of time, which varies between the devices. For example a smart phone which is being used to stream videos generates more traffic than a temperature sensor which periodically submits the readings. Then, the device is considered idle if some traffic can be observed. Lastly, a device is active if it sends significantly more traffic than during its idle state. Figure \ref{fig:state_class} visualizes this categorization workflow. 

\begin{figure}[h]
	\centering
	\includegraphics[width=\linewidth]{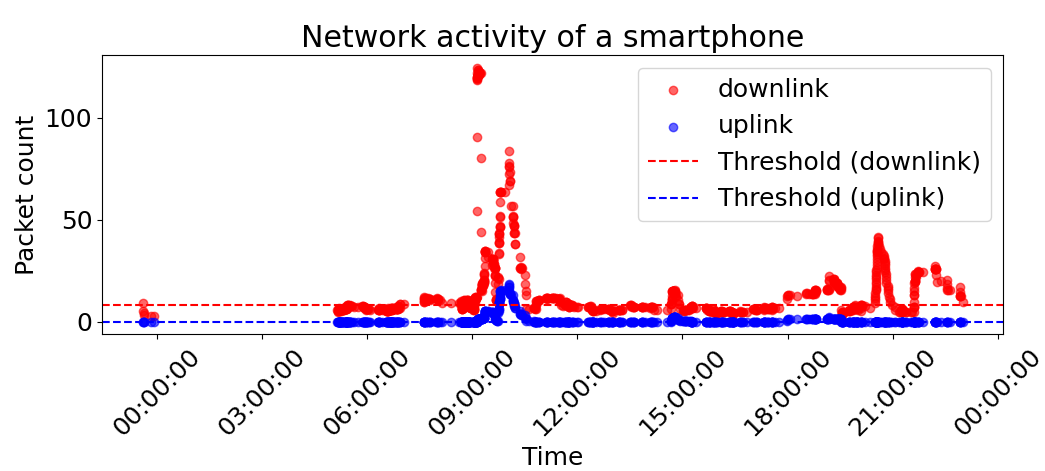}
	\caption{Packet count of sent (blue) and received (red) packets by the smartphone. Threshold represents the median packet count during that day. }
	\label{fig:phone_traffic}
\end{figure}

Network traffic of a device can be derived from the amount of the sent and received data packets, i.e. the uplink and downlink traffic, per time interval. Other network traffic parameters such as packet size, transmission time and i/o ratio correlate with the amount of the packets in a high degree, so traffic throughput alone suffices to describe different device states. While other parameters are relevant for determining the specific traffic type, such as streaming, listening to music or reading sensor values, they are not relevant to determine the state of a device using our classification. Figure \ref{fig:phone_traffic} shows the packet count in a ten second interval. It is visible with the naked eye when the device is off, idle or active. 

\begin{figure}[h]
	\centering
	\includegraphics[width=\linewidth]{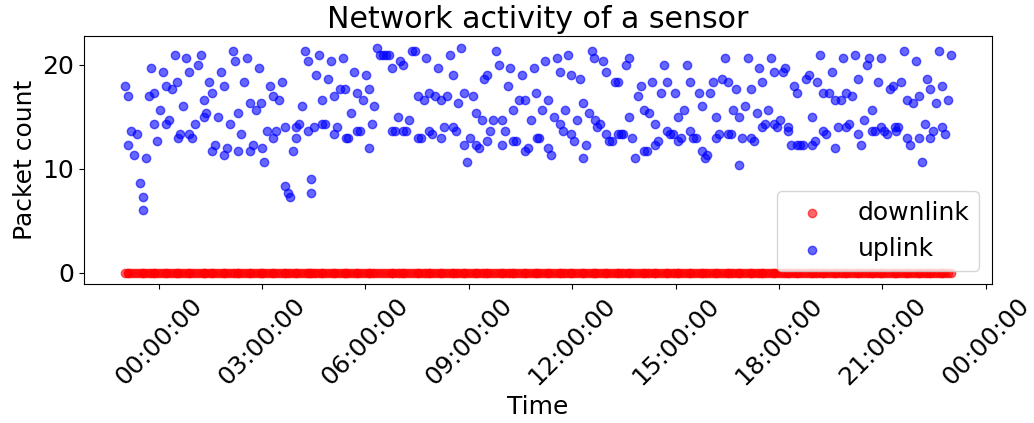}
	\caption{Packet count of sent (blue) and received (red) packets by the air quality sensor. }
	\label{fig:sensor_traffic}
\end{figure}

Examining the different device states throughout a day can give insight into the device type. For example, devices such as sensors submit their readings to the cloud at a constant rate. Figure \ref{fig:sensor_traffic} visualizes the traffic generated by a smart air quality sensor. There are no clear patterns, because the sensor sends its data in regular intervals around the clock. It is worth noting, that the traffic consists of uplink packets. This might be explained by packet loss in the sniffing system, but it is also reasonable to assume that a smart sensor would not download much data. 

\begin{figure}[h]
	\centering
	\includegraphics[width=\linewidth]{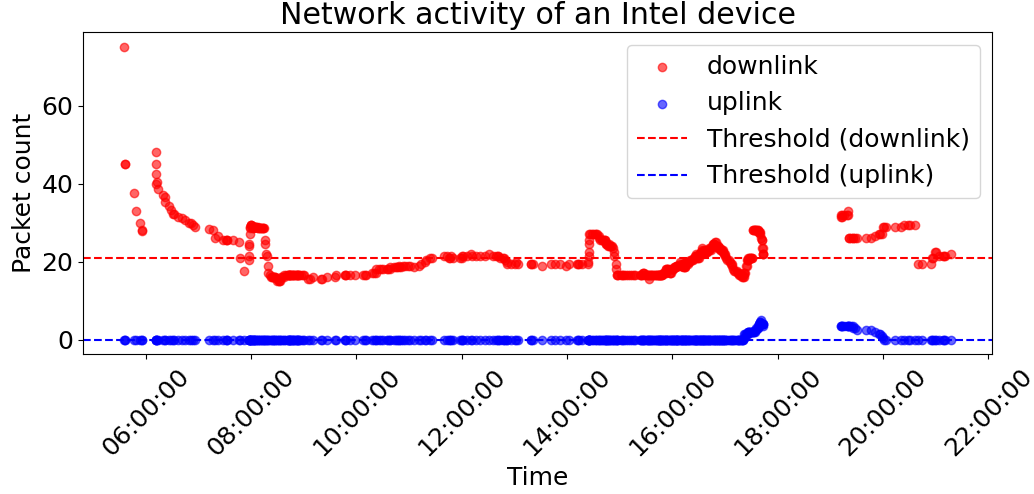}
	\caption{Packet count of sent (blue) and received (red) packets by an Intel device. Threshold represents the median packet count during that day.}
	\label{fig:tux_traffic}
\end{figure}

Then, the device state together with other background information can be used to approximate the type of a device. We could determine that the device with the mac address beginning with\textbf{ 9c:fc} might be a multimedia device, because it is used during daytime. Analysing its state throughout the day confirms this claim, since its traffic patterns are similar to the smartphone's. Figure \ref{fig:tux_traffic} shows that this device is active during daytime, with various activity states, which suggests traffic generated by different media types. For example, during its idle state between 8:00 and approx. 11:30, this device did not generate much traffic. Then, at around 14:00 the user might have watched a video on this device, generating more traffic then before. Given that the device's vendor is likely Intel, a popular computer manufacturer, we can assume that this device is a PC or laptop. The traffic burst in the morning, i.e. shortly after this device was online, speaks for this hypothesis. Operating systems and software often check and download updates shortly after booting, which is detectable as such downlink traffic spike. Research suggests that traffic-based OS recognition is possible \cite{7969609, 7444941, 8013420}.

\begin{figure}[h]
	\centering
	\includegraphics[width=\linewidth]{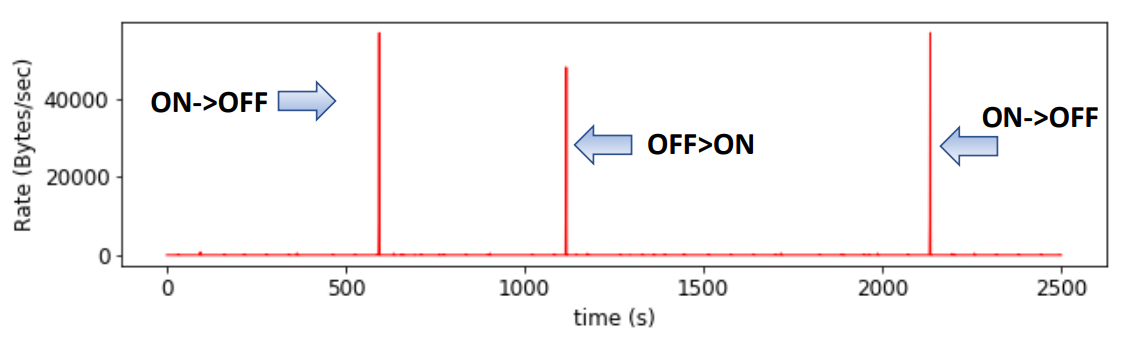}
	\caption{Usage state recognition of a WiFi light switch. Adopted from \cite{10.1145/3395351.3399421}. }
	\label{fig:peekabo_traffic}
\end{figure}

Certain types of smart home devices which are operated by humans can potentially reflect human activity in their traffic. For example, devices which activate on human interaction like light bulbs or plugs, can generate different types of traffic when being activated than when they're in use. Researchers in this paper \cite{10.1145/3395351.3399421} observed this phenomenon. Figure \ref{fig:peekabo_traffic} shows how the on/off state of a smart light switch can be detected in the traffic. Unfortunately, this phenomenon could not be observed any smart devices of this experiment, except the Smart TV. Neither the packet count, nor any other mentioned network traffic parameters reflected the usage of devices. It depends on the specific implementation of the firmware whether device state is visible in the traffic. For example, the state can be represented as an integer, 1 being active and 0 being idle. This information can then be transmitted in the keep-alive communication, which happens regularly. In such case, the device state doesn't affect the observed traffic.

In conclusion, a nosy neighbor could determine which devices are connected to its victim's network. Then, using information such as mac address, BLE advertisement, or unencrypted traffic of a WiFi device, they could determine what kinds of devices are installed in the smart home. Finally, we observe that the states of certain devices are reflected in the 802.11 traffic. Activity states of multimedia devices are visible with the naked eye, but no analyzed smart home devices revealed their state in the traffic. However, research suggests, that it is possible for some devices. 

\section{Can your neighbor know where you are?}
As research suggests, it is possible to locate devices based on their signal strength. This section presents what a nosy neighbor could learn about the spatial layout of a smart home from behind the wall. First, we determine if stationary devices can be effectively distinguished from mobile devices. Next, we analyze their fingerprints to detect interesting areas in the house.  Then we combine the gathered knowledge to approximate a floor plan of the apartment using context clues from the previous section. Finally, we present to what extent a device's location in the flat can be tracked in real time using our sniffing setup.

\subsection{Locomotion}
\label{sec:localization}
\begin{figure}[h]
	\centering
	\includegraphics[width=\linewidth]{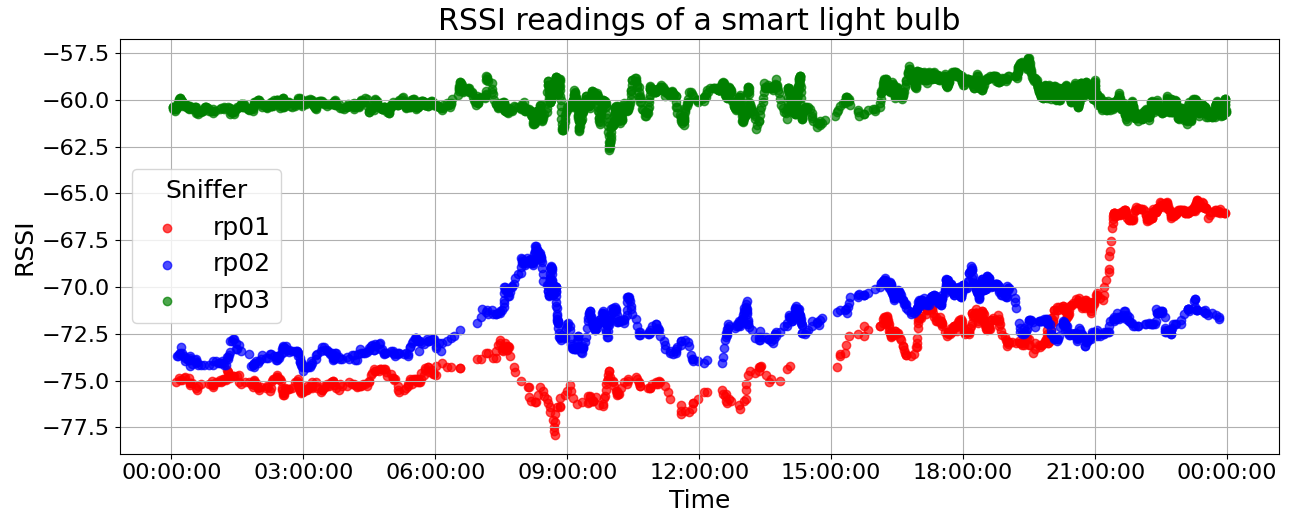}
	\caption{Signal strength readings of a smart light bulb throughout the day received by each sensor. }
	\label{fig:rssi_lightbulb}
\end{figure}

An adversary can infer whether a certain device is stationary or mobile based on its type. For instance, a smart light bulb or plug is typically installed in the ceiling socket and remains stationary for the vast majority of its lifetime. This immobility is reflected in the signal strength readings captured by the sniffing setup. If a device is not moving, the RSSI values at each sniffer should remain constant or at least stable. In reality, received signal strength fluctuates throughout the day due to interference with other radio signals and human presence. Figure \ref{fig:rssi_lightbulb} shows the RSSI readings of a smart light bulb received by the three sniffers. We see that the readings for each sniffer are relatively static. 

Interestingly, human activity is detectable in these RSSI readings as noise, even if the device is remains stationary. At night, the readings are almost perfectly constant during until 6:00. Then, readings begin to be less linear and more variable until they stabilize again at around 21:00. This phenomenon can be explained by the obstructions which temporarily occur during that time due to human activity. It is possible that a person was between the device and sniffer which distorted the readings. Later that day, we see that sniffer \textit{rp01} received a higher signal amplitude. Although the exact cause is unclear, given the timing of this event, it might correlate with some human activity. Or, if this light bulb was mounted in a lamp rather than a ceiling socket, it could have been moved.

\begin{figure}[h]
	\centering
	\includegraphics[width=\linewidth]{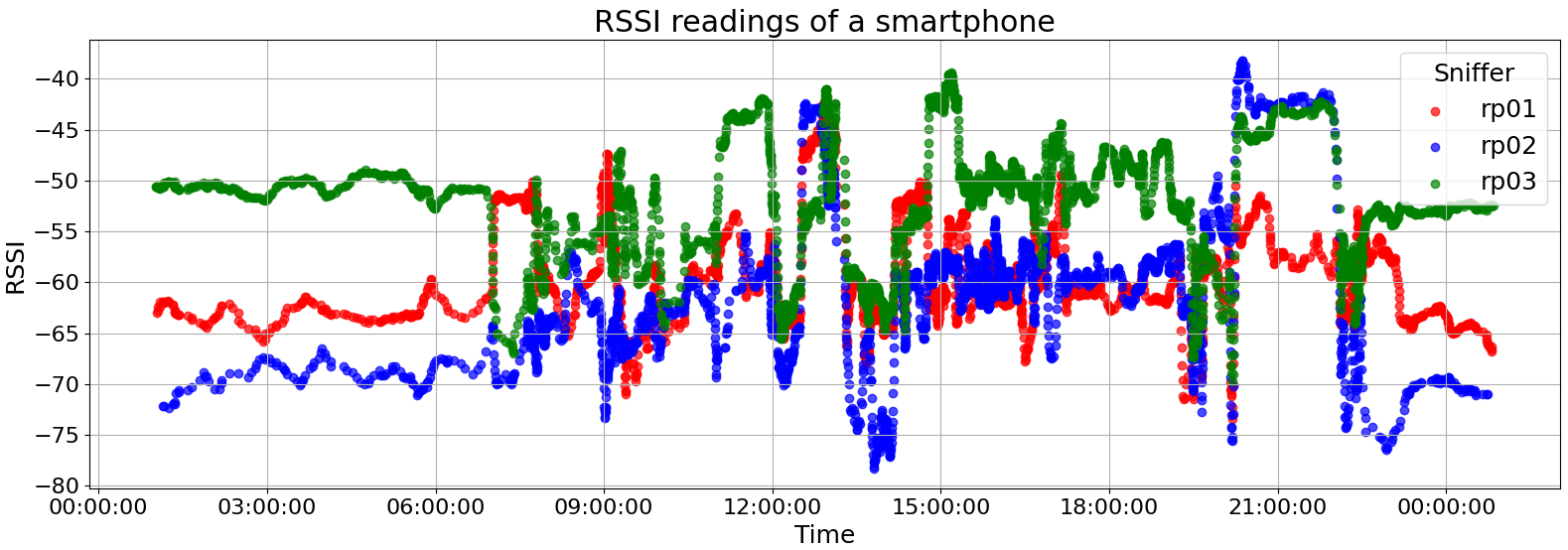}
	\caption{Signal strength readings of a smartphone throughout the day received by each sensor. }
	\label{fig:rssi_phone}
\end{figure}

Readings of mobile devices follow a different patterns, as presented in Figure \ref{fig:rssi_phone}. Similar to the smart light bulb, readings during night are stable. Then, likely after the phone's owner woke up, readings become more chaotic. Notably, between 15:00 and 20:00 and 20:00 to 22:00, the phone was immobile, yet the RSSI fingerprints differ significantly. This indicates that the phone was in different areas of the house. However, lacking any contextual information about these phases beyond the time of day, we can not determine the phone's exact location during these intervals. 

%

Table \ref{tab:stationary_devices} summarizes the results of our locomotion analysis. A passive eavesdropper equipped with spatially separated sniffers is be capable of distinguishing stationary from mobile devices. For example, as presented in Figures \ref{fig:rssi_lightbulb} and \ref{fig:rssi_phone}, the difference between these fingerprints is visible with the naked eye and it is trivial to classify the devices automatically.

\begin{table}[h]
	\centering
	\renewcommand{\arraystretch}{1.3} 
	\captionsetup{skip=1ex}
	\setlength{\tabcolsep}{8pt}
	\begin{tabular}{|c|c|c|}
		\hline
		\textbf{Mac} & \textbf{Type} & \textbf{Stationary} \\ \hline
		a4:45 & Smartphone & No \\ \hline
		9c:fc & Laptop & Yes \\ \hline
		d8:f1 & Tuya light bulb & Yes \\ \hline
		08:b6 & Air sensor & Yes \\ \hline
		6c:5a & Tapo light bulb & Yes \\ \hline
		54:af & Tapo smart plug & Yes \\ \hline		
		8c:f6 & Motion sensor & Yes \\ \hline
	\end{tabular}
	\vspace{1em}
	\caption{Classification as stationary or mobile device using their RSSI fingerprints.}
	\label{tab:stationary_devices}
\end{table}

\subsection{Room recognition}

In this section, we demonstrate how contextual information can help a nosy neighbor detect interesting areas within a smart home. First we show how an attacker can determine if two devices are close to each other. Then, we use information about the device types to classify areas in the house. Finally we present how movement patterns of mobile devices to aid the room recognition. 

\begin{figure}[h]
	\centering
	\includegraphics[width=\linewidth]{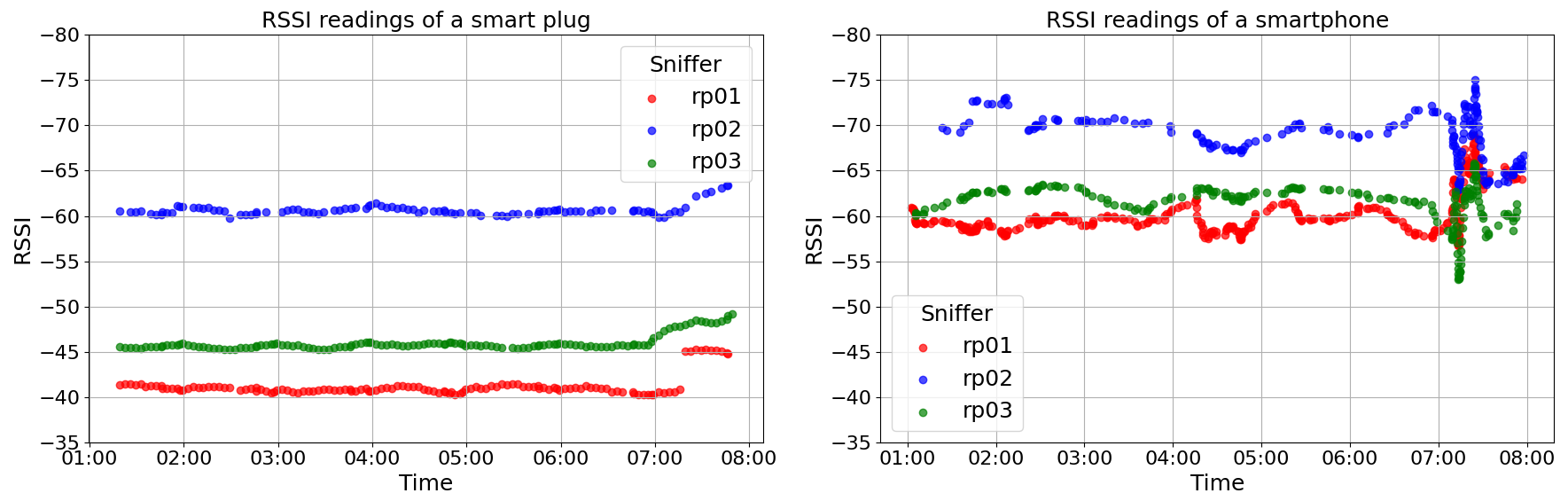}
	\caption{RSSI fingerprints of a smart plug and smartphone in the same room. Individual readings have the same vertical ordering and are proportionally spaced on both figures.}
	\label{fig:compared_rssi}
\end{figure}

Data shows that it is possible to approximate whether devices appear close to each other using their RSSI fingerprints. Precise identification is impossible due to unknown path loss factors, i.e. how much signal strength is lost between the transmitter and receiver due to obstructions. Then, devices possibly transmit their radio communication at different power levels, so they appear further away if power of one emitter is significantly lower. It is visible in Figure \ref{fig:compared_rssi}, where we observe that although their RSSI readings have roughly the same shape, they are scaled differently. In reality, the devices were quite close to each other and if their transmission power was equal, their RSSI fingerprints would almost overlap. While an attacker would likely deduce that the devices use varying TX, they would not know the precise values and could not effectively calibrate their algorithms. 

\begin{figure}[h]
	\centering
	\includegraphics[width=\linewidth]{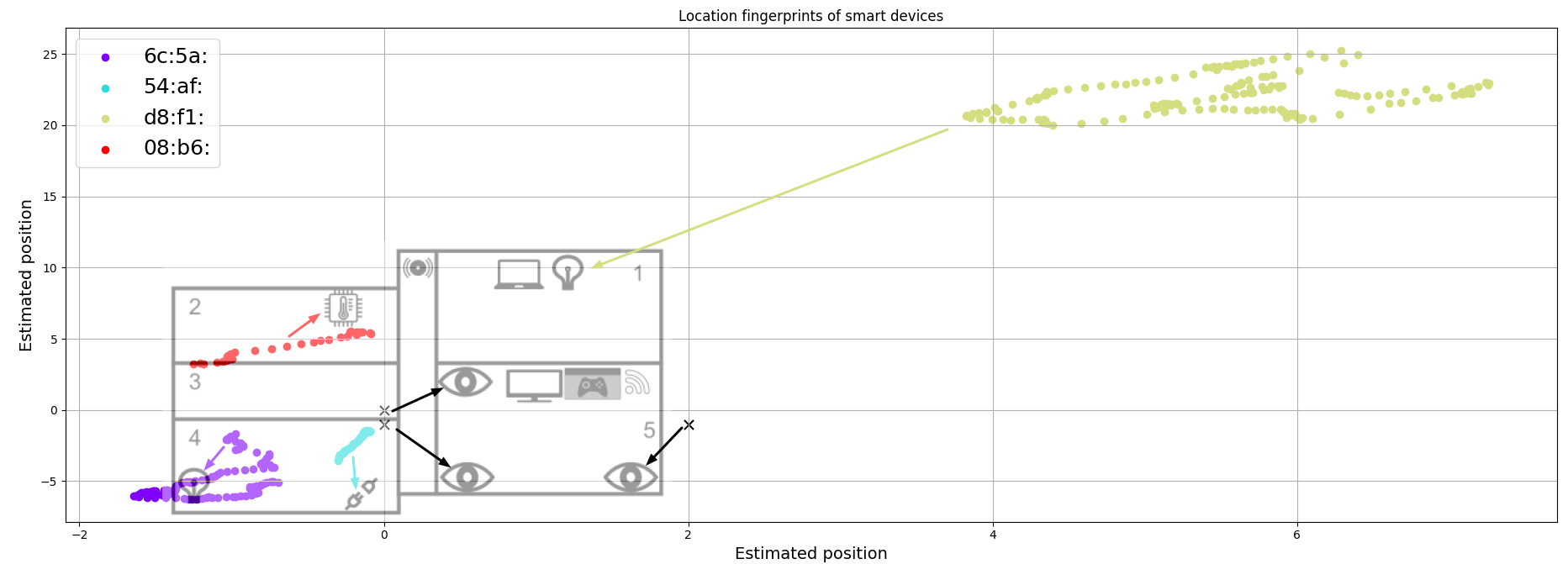}
	\caption{Estimated position of smart devices using their RSSI fingerprints and trilateration. Dots represent the calculated position of  devices. X represents location of the sniffers.}
	\label{fig:rssi_position}
\end{figure}

Despite the RSSI not reflecting the distance accurately, we can still apply trilateration \cite{LI2020103309, fang1986trilateration} to project the RSSI readings onto a 2D plane, i.e. a floor plan. It is important to note that scaling is unrealistic and devices appear closer or further away than they actually are, due to varying TX values and path loss. However, one crucial element of their position remains accurate: the direction of their signal. In order to calculate their approximated position, we first need to map the RSSI values to euclidean distance. This can be done using the formula introduced in Section \ref{cha:rssi_localization} with an arbitrary path loss coefficient (here: four), as it can not be accurately determined. Precise value of this number is not relevant for this scenario, as it mostly affects scale, not direction. Figure \ref{fig:rssi_position} shows the approximated positions of selected stationary devices. We observe that some calculated locations coincidentally overlap with their true location. More importantly, each signal comes from varying directions, which allows differentiating between different areas in the house. An attacker would not know that the red device (temperature sensor, see Table \ref{tab:smart_home_devices}), is in Room 3, (kitchen, see Table \ref{fig:sniffer_setup}, but could consider it a distinct sector of the house. Then, using this estimation, it would be reasonable to assume that cyan and purple devices (Tapo smart bulb and smart plug) are located in the same room. 

\begin{figure}[h]
	\centering
	\includegraphics[width=\linewidth]{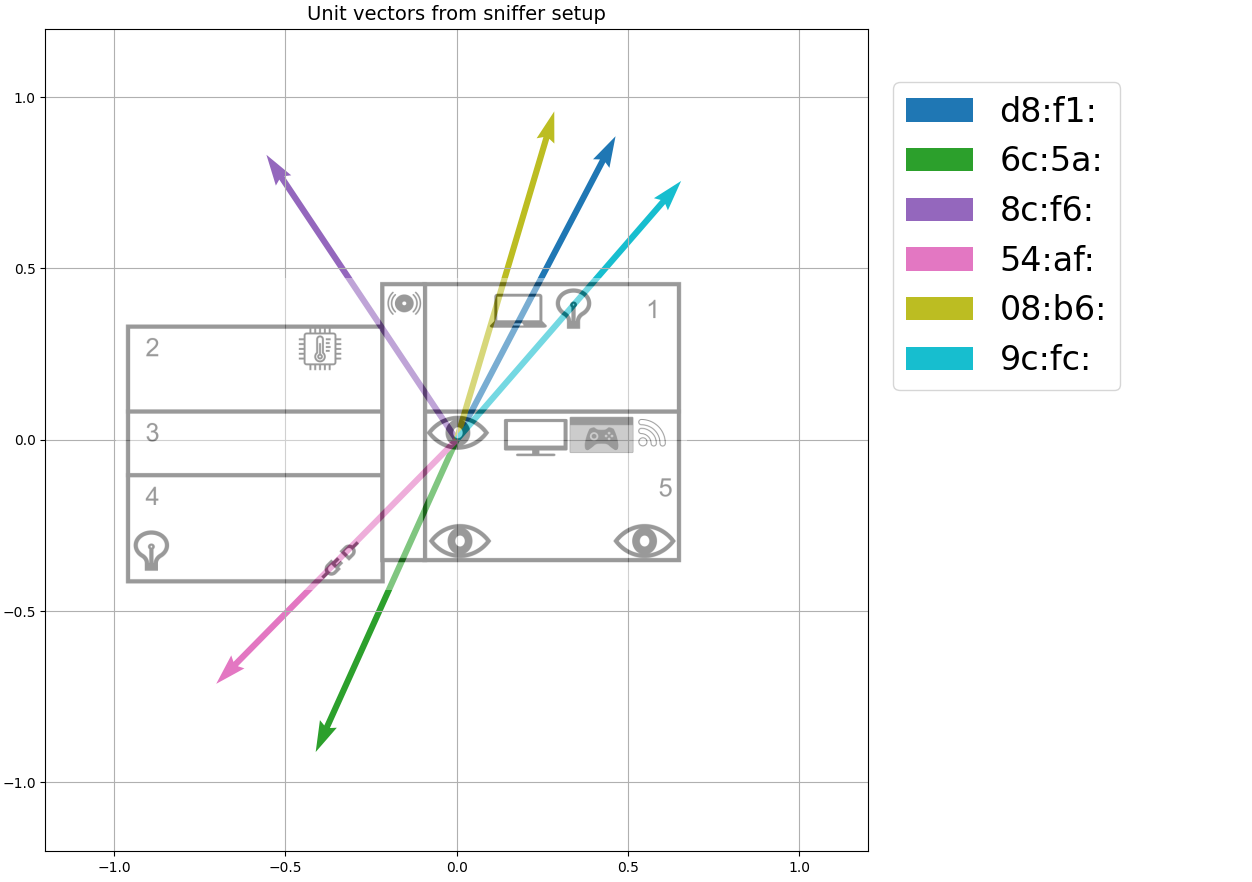}
	\caption{Direction vectors of the RSSI signals of all smart devices.}
	\label{fig:unit_vectors}
\end{figure}

Figure \ref{fig:unit_vectors} visualizes the direction vectors to the position of each stationary device. Vectors are normalized for visualization by transforming them into unit vectors. As we see, the arrows point quite accurately into the correct areas of the house where the devices are located. Only the motion sensor's and air sensor's (gold and purple, see Table \ref{tab:smart_home_devices}) location is not correctly represented. This can be explained by physical obstructions between these devices and the sniffers. 

Using this results, it is possible to draw conclusions about a floor plan of the adjacent apartment. While it would not be possible to visualize walls and individual rooms, we can still give some context to areas. For example, we see in Figure \ref{fig:unit_vectors} that device with mac address beginning with \textbf{9c:fc} is located in the north-west quadrant of the apartment. Previous device identification analysis reveals that it is most likely a laptop. Hence, we can deduce that user's office is located in that area. Under the assumption that security-related devices are typically mounted near the entrance, we can also deduce that the office is next to the entrance. We see that the vector for device with address \textbf{8c:f6}, identified as a motion sensor, points in a similar direction to the laptop. At this point, no other meaningful conclusions can be drawn from the devices, since they are generic and have no typical room associated with them. If room specific devices, such as cooking utensils like Thermomix, smart exercise devices or gaming consoles are present, an attacker has more context clues to better identify the areas.

\begin{figure}[h]
	\centering
	\includegraphics[width=\linewidth]{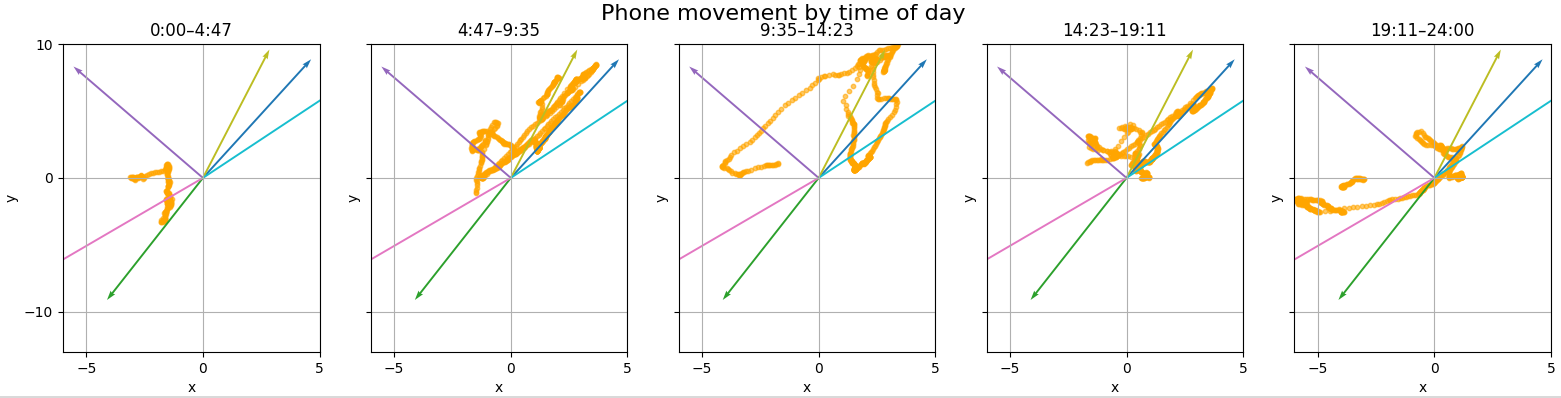}
	\caption{Estimated location of the phone, represented by orange dots, at different times of the day.}
	\label{fig:phone_track}
\end{figure}

Room identification can be improved if we consider the movement patterns of mobile devices throughout the day. Figure \ref{fig:phone_track} visualizes estimated position of the smartphone at different times of day. Each dot on the graph represents position during a 10 second time window. It is reasonable to map the phone's location in 2D, instead of reasoning about its position as direction vectors. While the distance is not realistic, we can still effectively differentiate between areas where the phone is closer or further away from the sniffers. Analyzing the phone's location over multiple days reveals that its transmission power is constant. Assuming that the phone is constantly generating network traffic, this kind of movement tracking can be done nearly in real-time. Following conclusions can be drawn from each subfigure: 

\begin{itemize}
	\item We observe that phone's position at night comes from a similar direction where the vectors of the Tapo plug and Tapo bulb point. We can conclude that these devices are located in the inhabitant's bedroom.
	\item Location in the early morning reveals the phone was in two distinct areas. First location fingerprint might indicate bathroom or kitchen, where the inhabitant would get ready or prepare breakfast. Then, they entered the area we classified as office. 
	\item Middle subfigure visualizes the position during lunchtime. As with the previous figure, we again see two distinct areas. One to the left most likely indicates the location of the kitchen, as the user likely prepared their lunch during that time. Reminder of the time was spent in the office. 
	\item Afternoon readings are most stable of all visualized times of the day. Here, the target was in the office. 
	\item Last subfigure shows where the user was at the evening. We see presence in three areas - office, bedroom and kitchen or bathroom. Readings in the left half of the figure are ambiguous and it is hard to tell if this which room it exactly is. Nevertheless, a typical daily schedule is reflected - the inhabitant finished their work and spent some time either in the bathroom, kitchen or in the bedroom, all of which is plausible. 
\end{itemize}

Smartphone's location fingerprints can be used as presented in Section \ref{cha:rssi_localization} to perform localization analysis based on detected reference points. As pointed out at the beginning of this section, estimating the true location of the phone using this data is infeasible and precise position of a smart device is irrelevant for this scenario. Instead, an adversary could use the interesting areas observed in Figure \ref{fig:phone_track} as reference points and apply clustering algorithms to classify the user's position. 

\begin{figure}[h]
	\centering
	\includegraphics[width=\linewidth]{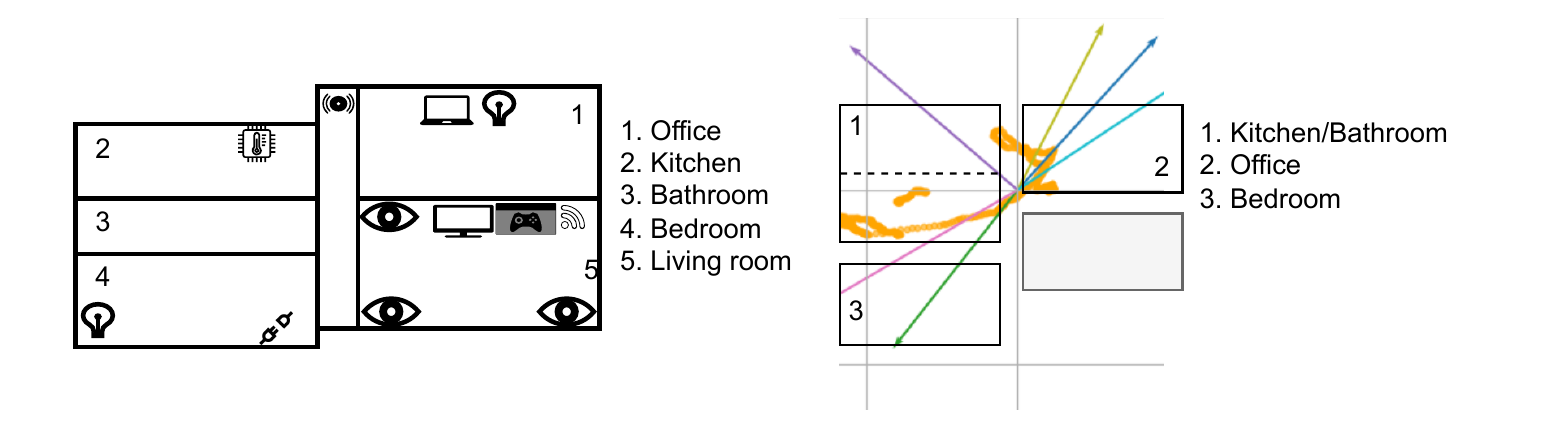}
	\caption{Sketch of a possible room layout of the victim's apartment. Gray box represents the sniffing room.  }
	\label{fig:interp_local}
\end{figure}

A nosy neighbor is able to confidently distinguish stationary from mobile devices. They can do it either based on their types or by analyzing fingerprints of their RSSI readings. Due to variations in the hardware of smart home devices and obstructions, an adversary might not accurately triangulate the position of all devices, but they could determine general area of the house where each device is likely installed. Then, if the types of localized devices are known, those areas can be classified. For example: areas where cooking devices are identified can reveal the position of the kitchen. Analogously room specific devices such as a baby camera reveals the location of a nursery. This method relies on the success of device identification, which is not guaranteed especially with WiFi-only devices. Alternatively, position of wearable or mobile devices at different times of the day can be used to develop an understanding of the apartment's layout. Here, the location of a smartphone at night revealed the RSSI fingerprint of the bedroom. Then, an area which could not be classified based on devices within it, could be determined based on timing patters in the smartphone's present as kitchen or bathroom. Figure \ref{fig:interp_local} provides an interpreted visualization of the identified areas in a smart home, compared to the true layout of the flat visualized in the same way.

\section{What can your neighbor learn about you?}

This section presents what a nosy neighbor can infer using information and methods presented in the previous sections. First an analysis of probe request based tracking will be presented. After that we aggregate the gathered context information to evaluate patterns in device activity and movements to derive weekly and daily schedules of the inhabitants. We conclude the results with a case study where we reconstruct a sleepover guest visit at the victim's house. 

\subsection{Probe requests}

Most WiFi-capable devices are configured to search for trusted networks if they are not connected to any WiFi network. As presented in Section \ref{cha:probe_request}, the name of the probed network is sent in plain text. WiGLE database can be used to lookup GPS coordinates of locations where WiFi networks with the same name were observed. This gives the attacker insight into the inhabitant's life beyond the smart home. This section presents which probe requests were recorded in the victim's smartphone's network traffic and which observations be made from this data. The true names and locations are randomized for privacy and contextually equivalent names are presented. Name of the network used in the experiment is represented as "EXPERIMENT". 

In our presumed scenario, where a nosy neighbor sets up network sniffers in an adjacent flat, no meaningful probes were observed. Table \ref{tab:probe_request_home} presents the recorded probe request of all smart home devices. All devices probed for the "EXPERIMENT network and the borrowed devices pinged networks of their previous owners. Overall this data is very minimal, which is an expected result.

\begin{table}[h]
	\centering
	\renewcommand{\arraystretch}{1.3} 
	\captionsetup{skip=1ex}
	\setlength{\tabcolsep}{8pt}
	\begin{tabular}{|c|c|}
		\hline
		\textbf{Mac} & \textbf{SSID} \\ \hline
		\multirow{2}{*}{a4:45}  & 123456789 \\ 
		& VodafoneMobileWiFi-A123456 \\ \hline
		\multirow{2}{*}{6c:5a}  & OpenRouter \\ 		
		& easy\_network \\ \hline	
		\multirow{2}{*}{54:af}  & OpenRouter \\ 		
		& easy\_network \\ \hline									
		d8:f1 & OpenRouter \\ \hline
		08:b6 & abc \\ \hline
	\end{tabular}
	\vspace{1em}
	\caption{Probe requests of all smart devices. Experiment network is excluded, as it was probed by all devices.}
	\label{tab:probe_request_home}
\end{table}

\begin{figure}[h]
	\centering
	\includegraphics[width=\linewidth]{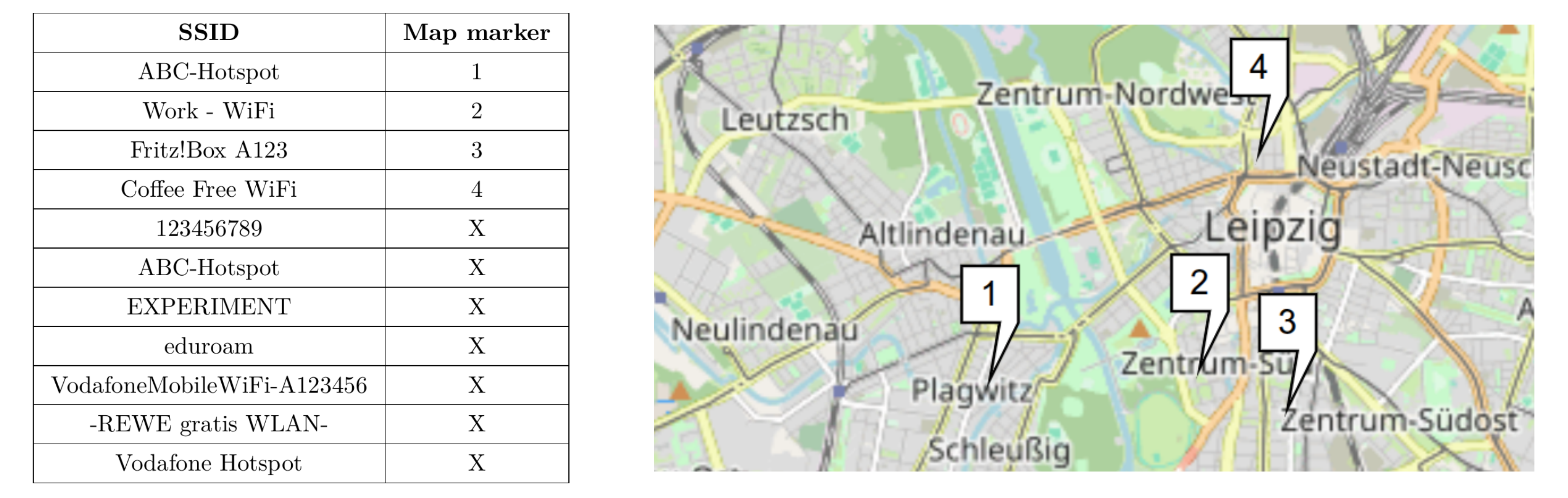}
	\caption{Probe requests captured during the tailgating attack on a smartphone. Map represents the (randomized) locations of these SSIDs provided by the WiGLE database.}
	\label{fig:probe_smartphone}
\end{figure}

During the preliminary expertiment, a tailgating attack was performed on the smartphone with mac address beginning with \textbf{a4:ff}. Figure \ref{fig:probe_smartphone} presents the observed probe requests and their GPS coordinates, retrieved from the WiGLE database. Not all probed networks were present in the wardriving dataset. Interestingly, the experiment network could not be localized, although its location is known beforehand. Some observed names are generic and can be correlated to multiple locations, for example "-REWE gratis WLAN-" is used by APs in various branches of a large grocery store chain with multiple locations in the city. Here, the attacker may assume that the neighbor visited the closest one. Additionally, we observed probes for eduroam, a popular university WiFi service. This information implies neighbor's association with the local university - they may be a student, scientific staff member or own a library card. We also observed probes for a network named after a café, where the true SSID contained the unaltered name of the place. WiGLE provided the correct coordinates of that location, but also its name alone was sufficient to identify the location through an internet search. One generic network name, i.e. Fritz!Box A123, was present in the probe request traffic. Its location could be looked up using the WiGLE database, however no additional meaning can be derived from the SSID alone. From the returned coordinates we see that the AP is positioned in a residential area, which implies that the victim could have visited someone in that area.

The most informative SSID observed is the network name of the victim's possible workplace. Its true name is very descriptive and the company appears as the first internet search result. The victim's possible workplace is a highly specialized IT company, which implies what education and occupation the victim might have. For example, if the company develops Internet of Things solutions, an attacker can hypothesize that the neighbor is employed as a software developer. 
 
Again, it is important to note that probe requests do not yield concrete information, as opposed to previous analyses, but only possible indicators. The fact that the victim's phone pings an IT company's network, does not confirm that they work there. Additionally, observed probe requests for a certain cafe or other public place does not reflect how often the victim visits these venues. However, as research suggests (see Section \ref{cha:fingerprinting}), even such weak evidence is sufficient to profile and draw possible conclusions or traits about the neighbor.

A nosy neighbor can learn information about the life of their victim beyond the smart home from observed probe requests. While acquiring a meaningful dataset of probed networks is challenging and requires alteration of the sniffing setup, the payoff can be great. In this experiment we could deduce that the victim is likely associated with the local university, we learned their possible occupation and using WiGLE database we could localize some places which the victim has visited.

\subsection{Schedule recognition}

Recording our gathered context information such as device activity and location over time, reveals inhabitants' daily and weekly schedules. In this section we present how a nosy neighbor can visualize and interpret patterns in device activity over time. We preset a reconstructed weekly schedule of an inhabitant during the preliminary experiment. 

\vspace{1em}
\begin{figure}[h]
	\centering
	\includegraphics[width=\linewidth]{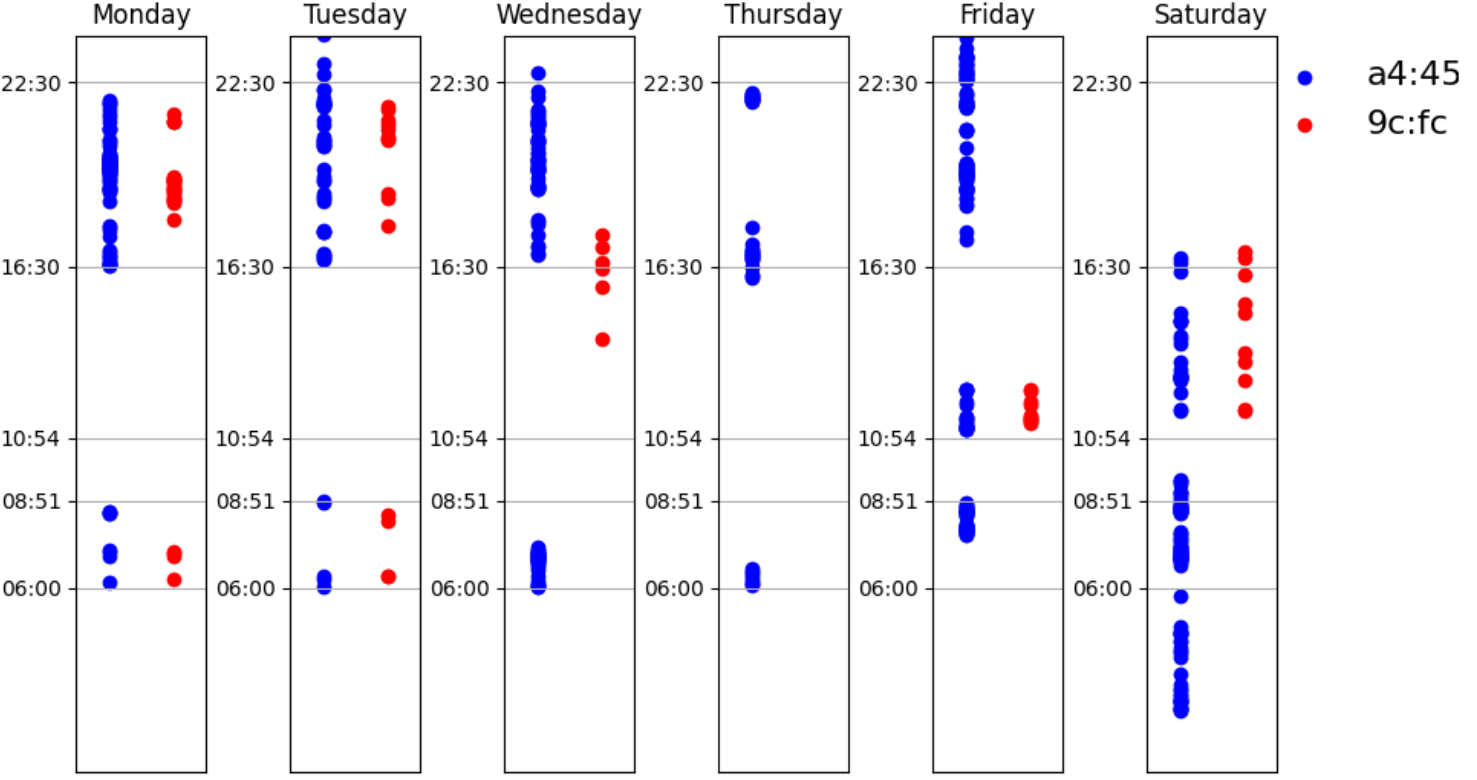}
	\caption{Smartphone's (blue) and laptop's (red) daily activity over one week during the preliminary experiment. }
	\label{fig:weekly_traffic}
\end{figure}

During the preliminary experiment (see Section \ref{sec:wifi-setup}), inhabitants of the smart home lived according to their regular schedule which involved attending to their activities outside of the apartment. Figure \ref{fig:weekly_traffic} presents the times at which the smartphone was in the network, i.e. it sent data packets. While this figure only shows the data of one week, these patterns are consistent over the entire monitoring period of three weeks. It is visible with the naked eye how the activity readings reflect similar patterns we identified in Section \ref{sec:dev-id}. We see that the device is inactive at night and that the activity is not distributed evenly throughout the day.

In this case, the time intervals with missing readings from the smartphone indicate that the inhabitant has left the house. This can be confirmed by the lack of activity of other devices. For example, if we know that the neighbor lives alone and we observe activity from other smart devices, it implies that the phone is turned off, but the inhabitant is still in the house. Preliminary dataset is very minimal and does not contain any smart devices which might reflect the inhabitant's presence, so we must rely on the laptop's traffic. 

The anomaly on Wednesday likely reflects some background tasks of the laptop. For example, the inhabitant didn't turn off their computer before leaving and it performed some task such as download of an update or file synchronization. Analysing the traffic fingerprint, as presented in Section \ref{sec:dev-state}, can confirm it, but this data is not available in the preliminary dataset. Lastly, the phone and computer activity on Sunday is not displayed on Figure \ref{fig:weekly_traffic}, because it is missing in the dataset entirely. This is due to the limitations of the setup, but also because the victim was outside of the house for the entire day. A better setup would have picked up some traffic in the early morning and then late at night.

Despite the minimal and incomplete dataset, clear patterns emerge. We notice that on most weekdays, the inhabitant's phone and computer begin to generate traffic at almost exactly 6:00. This is a strong indicator that the inhabitant wakes up at that time. Then again, for most weekdays we observe a gap from around 9:00 till 16:30. This implies that the inhabitant is outside. At this point we can utilize the inferred information about the victim to reason about what they are doing during that time. 

From the received probe request, we learned that the victim is likely associated with the local university. We can reason that the victim is a student and was attending their lectures. For example, if they had a lecture at 9:15 on the Leipzig University's Augustusplatz Campus, leaving at around 8:50 allows enough time to arrive using public transport or a bike. Augustusplatz is the largest university campus which hosts lectures for many programs. The same logic applies to their arriving time. If their last lecture of the day ends at 15:45, as per the usual schedule, it is plausible that they would get back at around 16:30. 

The observed device activity patterns look different on Friday. There, the inhabitant likely doesn't have lectures in the morning which allows them to sleep longer. Later, we see that they left the house at around 9:00 and then after 12:40. The duration of their first absence is too short for a lecture, so it is some other event which they attend regularly. Then, the second absence could be another lecture, given its duration and the returning time overlaps with other possible lecture attendance events. 

Lastly, two more anomalies in the schedule over three weeks, are visible in Figure \ref{fig:weekly_traffic}. Both occur in the evening, one on Thursday and the other on Saturday. Given the time of day, it is likely that these deviations are due to attendance at social events.

By observing the state of smart devices in the network, we can approximate when the inhabitants of a smart home are outside. Then, our gathered context information from the probe requests helps us fill the gaps in observed traffic and provides a basis for educated guesses about what the neighbor might be doing outside of the house. For example, if we deduce that the victim is associated with the university and leaves early enough to reach the university before their lecture begins, we can assume that they attend a lecture. It is important to reiterate that the presented reconstruction of the schedule is based on educated guesses which take into account all previously gathered context information. There is no feasible way to confirm that the victim actually attends university at these times using the data we have at hand.

\subsection{Human Activity Recognition}

In this section, we finally apply all of the presented methods and knowledge which the nosy neighbor can derive from the wireless traffic. We present to which extent human activity recognition can be performed using this data by reconstructing activities of the victim during one day. Since the state recognition results for smart devices were minimal, we will only evaluate multimedia devices such as computers, smart TVs and gaming consoles.

\begin{figure}[h]
	\centering
	\includegraphics[width=\linewidth]{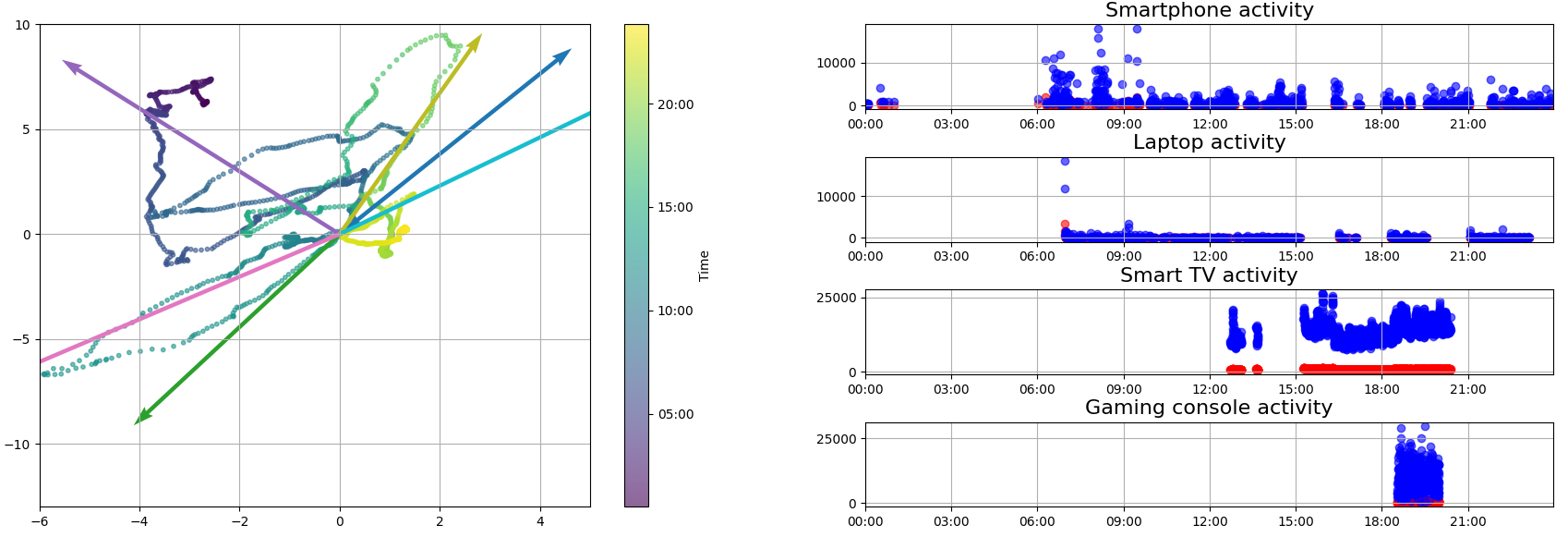}
	\caption{Smartphone position (left) and activity of multimedia devices (right) during one day. Blue dots represent downlink traffic and red dots represent uplink traffic. The y-axes of the traffic graphs denote the number of packets per second.}
	\label{fig:har_all_day}
\end{figure}

Figure \ref{fig:har_all_day} presents the observed position and activity throughout the entire day. Since this figure is complex, we first explain how to interpret the presented graphs. The graph on the left hand side represents the location of the victim's smartphone in real time, as discussed in Section \ref{sec:localization}. The position is color-coded to show the location at different times. It is important to reiterate that while we focus the location of the smartphone, which is not guaranteed to reflect the person's location, any other device which potentially reflects its owner's position more precisely can be analysed. Right hand side represents the network traffic of each device, like discussed in Section \ref{sec:dev-state}. The network traffic of the smart TV has been aggregated by applying a rolling window averaging algorithm. This is to highlight the different states of the TV. This processing is only necessary for the TV, as here the state change is not clearly visible in the raw data. 

As in previous analyses, it is immediately visible when the inhabitant was active. Similarly to the results in Section \ref{sec:dev-state}, the device activity indicates that their day begins roughly at 6:00 and ends in the late evening, likely around midnight. While the smartphone's position is difficult to interpret from this figure, we still identify distinct areas of the house where the phone spends time. Comparing this position estimate to Figure \ref{fig:phone_track}, we recognize that the presented location fingerprints overlap with the derived floor plan of the house. 

In order to gain more insight into the inhabitant's daily routine, we need zoom in on the time scale. We will discuss activity states of multimedia devices and the phone location at specific times of day to recognize human activity.

\begin{figure}[h]
	\centering
	\includegraphics[width=\linewidth]{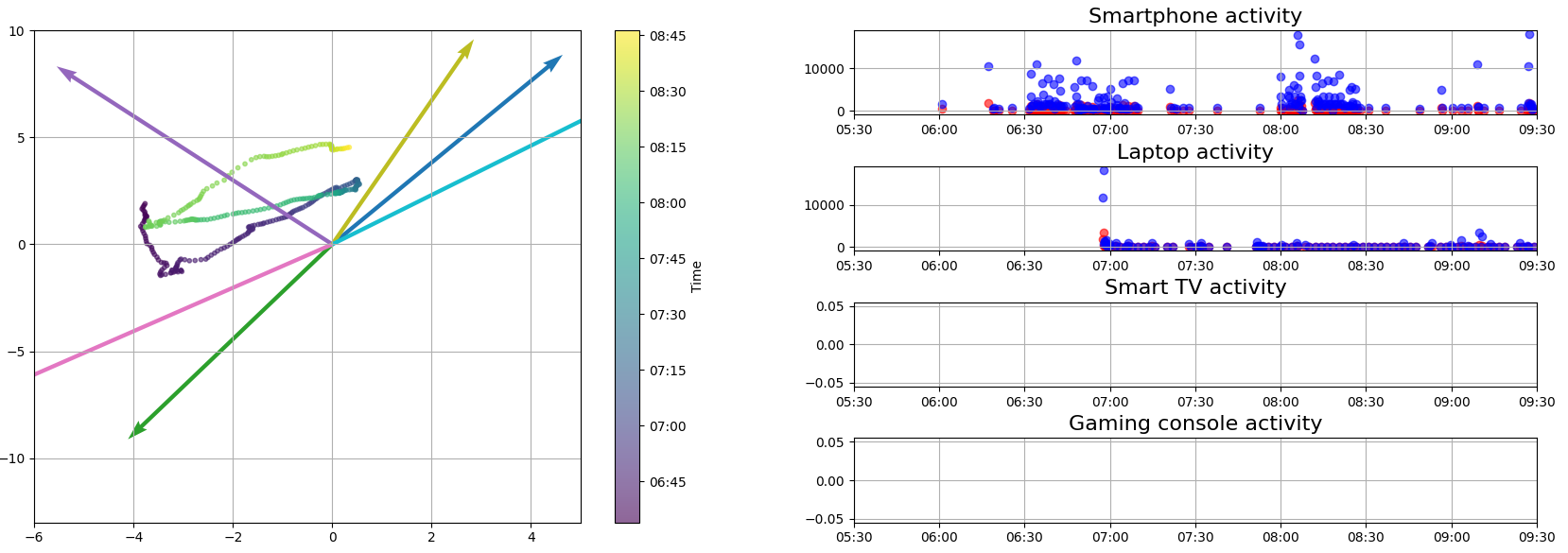}
	\caption{Smartphone position (left) and activity of multimedia devices (right) at the morning. Blue dots represent downlink traffic and red dots represent uplink traffic. The y-axes of the activity graphs denote the number of packets per second.}
	\label{fig:har_morning}
\end{figure}

First time interval analyzed is the morning. Figure \ref{fig:har_morning} shows the position and device's activities from 6:00 til 9:30. The estimated location coincides with the interpreted floor plan. During the first hour of the day, we see from the earliest readings that the phone is located in the general direction of the bedroom, later in estimated area of the kitchen/bedroom. From this we can interpret that the victim woke up, then went to the bathroom or kitchen to get ready and prepare breakfast. During this time, we see that the phone was generating mostly downlink network traffic. This can be interpreted as listening to music while preparing breakfast.

Shortly before 7:00, the phone's position shifts to the area interpreted as the office. At around the same time, we observe first network activity from the laptop. As discussed in Section \ref{sec:dev-state}, we see a typical spike in the traffic associated with booting of multimedia devices. These events suggest that the victim turned on their laptop and likely ate their breakfast while sitting at the computer. The fact that the throughput of the phone diminishes shortly after the computer turns on speaks for this hypothesis, suggesting that the user stopped using it. 

Reminder of the morning appears uneventful, except a 30 minute window after 8:00. Here, the phone was located in the general area of the bathroom or kitchen and produced a similar traffic pattern to its earlier active session. This can be interpreted as the victim cleaning up after breakfast or using the bathroom while streaming media.  

We observe no activity of other multimedia devices, which suggests that they weren't used during this time. It is plausible, as the estimated position of the victim did not overlap with the placement of these devices. Additionally, based on the reconstructed morning schedule, the victim did not have time to use them.

\clearpage

\begin{figure}[h]
	\centering
	\includegraphics[width=\linewidth]{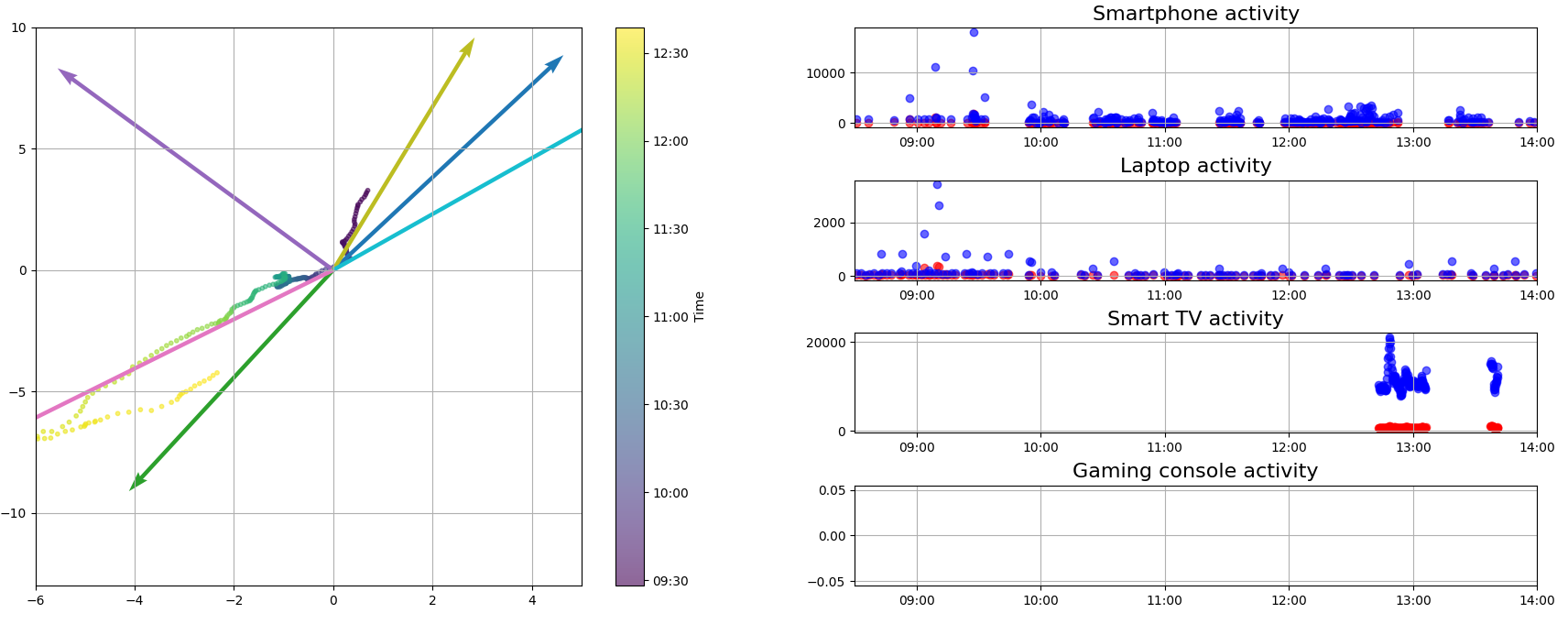}
	\caption{Smartphone position (left) and activity of multimedia devices (right) at noon. Blue dots represent downlink traffic and red dots represent uplink traffic. The y-axes of the activity graphs denote the number of packets per second.}
	\label{fig:har_mittag}
\end{figure}

Next relevant time interval describes the later morning and noon. It begins at 9:00 and ends at 14:00. Figure \ref{fig:har_mittag} It is important to note that during this time, the phone was located in the room which was excluded from the localization analysis (see Section \ref{sec:smart_home_setup}). This results in inaccurate position estimates, since the trilateration algorithm is not calibrated for that area. Smartphone position in this figure which appears to be in the bedroom, is actually in the sniffing room. 

The inhabitant's activity during this period differs visibly from the early morning. By examining the location and activity of the both phone and laptop at the beginning of this period, we observe relatively unchanging readings. The location fingerprint suggests that the user was most likely in the office. The laptop shows consistent network activity, however not typical for network-intensive activities. A plausible interpretation of these observations can be that the user was working on their computer. Minimal phone activity supports this claim, as it could just represent background processes of the smartphone, rather than active usage. At around 9:00, both the laptop and smartphone produce a spike in downlink network traffic. Its exact cause of, is difficult to explain, however it is most likely an anomaly or unrelated event. 

Around noon, the activity of the phone stops and the laptop shows reduced network usage. At the same time we observe a big spike in the smart TV's downlink traffic. Additionally, the smartphone's estimated location shifts towards the kitchen or bathroom area. Together, it is evident that the victim prepared their lunch and then consumed it in the excluded room while watching TV. 

The laptop's reduced network traffic persisted until the end of this period, which implies that the inhabitant remained in that room. This is also evident from the location fingerprint where we do not observe any significant movement into other areas. We see a brief pause in the TVs activity. A possible explanation can be that the inhabitant finished eating and took a break. Likely they washed the dishes before resuming their media consumption. We do not see this activity in the smartphone traffic and location readings, because the inhabitant likely didn't take their phone to the kitchen. 

\clearpage

\begin{figure}[h]
	\centering
	\includegraphics[width=\linewidth]{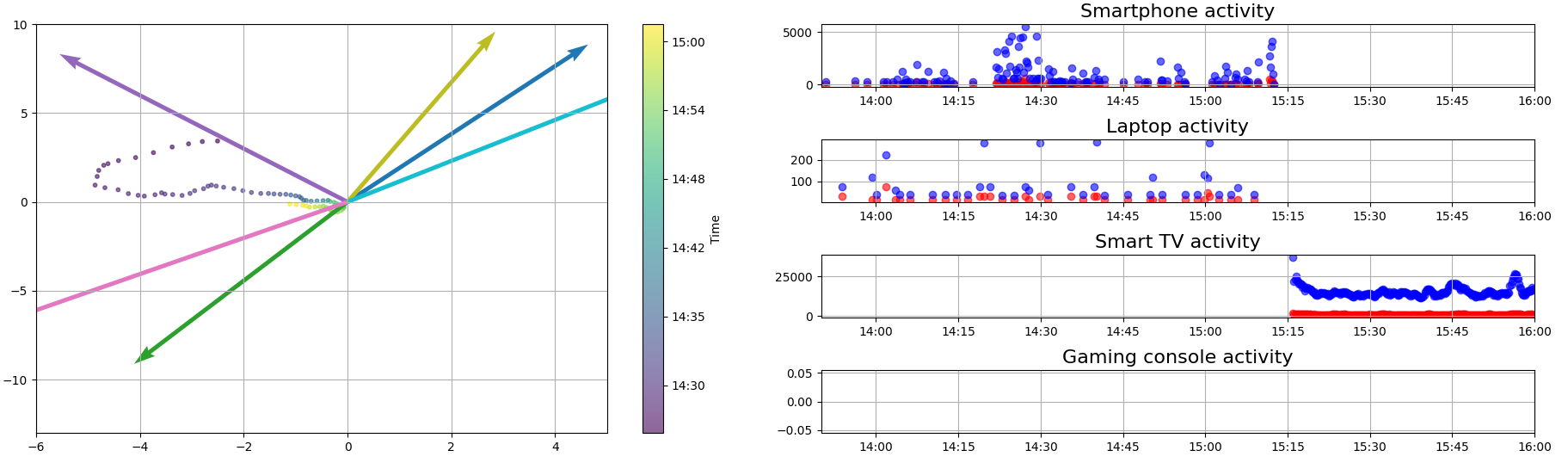}
	\caption{Smartphone position (left) and activity of multimedia devices (right) after lunch. Blue dots represent downlink traffic and red dots represent uplink traffic. The y-axes of the activity graphs denote the number of packets per second.}
	\label{fig:har_nachmittag}
\end{figure}

The most eventful period of the day in this case is undoubtedly the afternoon, as shown in Figure \ref{fig:har_nachmittag}. Here we observe the estimated position and traffic of multimedia devices from 14:00 to 16:00, i.e. after the lunch. As illustrated in Figure \ref{fig:har_all_day}, the phone's activity during the time from lunch until evening is incomplete, with visible gaps. This represents inactive periods of the phone. While this is an indicator of human activity, it is a limitation of our system as we can not reason about the inhabitant's location. 

Traffic readings of the laptop remains the same as before, indicating its working state. Phone's traffic fingerprint resembles the activity from the morning, i.e. we see a spike in the downlink traffic. By including the phone's location at that time we conclude that the inhabitant was most likely working while the location readings were unavailable and then entered the kitchen or bathroom area where they consumed some media on their phone. 

Afterwards, the inhabitant appears to return to the office, where they most likely continued their work until around 15:15. At that point, the phone and laptop stop generating traffic and the smart TV becomes active again. Due to the lack of 802.11 traffic from the smartphone, we lose the insight into the location of the user. However, by comparing the throughput of the TV to the earlier patter in Figure \ref{fig:har_mittag}, we see that it most likely streaming content again. This together describes the following possible scenario: after finishing their work, the inhabitant turned off their computer and phone and watched TV to relax. 

\clearpage

As stated in the previous paragraphs, the phone's network traffic is sporadic until the early evening. This suggests that the inhabitant was performing activities which do not involve their phone. Because of that, we omit the phone's estimated location for the time period between 16:00 and 21:00. During this time, the activity states of the other devices provide enough insight into the inhabitant's behavior. Moreover, due to the irregular nature of the readings, location estimates may be inaccurate or misleading. Jumps between different locations without interpolation make the reconstruction of events difficult.

\begin{figure}[h]
	\centering
	\includegraphics[width=\linewidth]{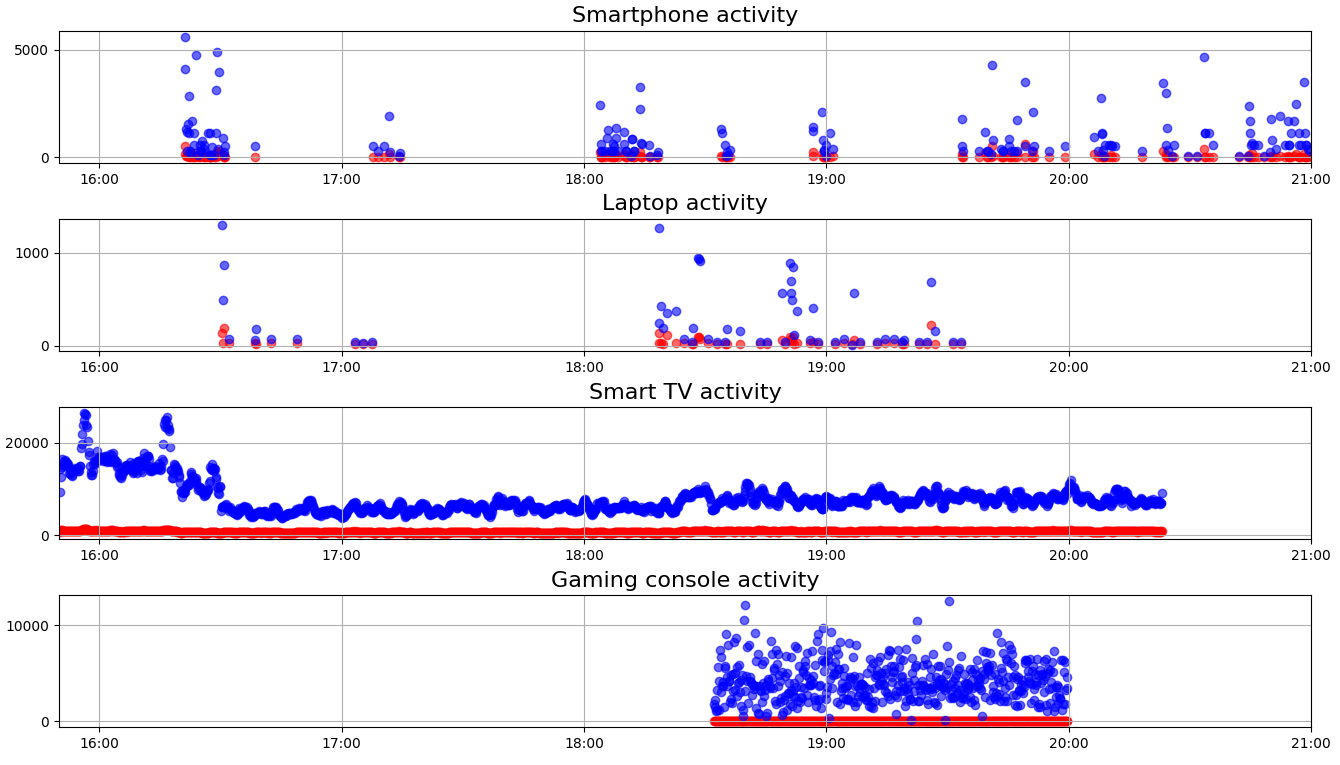}
	\caption{Activity states of multimedia devices (right) in the early evening. Blue dots represent downlink traffic and red dots represent uplink traffic. The y-axes denote the number of packets per second.}
	\label{fig:har_nachmittag_1}
\end{figure}

Figure \ref{fig:har_nachmittag_1} shows the network traffic of the multimedia devices during this period. At the beginning, we observe that the smart TV continues to actively generate downlink traffic like before, which can be interpreted as streaming content. Both the laptop and smartphone remain inactive until around 16:30, when a spike in their downlink network traffic is observed. At the same time, the throughput of the smart TV diminishes noticeably. 

The following scenario could replicate a similar traffic fingerprint: the inhabitant stopped actively watching the TV, but did not turn it off so it remained in an idle state. Afterwards, they briefly checked their phone and laptop shortly after. It is difficult to determine what the user was doing while all device were inactive. Better results of smart device state recognition, as presented in Section \ref{sec:har}, could provide more insight into what the user was doing at that time. For instance, we could use the activity of motion-activated devices such as motion sensors or cameras to approximate the user's location. Devices which reflect human activity in their network traffic, could also give insight into their activity. For example, if we detect a smart lightbulb being active, it is plausible that the inhabitant is in that room.  

This period of inactivity persisted until around 18:00. We still observe minor network activities of the smartphone and laptop, but they likely result from background processes of these devices. After 18:00, we see network activity from the phone and shortly after from the laptop. Next, we observe a slight increase of the network traffic of the smart TV, followed by activity of the gaming console. This is evidence that the inhabitant used their gaming console which was connected to the TV. 

Interestingly, during that time, the laptop produced a similar network traffic fingerprint to when it was being actively used in the morning. Together, the activity of these three devices, suggests the following scenario: the inhabitant plays a game on their Nintendo console, which is connected to the TV. Simultaneously, we see activity of the laptop possibly produces by accessing a tutorial or game wiki while playing. Finally, around 19:30 the inhabitant used their phone and stopped playing at about 20:00. The TV was turned off shortly after.

\begin{figure}[h]
	\centering
	\includegraphics[width=\linewidth]{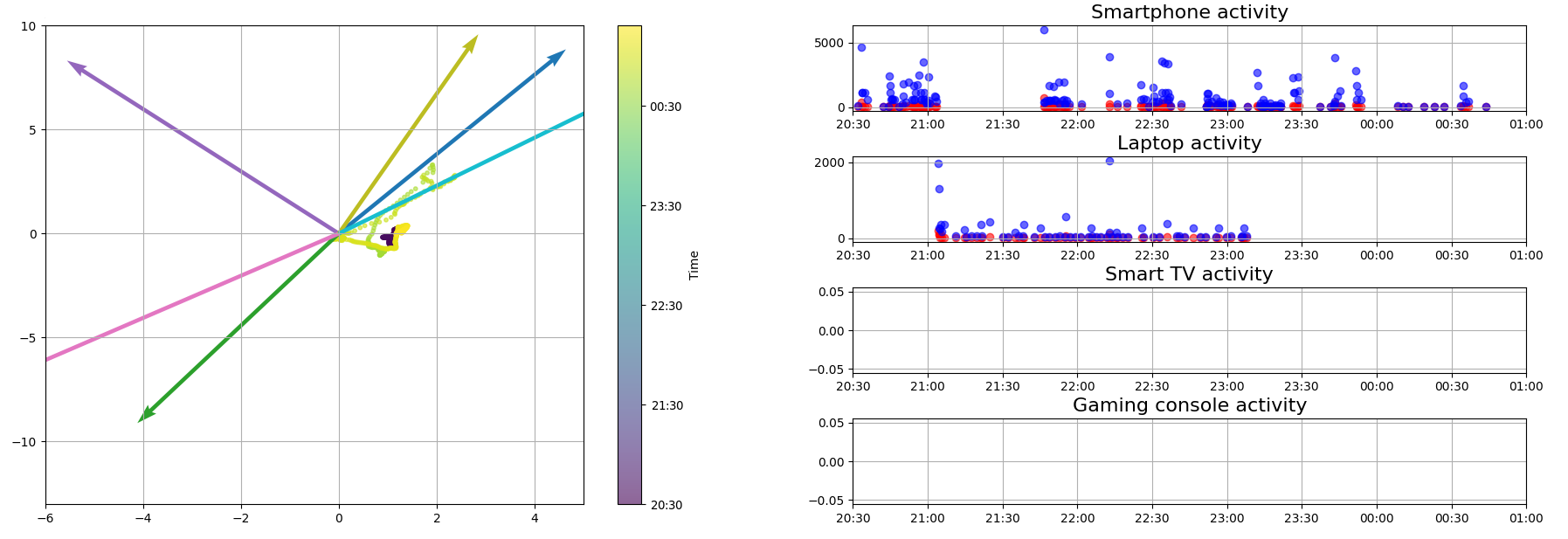}
	\caption{Smartphone position (left) and activity of multimedia devices (right) in the late evening. Blue dots represent downlink traffic and red dots represent uplink traffic. The y-axes of the activity graphs denote the number of packets per second.}
	\label{fig:har_night}
\end{figure}

The final period of the day to analyze is the late evening, beginning at 21:00 lasting until the inhabitant likely went to sleep. Figure \ref{fig:har_night} presents the phone's location and devices' activities. During this period, the phone readings are more consistent than before, which allows meaningful estimation of its location. Again, the phone was present in the room which was excluded from the location analysis. 

It is immediately noticeable that not much movement is detected by the sniffing setup. Only around 20:00 did the inhabitant leave the room, but remained in the excluded room for the remainder of the time. Interestingly, we don't see the device enter the kitchen or bathroom area in the evening, where we would expect the inhabitant to prepare their dinner or take a shower. Possibly, the victim either did not have their phone with them during those activities or it happened in the previous period where we did not include the localization features. 

Overall this time period is uneventful. We observe occasional spikes in the smartphone's activity and constant traffic from the laptop. A notable spike at the beginning of the laptop's network traffic fingerprint, implies that the laptop was turned on after being idle. It is likely that the user was using their computer from 21:00 until after 23:00, while checking their phone sporadically. The phone's readings cease after midnight, indicating when the inhabitant went to sleep. In contrast to previous findings presented in Section \ref{sec:localization}, the last recorded location of the phone was not the bedroom, but the sniffing room. This suggests that the user either left their phone there or possibly fell asleep in that room.

A nosy neighbor with access to inhabitant's location based on RSSI fingerprints of their mobile devices and states of other devices is able to reconstruct the daily schedule of their neighbor. We observe that in our experiment, multimedia devices reveal the most insight into activities performed by the inhabitants of a smart home. If state analysis of smart devices was more successful, methods for human activity recognition presented in Section \ref{sec:har} could be applied to obtain an even better resolution of the activities. Especially during times where the phone's network traffic was missing.  

Using the data which a nosy neighbor in this scenario has at hand, device-oriented activities are reliably detectable. Specifically, we observed when the victim was likely working on their computer, when they took breaks from work and, in most cases, what they did during their breaks. In addition to device activity, room-specific activities such as cooking or using the bathroom can be derived, however it is often challenging to estimate what exactly the victim is doing in these rooms with the data at hand.

\subsection{Case study - Guest Visit}
\label{sec:guest}

During the experiment, a two guests visited the apartment. They arrived in the evening, stayed overnight and left in the late morning of the following day. All inhabitants spent the time together performing different activities. This section will present a reconstruction of these events based on the observed traffic and known context information. We present how anomalies and deviations from the usual schedule manifest themselves in the dataset and what can be derived from previously unseen data. 

\begin{figure}[h]
	\centering
	\includegraphics[width=\linewidth]{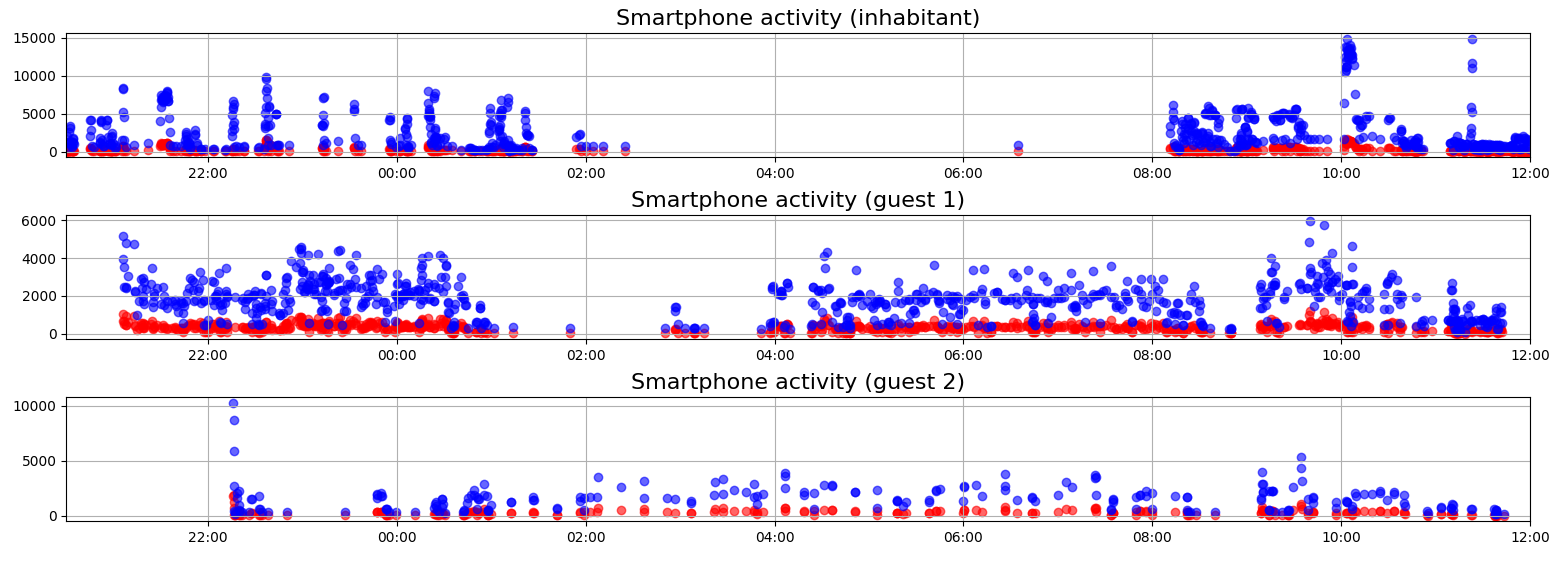}
	\caption{Network activity of the smartphones of the inhabitant and their guests. Blue dots represent downlink traffic and red dots represent uplink traffic. The y-axes denote the number of packets per second.}
	\label{fig:har_guests}
\end{figure}

Specifically, we demonstrate how a nosy neighbor could detect the presence of other persons in the victim's home. Then, we perform a human activity analysis similar to the previous section where we interpret both location information and device states to infer the behavior of the victim and their guests. Similar to the previous section, this analysis will be divided into two periods: the evening after the guest's arrival and the morning until the guests leave.

An attacker could infer that someone visits their neighbor from to the presence of new devices in the network. This is visible in the raw 802.11 traffic, but also in the probe request packets. Both guests' smartphones pinged the experiment network shortly after their arrival and the router's mac address was in the BSSID field of the data packets. No other meaningful probe requests from their phones could be observed. Unfortunately, due to the fact that both guests had an Apple device, the mac address resolution was unsuccessful. Apple typically randomizes the mac addresses of their devices such that the OUI field is not present within address. Figure \ref{fig:har_guests} presents the network activities of the inhabitant's and guests' smartphones. The traffic fingerprints of the new devices resemble those of the inhabitant's smartphone in previous analyses. Specifically, we observe irregularities such as short bursts of downlink traffic and non-continuous network activity. 

From Figure \ref{fig:har_guests}, it is noticeable with the naked eye when the guests arrived. Guest 1 arrives around 21:00 and Guest 2 joined later after 22:00. This figure also reveals the exact boundries of two periods in which the smart home residents were active. We see that typical smartphone activity stops at around 2:00. After that, some network activity is visible in the figure, however the fingerprints differ from the active phone usage. The traffic of Guest 1 is more uniform than during the active period and traffic fingerprint of Guest 2 is more sparse has less throughput. These differences likely reflect background processes of the smartphones. Different activities between the two devices, who's exact type or manufacturer is unknown to the nosy neighbor, can be explained by different models or apps installed on these devices. Throughout the entire visit, no activity in devices which typically reflect human activity was observed. This indicates that the residents' activities did not involve the multimedia devices.

Then, all three devices became active between 8:00 and 9:00. This shows when the residents most likely waked up. Later we see that the guests' network traffic stops before 12, which reveals when they left the apartment. Compared to the evening, the morning, network traffic fingerprints of the inhabitant and Guest 1 appear less active. This indicates that they used their phones less actively. 

Overall, the traffic patterns do not reveal much about the activities which were performed during this event. We observe occasional spikes in the network traffic of all devices, which typically indicate media streaming. One plausible interpretation is that the residents were showing each other videos on their smartphones. However, beyond this, it is unfeasible to draw any other meaningful conclusions from the traffic alone. Hence, we can examine the device's estimated location to acquire more context about the events which occurred during the visit. Due to sparse network activity of Guest 2's device, it will not be presented in this localization analysis. However, the sparse location clusters of the second guest tightly overlap with first guest's location.  
\clearpage

\begin{figure}[h]
	\centering
	\includegraphics[width=\linewidth]{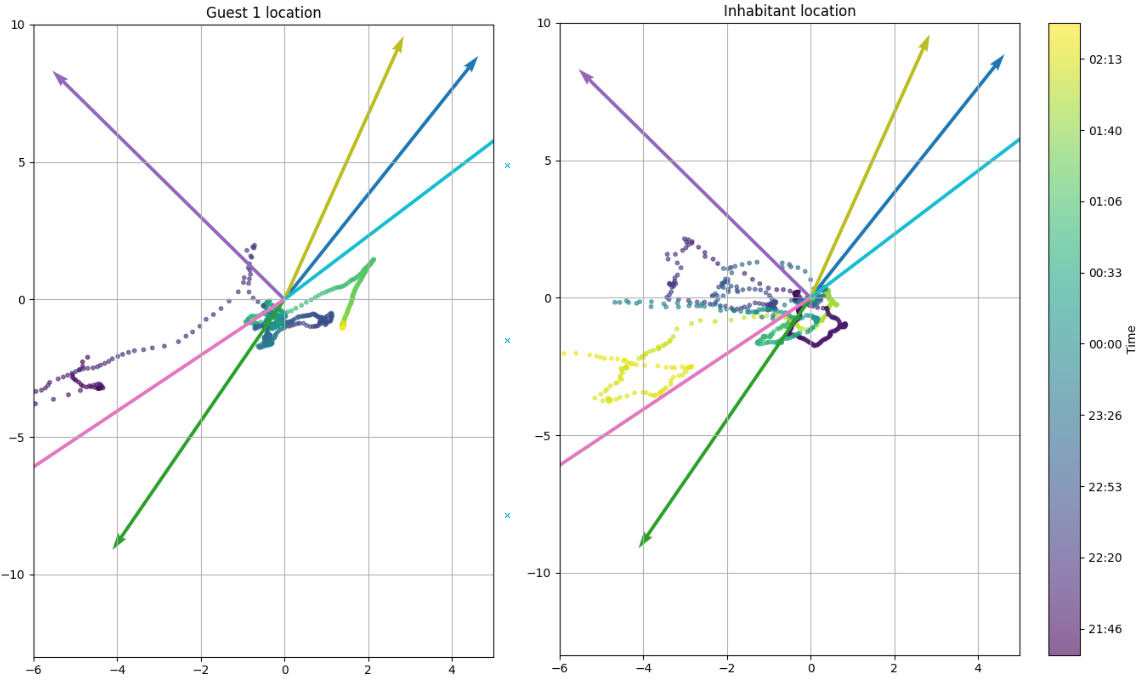}
	\caption{Location of the smartphones of Guest 1 (left) and inhabitant (right) during the evening of the visit.}
	\label{fig:har_locali_guest_evening}
\end{figure}

Figure \ref{fig:har_locali_guest_evening} shows the estimated locations of the smartphones belonging to Guest 1 and the inhabitant during the first period of the visit, specifically, between 21:30 and 2:30. As in previous HAR analyses, the both smartphones were in the area which was excluded from location analysis, i.e the living room.

Guest's smartphone was observed in three distinct areas after arrival. First, it entered the bedroom, where we see a relatively stationary location fingerprint. While we observe some movement, the majority of location estimates is concentrated around the coordinates (-5,-3). Later around 22:00, it briefly entered the kitchen. For the majority of that period, the device was in the living room, except around 1:00 where it entered the office for some time. Finally, it headed back to the living room. Based on the fact that it was its last observed location, we deduce that the guest slept in the living room. 

The inhabitant's location fingerprint clearly resembles that of their guest. We observe that the inhabitant spent more time in the kitchen at the beginning of the visit. But later we identify a similar pattern where the inhabitant was in the living room for the majority of the time and then went to sleep in their bedroom. 

In summary, considering the traffic patterns and localization analysis during the first period of the visit, we can reconstruct the following sequence of events. The inhabitant was in the office before their guests arrived. After that, the inhabitant was in the kitchen area and the guest's device was located in the bedroom. Given the stationary fingerprint of the guest's device, it was likely left in the bedroom, possibly to charge. Given the time of day and activity in the kitchen, we can interpret this as the residents preparing a meal. After that, they likely ate their meal in the living room while socializing. Network traffic reveals that they likely shared media on their phones. Finally, around 2:00 the inhabitant went to sleep in their bedroom and their guest slept in the living room.

\begin{figure}[h]
	\centering
	\includegraphics[width=\linewidth]{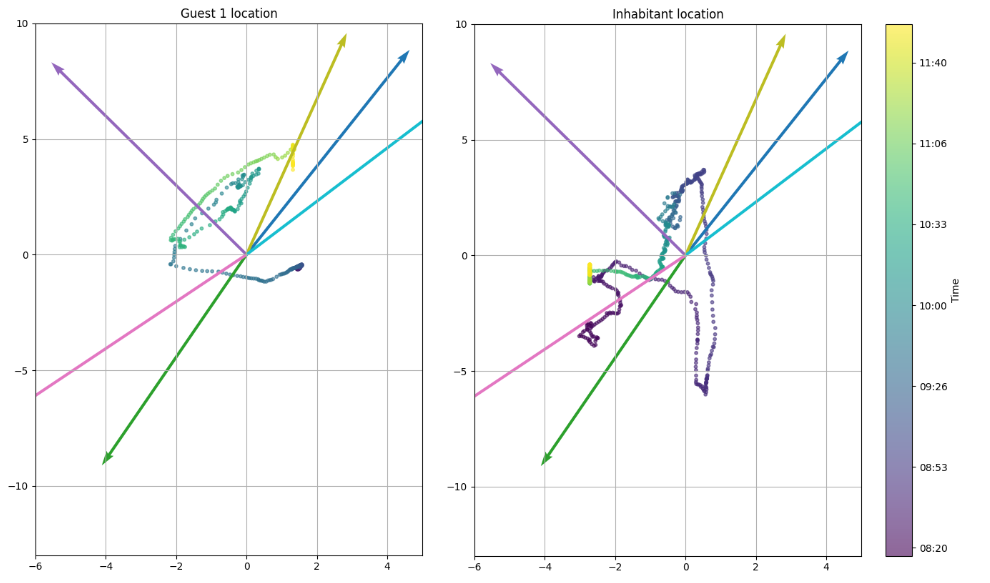}
	\caption{Location of the smartphones of Guest 1 (left) and inhabitant (right) at the morning of the visit.}
	\label{fig:har_locali_guest_morning}
\end{figure}

The location fingerprint of the guest's smartphone in the second period of the visit is comparatively less eventful. We observe that they woke up in the living room and remained there until around 10:00. After that they were observed in the kitchen/bathroom area where they helped prepare the breakfast or got ready. At the end, we see that they left at around 12:00, as their smartphone approached the location of the entry door. 

The inhabitant's location pattern is more complex. Initially, their device is detected in the bedroom, where they slept, followed by brief activity likely in the bathroom area. Around the time when their guest wakes up, the inhabitant's device is observed in the same room where their guest resides. This suggests socializing and using their phones, which is supported by the network traffic fingerprints in Figure \ref{fig:har_guests}. After that, we observe a stationary fingerprint, similar to the one of the guest's from the previous period. This suggests that the inhabitant left their phone in their office, likely for charging. At the end of the period, we detect the device in the similar area, i.e. kitchen/bathroom, as the guest.

While no concrete activities could be detected during the guest visit, it is possible to reconstruct a rough sketch of the events during that time. Data shows that the inhabitant and their guests most likely prepared a meal together in the kitchen and consumed it in the living room. All three parties remained in that room where they most likely socialized and consumed media on their smartphones. Changes in network traffic patterns of each devices disclose when each resident likely went to sleep and woke up. We observe that the inhabitant woke up first, followed by both guests waking up around the same time. This is plausible, as location analysis suggests that both guests slept in the living room. The events in the morning were typical - we observe that some point all residents were in the general kitchen/bathroom and spent time together socializing, as in the evening before. We finally see both guests leaving at the same time.

A nosy neighbor is capable of learning a surprising amount of information about their victim by passively observing their 802.11 traffic. Beginning with the plain-text elements of WiFi, i.e. probe requests, an attacker can get insight into the life of the victim beyond the smart home. The obtained information, such as association with public places, can be used to fill in contextual gaps in the analysis. By observing device presence over time, an attacker is capable of approximating a weekly schedule of their victim. Then, using the information from probe request analysis, we can reason what the neighbor is doing when they're absent. Furthermore, putting all of the gathered context information such as installed devices, their activities and an approximate floor plan of the apartment, opens possibilities for human activity recognition. Using key information such as position, time of day and network throughput, it is possible to detect activities such as cooking, working, consuming media or using the bathroom. Finally, the generated contextual framework is applicable on previously unseen devices. It is possible to detect visits and approximate which activities were performed by the inhabitants.

\chapter{Discussion}

\section{Summary of Results}
In this work we explored what a nosy neighbor in an adjacent apartment can learn about their victim by monitoring their wireless network traffic. We focused on WiFi and Bluetooth Low Energy, two popular communication protocols used in smart home environments. To emulate what the adversary can see, several smart home and multimedia devices were installed in an apartment and one room was equipped with spatially separated WiFi and BLE sniffers. 

We found multiple opportunities for device identification. While enumerating which devices are deployed in the victim's apartment is relatively uncomplicated, however determining their specific types requires more detailed analysis. For one, if the mac address of a device is not randomized, its OUI element reveals the manufacturer. In some cases, it is enough to classify the type of the device. Then, an adversary can analyze the wireless traffic which the device generates. Nearly all Bluetooth-capable devices which were considered in this work revealed their full name in BLE advertisements which led to complete identification. Patterns in WiFi traffic also reveal insights into the device type, because smart devices typically operate around the clock, while multimedia devices such as smartphones, are active at daytime. Finally, during the pairing and installation process, some devices open unencrypted WiFi networks. Listening on the traffic of these networks, identifying information about the device as well as the companion smartphone were revealed in all cases. 

By analyzing the network throughput of each device, different states of multimedia devices could be detected. This allows an attacker to see whether the devices are used and at what rate. This analysis was not successful for smart devices, as we could not detect any significant change in network traffic while the device was actively used. However for multimedia devices such as smartphones and smart TVs, we could classify their state as off, idle and active. 

The spatially separated setup opens possibilities for signal strength based tracking and localization analyses. We found that it is possible to distinguish between stationary and mobile devices. When trilaterating the signal strength readings onto an approximated 2D plane, we can estimate the general areas in which the devices are located. Using the gathered context knowledge, we can assign meaning into these areas and classify them as rooms. We find that this technique can be used to track the movement of mobile devices nearly in real time.

Then, by examining the subset of trusted networks of the victim's smartphone, we can gain an insight into their life beyond the smart home. We analyze probe request of their smartphone, which correlate with places they visit. Using this information as well as regularities in the absence of wearable devices, we derive a weekly schedule of their victim. Using the results of the probe request analysis, we reason about the possibilities of what the victim likely was doing during their absence. 

Finally we put all of the knowledge about the victim and their smart home together and perform human activity recognition by interpreting the observed events of one day. We find that activities which are related to WiFi-capable devices and behaviors which correlate with a specific area of the apartment can be detected.

\section{Practical Implications}

A nosy neighbor is able to peek inside a smart home and violate its privacy. While this attack requires patience, as most meaningful findings emerge from patterns observed over time, it is not complex. There are no prerequisites in the neighbor's smart home setup and we are not reliant on any specific devices which have an exploitable vulnerability. In this research we find that multimedia devices disclose most information about their owners, since their traffic patterns are the easiest to classify and, in this experiment, their activity correlate to user's behavior in a strong degree. While a smart lightbulb is usually operated a handful times per day, the inhabitants smartphone was in constant use.

The hardware needed for this kind of attack is cheap and readily available. In most cases, if the attacker does not intend to analyze the localization features of their victim's apartment, one WiFi capable modern computer suffices to monitor the traffic. While this experiment produced a big dataset of over 150 GB, an attacker with a dedicated configuration tailored to one environment which does not store the raw payloads can  vastly reduce their data footprint. This results in a relative and affordable minimum monitoring technical setup. The software needed for packet capturing and filtering is freely available and not complex to operate and configure according to the monitoring strategy. It makes the attack, from the technical perspective, relatively simple and accessible. 

However, the raw traffic dumps, are virtually meaningless for an attacker without pre-processing and visualizations. While again, the software and technologies are freely available and relatively accessible for attackers with programming experience, data processing foundations are necessary to extract meaningful information from the raw data. Furthermore, as mentioned in the previous paragraph, this analysis is multi-modal and relies heavily on context. For example, device presence information over three weeks was necessary to recognize repeating patterns in the victim's schedule. An attacker who wants to perform this analysis must be able to read and process raw pcap files, understand 802.11 addressing system to correctly identify the uplink and downlink data flows of devices and finally aggregate and present the data. For more complex analyses, machine learning methods for pattern recognition may be necessary. 

Then, localization analysis is a multi-step process involving multiple non-trivial techniques from the domain of linear algebra. Then, due to the black-box nature of this scenario, the attacker can only assume the possible positions of the devices, since any obstructions and exact signal parameters are not known. While in our case it was sufficient to acquire the general area in which the devices were located, it is not guaranteed that even this simplified localization recognition is successful or might require more advanced techniques.

Overall, the nosy neighbor attack is relevant and smart home owners should be aware of this possibility. While the technical setup is not particularly challenging to deploy, the inference phase of this attack is not trivial. The attacks on victim's privacy vary in difficulty, with the simpler ones being device identification and state recognition, to most difficult being probe request analyses and multi-modal behavioral fingerprinting. Despite the difficulty, a successful attack can be devastating to the users privacy. Then, due to the undetectability of the attack, the victim is unaware about the attacker knowing the intimate details about their life.

\section{Limitations}

In this study we emulated an environment of a smart home with spatially separated WiFi sniffers behind a wall. Despite a realistic obstruction in the form of a wall with electrical installations within it, the setup was still located in the same apartment. This could have an effect on the quality of the positioning analysis, resulting in better accuracy. While this setup had an advantage, it does not dispute the claim that such localization is possible. Since it relies highly on the geometry of both apartments and the area of the adjacent wall, an attacker might have better results by strategically distributing the sniffers over a greater area. This work did not evaluate the performance of this monitoring setup over other, more or less optimal architectures. 

To maximize the packet capture rate, all WiFi traffic was configured to occur in one pre-defined channel. In reality, the WiFi traffic of the victim would dynamically switch its communication channel to gain the best signal quality. An adversary would have to apply channel-hopping to follow the packets within the WiFi channel spectrum. They can use dedicated tools \cite{airodump} which tune the interfaces to currently utilized channel or use more antennas, each tuned to a different channel. 

Furthermore, for inhabitant localization, only their smartphone was used as a marker for their location. It does not necessarily reflect the true position in all smart home environments. As observed in Section \ref{sec:guest}, if the phone is left in an other room, an attacker does not have any possibility of determining the position of their victim. Alternatively, wearable devices such as smart watches, hearing aids or insulin patches can be used, as they can potentially reflect the true location more accurately.

Then, device state recognition of used smart devices was not successful. This is likely due to their type, such that the used devices do not necessarily reflect their state in the network traffic. Alternatively, it could be caused by a flaw in the data monitoring or processing pipeline. 

Many results in this work, as well as presented methods, rely on successful device identification. While the knowledge about each installed device was available to the attacker, it was not utilized in the analyses and only context acquired through the device identification methods was used. However, the experiment captured the traffic during the installation process of all WiFi-capable smart devices. As elaborated in Section \ref{sec:dev-id}, it is a rare event which requires a more sophisticated sniffing setup to capture. Had this information not been included in the experiment, device enumeration as well as localization analysis would be less successful.

\section{Future Work}

This work lays foundations for inference attacks on smart homes by listening to obstructed traffic. There are many variables in this setting which deserve to be explored and evaluated in detail. 

Literature review showed that localization analysis in obstructed, unknown environments is underrepresented. We find many works which present methods and framework for RSSI based localization, however all of them are based on a white-box pre-mapped scenarios. In the context of this work, such prerequisite is impossible to satisfy. One branch which needs further research is different sniffer setup architectures. Spatial separation of the sniffers, path-loss equation parameters and positioning techniques presented here did yield realistic results, however their efficiency was not evaluated. Then, this work only considered the scenario of analyzing the traffic of adjacent flats. We hypothesize that these techniques are also applicable to free standing houses. Lastly, potential of channel state information for localization analyses may be applicable in this setting. Since it can potentially yield results which are more accurate by orders of magnitudes, it is worth considering. 

Then, device state recognition from obstructed encrypted WiFi traffic in this work was reduced to distinguishing between active and passive. Research presented in Section \ref{cha:fingerprinting} suggests that it is possible to detect the specific traffic source, i.e. file download, media stream or chatting from traffic fingerprints. Device state recognition of multimedia devices can potentially be enhanced by applying these techniques. That way, an attacker would see beyond that the user is using their smartphone and could tell if the user is watching videos or video calling someone. 

Furthermore, defense strategies for this kind of attack should be evaluated. Given the relevance of this scenario, it is important to reason about possible countermeasures. Since this attack is multi-modal and attacks the traffic on different layers, its defense must be equally multi-modal and protect all presented inferred information.

\chapter{Conclusion}

This work investigates the possibilities for privacy-invasive inference attacks which a nosy neighbor behind a wall can perform on broadcasted traffic of a smart home. To do so, a smart home environment with various smart devices such as plugs and bulbs, as well as multimedia devices such as laptops and smartphones are installed in an apartment. Then, in a room behind a wall, WiFi and Bluetooth sniffers are installed which monitor the traffic. 

Using patters in the traffic, broadcasted protocol information and signal strengths, we are able to identify all of the used device at least to their general type of being a smart device or manually-controlled. Most devices can be classified further to their more specific kinds by analyzing their Bluetooth traffic. Having an overview of the installed devices, we examine their network traffic more closely to detect different states. Then, using the signal strength of these devices, we are able to calculate general areas where these devices are installed and classify them based on context information such as device activity at different times of day. Finally, we examine what can be learned about the inhabitants of a smart home. To do so, we present possibilities of probe request analyses and how it can be used to fill gaps in their weekly schedules. Lastly, we put the device and localization results together to present to which extent human activity can be observed in this setup.


\printbibliography

\addchap{Erklärung}
Ich versichere, dass ich die vorliegende Arbeit mit dem Thema:

\begin{center}
\textit{\glqq\titel\grqq}\\[1em]
\end{center}
			
selbständig und nur unter Verwendung der angegebenen Quellen und Hilfsmittel angefertigt habe, insbesondere sind wörtliche oder sinngemäße Zitate als solche gekennzeichnet. Mir ist bekannt, dass Zuwiderhandlung auch nachträglich zur Aberkennung des Abschlusses führen kann. Ich versichere, dass das elektronische Exemplar mit den gedruckten Exemplaren übereinstimmt.
\par
\ort, den \eingereicht

\rule[-0.2cm]{5cm}{0.5pt}

\textsc{\autor} 

\appendix
\clearpage
\renewcommand*{\thesection}{\Alph{section}} 
\pagenumbering{Roman}


\end{document}